\documentclass{dune} 
\pdfoutput=1            
\graphicspath{ {figures/}{executive-summary/figures/}{generated/} }

\def\expshort{DUNE\xspace}
\def\explong{The Deep Underground Neutrino Experiment\xspace}

\def\thedocsubtitle{Deep Underground Neutrino Experiment (DUNE)} 

\def\voltitlespfd{Volume 2: Single-Phase Module\xspace}
\def\voltitledpfd{Volume 3: Dual-Phase Module\xspace}

\def\volexecsumm{\textbf{Physics, Technologies, and Strategies\xspace}}

\newcommand{\numu}{$\nu_\mu$\xspace}
\newcommand{\nue}{$\nu_e$\xspace}

\newcommand{\anumu}{$\bar\nu_\mu$\xspace}
\newcommand{\anue}{$\bar\nu_e$\xspace}

\newcommand{\dm}[1]{$\Delta m^2_{#1}$\xspace} 

\newcommand{\sinst}[1]{$\sin^2\theta_{#1}$\xspace} 
\newcommand{\sinstt}[1]{$\sin^22\theta_{#1}$\xspace}  

\newcommand{\deltacp}{$\delta_{\rm CP}$\xspace}   
\newcommand{\mdeltacp}{$\delta_{\rm CP}$}   

\newcommand{\hyperk}{Hyper--Kamiokande\xspace}

\newcommand{\microboone}{MicroBooNE\xspace}
\newcommand{\minerva}{MINER$\nu$A\xspace}
\newcommand{\nova}{NO$\nu$A\xspace}

\newcommand{\lariat}{LArIAT\xspace}

\newcommand{\lartpc}{LArTPC\xspace}

\newcommand{\larsoft}{LArSoft\xspace}

\def\argon40{$^{40}$Ar}  
\def\Ar39{$^{39}$Ar}
\def\Cl40{$^{40}$Cl}
\def\K40{$^{40}$K}
\def\B8{$^{8}$B}
\newcommand\isotope[2]{\textsuperscript{#2}#1} 

\def\fdfiducialmass{\SI{40}{\kt}\xspace}

\def\larmass{\SI{17.5}{\kt}\xspace} 
\def\cryostatht{\SI{14.1}{\meter}\xspace} 
\def\cryostatlen{\SI{62.0}{\meter}\xspace} 
\def\cryostatwdth{\SI{14.0}{\meter}\xspace} 
\def\nominalmodsize{\SI{10}{kt}\xspace} 

\def\beamturnon{{2026}\xspace} 
\def\firstfdmodstartinstall{{2022}\xspace} 
\def\maincavernstartexc{{2019}\xspace} 
\def\pipiibeampower{\SI{1.2}{MW}\xspace}

\def\spmaxfield{\SI{500}{\volt/\centi\meter}} 
\def\spactivelarmass{\SI{10}{\kt}\xspace} 
\def\spmaxdrift{\SI{3.53}{\m}\xspace}

\def\sptargetdriftvoltpos{\SI{180}{kV}\xspace} 

\def\dpactivelarmass{\SI{12.096}{\kt}\xspace} 
\def\dpmaxdrift{\SI{12}{\m}\xspace} 
\def\dpnumcrpch{\num{153600}\xspace} 
\def\dpnominaldriftfield{\SI{500}{V/cm}\xspace} 
\def\dptargetdriftvoltpos{\SI{600}{kV}\xspace} 

\def\surffnalbw{\SI{100}{\Gbps}\xspace}
\newcommand{\fnal}{Fermilab\xspace}
\newcommand{\surf}{SURF\xspace}

\newcommand{\efield}{E field\xspace}
\newcommand{\lbl}{long-baseline\xspace}

\newcommand{\threed}{3D\xspace}
\newcommand{\twod}{2D\xspace}

\newcommand{\detmodule}{detector module\xspace}
\newcommand{\dual}{DP\xspace}

\newcommand{\single}{SP\xspace}

\newcommand{\dpmod}{DP detector module\xspace}
\newcommand{\spmod}{SP detector module\xspace}

\newcommand{\lar}{LAr\xspace}

\newcommand{\phel}{photoelectron\xspace}

\DeclareSIUnit \s {\second}
\DeclareSIUnit \MB {\mega\byte}
\DeclareSIUnit \GB {\giga\byte}
\DeclareSIUnit \TB {\tera\byte}
\DeclareSIUnit \PB {\peta\byte}
\DeclareSIUnit \Mbps {\mega\bit/\s}
\DeclareSIUnit \Gbps {\giga\bit/\s}
\DeclareSIUnit \Tbps {\tera\bit/\s}
\DeclareSIUnit \Pbps {\peta\bit/\s}
\DeclareSIUnit \kton {\kilo\tonne} 
\DeclareSIUnit \kt {\kilo\tonne}
\DeclareSIUnit \Mt {\mega\tonne}
\DeclareSIUnit \eV {\electronvolt}
\DeclareSIUnit \keV {\kilo\electronvolt}
\DeclareSIUnit \MeV {\mega\electronvolt}
\DeclareSIUnit \GeV {\giga\electronvolt}
\DeclareSIUnit \m {\meter}
\DeclareSIUnit \cm {\centi\meter}
\DeclareSIUnit \in {\inchcommand}
\DeclareSIUnit \km {\kilo\meter}
\DeclareSIUnit \kV {\kilo\volt}
\DeclareSIUnit \kW {\kilo\watt}
\DeclareSIUnit \MW {\mega\watt}
\DeclareSIUnit \MHz {\mega\hertz}
\DeclareSIUnit \mrad {\milli\radian}
\DeclareSIUnit \year {year}
\DeclareSIUnit \POT {POT}
\DeclareSIUnit \sig {$\sigma$}
\DeclareSIUnit\parsec{pc}
\DeclareSIUnit\lightyear{ly}
\DeclareSIUnit\foot{ft}
\DeclareSIUnit\ft{ft}
\DeclareSIUnit \ppb{ppb}
\DeclareSIUnit \ppt{ppt}
\DeclareSIUnit \samples{S}

\def\ktMWyr{\si[inter-unit-product=\ensuremath{{}\cdot{}}]{\kt\MW\year}\xspace}

\newcommand{\SIadj}[2]{\SI{#1}{#2}}

\newcommand{\ktadj}[1]{\SIadj{#1}{\kt}}
\newcommand{\kmadj}[1]{\SIadj{#1}{\km}}

\newcommand{\GeVadj}[1]{\SIadj{#1}{\GeV}}
\newcommand{\MWadj}[1]{\SIadj{#1}{\MW}}

\usepackage[toc]{glossaries}
\makeglossaries

\newcommand{\dshort}[1]{\glsentrytext{#1}}

\newcommand{\dlong}[1]{\glsentrylong{#1}}

\newcommand{\dword}[1]{\gls{#1}}
\newcommand{\dwords}[1]{\glspl{#1}}
\newcommand{\Dword}[1]{\Gls{#1}}

\newcommand{\newduneword}[3]{
    \newglossaryentry{#1}{
        text={#2},
        long={#2},
        name={\glsentrylong{#1}},
        first={\glsentryname{#1}},
        firstplural={\glsentrylong{#1}\glspluralsuffix},
        description={#3}
    }
}

\newcommand{\newduneabbrev}[4]{
  \newglossaryentry{#1}{
    text={#2},
    long={#3},
    shortplural={{#2}\glspluralsuffix},
    longplural={{#3}\glspluralsuffix{}},
    name={\glsentrylong{#1}{} (\glsentrytext{#1}{})},
    first={\glsentryname{#1}},
    firstplural={\glsentrylong{#1}\glspluralsuffix{} (\glsentrytext{#1}\glspluralsuffix{})},
    description={#4}
  }
}

\newcommand{\newduneabbrevs}[5]{
  \newglossaryentry{#1}{
    text={#2},
    long={#3},
    plural={#4},
    shortplural={{#2}\glspluralsuffix},
    longplural={#4},
    name={\glsentrylong{#1}{} (\glsentrytext{#1}{})},
    first={#3 (#2)},
    firstplural={#4 (\glsentrytext{#1}\glspluralsuffix{})},
    description={#5}
  }
}

\newduneword{dword}{DUNE Word}{A term in the DUNE lexicon}

\newduneabbrev{nd}{ND}{near detector}{Refers to the detector(s) 
 installed close to the neutrino source at \fnal }

\newduneabbrev{fd}{FD}{far detector}{Refers to the \fdfiducialmass fiducial mass DUNE detector 
  to be installed at the far site at \surf in
  Lead, SD, to be composed of four \nominalmodsize modules}

\newduneabbrev{sp}{SP}{single-phase}{Distinguishes one of the DUNE far detector technologies by the fact that it operates using argon in its liquid phase only
  }

\newduneabbrev{dp}{DP}{dual-phase}{Distinguishes one of the DUNE far detector technologies by the fact that it operates using argon 
 in both gas and liquid phases}

\newduneabbrev{pds}{PDS}{photon detection system}{The detector 
  subsystem sensitive to light produced in the \lar }

\newduneabbrev{tpc}{TPC}{time projection chamber}{The portion of each
  DUNE \gls{detmodule} that records ionization electrons
  after they drift away from a cathode through the \lar, and also
  through gaseous argon in a \dual module. 
  The activity is recorded by digitizing the waveforms of current
  induced on the anode as the distribution of ionization charge passes by
  or is collected on the electrode}

\newduneabbrevs{apa}{APA}{anode plane assembly}{anode plane assemblies}{A unit of the \single
  detector module containing the elements sensitive to ionization in the \lar. 
  It contains two faces each of three planes of wires, and interfaces to the cold
  electronics and photon detection system} 

\newduneabbrev{cro}{CRO}{charge readout}{The system for detecting
  ionization charge distributions in a \dual detector module}

\newduneabbrev{lro}{LRO}{light readout}{The system for detecting
  scintillation photons in a \dual  detector module}

\newduneabbrev{fe}{FE}{front-end}{The front-end refers a point that is
  ``upstream'' of the data flow for a particular subsystem. 
  For example the front-end electronics is where the cold electronics
  meet the sense wires of the TPC and the front-end DAQ is where the
  DAQ meets the output of the electronics}

\newduneabbrev{adc}{ADC}{analog-to-digital converter}{A sampling of a voltage
  resulting in a discrete integer count corresponding in some way to
  the input}

\newduneabbrev{daq}{DAQ}{data acquisition}{The data acquisition system
  accepts data from the detector FE electronics, buffers
  the data, performs a \gls{trigdecision}, builds events from the selected
  data and delivers the result to the offline \gls{diskbuffer}}

\newduneword{detmodule}{detector module}{The entire DUNE far detector is
  segmented into four modules, each with a nominal \SI{10}{\kton}
  fiducial mass}

\newduneword{detunit}{detector unit}{A \gls{submodule} may be partitioned
  into a number of similar parts. 
  For example the single-phase TPC \gls{submodule} is made up of APA
  units}

\newduneword{diskbuffer}{secondary DAQ buffer}{A secondary
  \dshort{daq} buffer holds a small subset of the full rate as
  selected by a \gls{trigcommand}. 
  This buffer also marks the interface with the DUNE Offline}

\newduneabbrev{om}{OM}{online monitoring}{Processes that run inside
  the DAQ on data ``in flight,'' specifically before landing on the
  offline disk buffer, and that provide feedback on the operation of
  the DAQ itself and the general health of the data it is marshalling}

\newduneabbrev{dqm}{DQM}{data quality monitoring}{Analysis of the raw
  data to monitor the integrity of the data and the performance of the
  detectors and their electronics. This type of monitoring may be
  performed in real time, within the \gls{daq} system, or in later
  stages of processing, using disk files as input}

\newduneword{dumpbuffer}{DAQ dump buffer}{This \dshort{daq} buffer
  accepts a high-rate data stream, in aggregate, from an associated
  \gls{submodule} sufficient to collect all data likely relevant to
  a potential \gls{snb}}

\newduneabbrev{etl}{ETL}{external trigger logic}{Trigger processing
  that consumes \gls{detmodule} level \gls{trignote} information
  and other global sources of trigger input and emits
  \gls{trigcommand} information back to the \glspl{mtl}}

\newduneword{trignote}{trigger notification}{Information provided by
  \gls{mtl} to \gls{etl} about \gls{trigdecision} its processing}

\newduneword{trigprimitive}{trigger primitive}{Information derived by
  the DAQ \gls{fe} hardware that describes a region of space (e.g.,
  one or several neighboring channels) and time (e.g., a contiguous set
  of ADC sample ticks) associated with some activity}

\newduneword{externtrigger}{external trigger candidate}{Information
  provided to the \gls{mtl} about events external to a
  \gls{detmodule} so that it may be considered in forming
  \glspl{trigcommand}}

\newduneabbrev{daqoob}{OOB dispatcher}{out-of-band trigger command
  dispatcher}{This component is responsible for dispatching a \gls{snb} dump
  command to all \glspl{daqfer} in the \gls{detmodule}}

\newduneabbrev{mtl}{MTL}{module trigger logic}{Trigger processing
  that consumes \gls{detunit} level \gls{trigcommand} information
  and emits \glspl{trigcommand}. 
  It provides the \gls{etl} with \glspl{trignote} and receives back any
  \glspl{externtrigger}}

\newduneword{octant}{octant}{Any of the eight parts into which 4$\pi$
  is divided by three mutually perpendicular axes. 
  In particular in referencing the value for the mixing angle
  $\theta_{23}$}

\newduneword{submodule}{subdetector}{A detector unit of granularity less
  than one \gls{detmodule} such as the TPC of either a \single or \dual 
  module}

\newduneword{trigcandidate}{trigger candidate}{Summary information derived
  from the full data stream and representing a contribution toward
  forming a \gls{trigdecision}}

\newduneword{trigcommand}{trigger command}{Information derived from
  one or more \glspl{trigcandidate}  that directs elements of the
  \gls{detmodule} to read out a portion of the data stream}

\newduneabbrev{tcm}{TCM}{trigger command message}{A message flowing
  down the trigger hierarchy from global to local context.  Also see \gls{tpm}}

\newduneword{trigdecision}{trigger decision}{The process by which
  \glspl{trigcandidate} are converted into \glspl{trigcommand}}

\newduneabbrev{tpm}{TPM}{trigger primitive message}{A message flowing
  up the trigger hierarchy from local to global context.  Also see \gls{tcm}}

\newduneabbrev{eb}{EB}{event builder}{A software agent servicing one
  \gls{detmodule} by executing \glspl{trigcommand} by reading out
  the requested data}

\newduneabbrev{cob}{COB}{Cluster On Board}{An ATCA motherboard housing four RCEs}

\newduneabbrev{rce}{RCE}{reconfigurable computing element}{Data processor located outside of the cryostat on a \gls{cob} which contains \gls{fpga}, RAM and SSD resources, responsible for buffering data, producing trigger primitives, responding to triggered requests for data and sinking \gls{snb} dumps}

\newduneabbrev{bow}{BOW}{Bump On Wire}{A working name for the front-end readout computing elements used in the nominal DAQ design to interface the \dual  crates to the DAQ front-end computers}

\newduneabbrev{atca}{ATCA}{Advanced Telecommunications Computing
  Architecture}{An advanced computer architecture specification developed for the telecommunications, military, and aerospace industries that incorporates the latest trends high-speed interconnect technologies, next-generation processors, and improved reliability, availability and serviceability} 

\newduneabbrev{utca}{$\mu$TCA}{Micro Telecommunications Computing Architecture}{The computer architecture specification followed by the crates that house charge and light readout electronics in the dual-phase module} 

\newduneabbrev{udp}{UDP}{user datagram protocol}{A simple,
  connectionless Internet protocol that supports data integrity
  checksums, requires no handshaking, and does not guarantee packet delivery}

\newduneabbrev{amc}{AMC}{advanced mezzanine card}{Holds digitizing
  electronics and lives in \gls{utca} crates}

\newduneabbrev{rf}{RF}{radio frequency}{Electromagnetic emissions
  that are within the (radio) frequency band of sensitivity of the detector
  electronics}

\newduneabbrev{fpga}{FPGA}{field programmable gate array}{An
integrated circuit technology that allows the hardware to be reconfigured to
execute different algorithms after its manufacture and deployment}

\newduneabbrev{felix}{FELIX}{Front-End Link eXchange}{A
  high-throughput interface between front-end and trigger electronics
  and the standard PCIe computer bus}

\newduneword{daqpart}{DAQ partition}{A cohesive and
 coherent collection of DAQ hardware and software working together to trigger and read out some portion of one detector module; it consists of an integral number of \glspl{daqfrag}. 
 Multiple DAQ partitions may operate simultaneously, but each instance operates independently}

\newduneabbrev{fec}{FEC}{front-end computer}{The portion of one
  \gls{daqpart} that hosts the \gls{daqdr}, \gls{daqbuf} and
  \gls{daqds}.  It is connected to the \gls{daqfer} via fiber optic. 
Each \gls{detunit} of a  certain granularity, such as two \single \glspl{apa}, has one front-end computer
  that receives data from the readout hardware, hosts the primary DAQ
  memory buffer for that data, emits trigger candidates derived from
  that data, and satisfies requests for producing subsets of that data
  for egress}

\newduneword{daqfrag}{DAQ front-end fragment}{The portion of one
  \gls{daqpart} relating to a single \gls{fec} and corresponding to an
  integral number of \glspl{detunit}.  See also \gls{datafrag}}

\newduneword{datafrag}{data fragment}{A block of data read out from a single \gls{daqfrag} that
span a contiguous period of time as requested by a \gls{trigcommand}}

\newduneabbrev{daqfer}{FER}{DAQ front-end readout}{The portion of a
  \gls{daqfrag} that accepts data from the detector electronics and
  provides it to the \gls{fec}. 
  In the nominal design it is also responsible for generating channel
  level \glspl{trigprimitive}}

\newduneabbrev{daqdr}{DDR}{DAQ data receiver}{The portion of the
  \gls{daqfrag} that accepts data from the \gls{daqfer}, emits
  trigger candidates produced from the input trigger primitives, and
  forwards the full data stream to the \gls{daqbuf}}

\newduneabbrev{daqbuf}{primary buffer}{DAQ primary buffer}{The portion
  of the \gls{daqfrag} that accepts full data stream from the
  corresponding \gls{detunit} and retains it sufficiently long for it
  to be available to produce a \gls{datafrag}}

\newduneword{daqds}{data selector}{The portion of the \gls{daqfrag}
  that accepts \glspl{trigcommand} and returns the corresponding
  \gls{datafrag}}

\newduneabbrev{femb}{FEMB}{front-end mother board}{Refers a unit of
  the \gls{sp} cold electronics that contains the front-end amplifier
  and ADC ASICs covering 128 channels}

\newduneword{asic}{ASIC}{application-specific integrated circuit}

\newduneword{lv}{LV}{low voltage}

\newduneword{coldata}{COLDATA}{a 64-channel control and communications ASIC}

\newduneword{larasic}{LArASIC}{A 16-channel front-end ASIC that provides signal amplification and pulse shaping}

\newduneword{cmos}{CMOS}{Complementary metal-oxide-semiconductor}

\newduneword{enc}{ENC}{equivalent noise charge}

\newduneword{dre}{DRE}{dynamic range enhancement}

\newduneword{sar}{SAR}{successive approximation register}

\newduneword{protodune}{ProtoDUNE}{Either of the two DUNE prototype detectors constructed at CERN and operated in
  a CERN test beam (expected fall 2018). 
  One prototype implements \single and the other \dual technology}

\newduneword{pdsp}{ProtoDUNE-SP}{The \single ProtoDUNE detector}

\newduneword{pddp}{ProtoDUNE-DP}{The \dual ProtoDUNE detector}

\newduneword{wa105}{WA105 DP demonstrator}{The \SI[product-units=power]{3x1x1}{m} WA105 dual-phase prototype detector at CERN}

\newduneword{rawevent}{DAQ event block}{The unit of data output by the
  DAQ. 
  It contains trigger and detector data spanning a unique, contiguous
  time period and a subset of the detector channels.}

\newduneabbrev{ssd}{SSD}{solid-state disk}{Any storage device that
  may provide sufficient write throughput to receive, both collectively and
  distributed, the sustained full rate of data from a \gls{detmodule}
  for many seconds}

\newduneabbrev{hlt}{HLT}{high-level trigger}{A source of triggering at
  the module level}

\newduneabbrev{pid}{PID}{particle ID}{Particle identification}

\newduneword{readout window}{readout window}{A fixed, atomic and
  continuous period of time over which data from a \gls{detmodule}, in
  whole or in part, is recorded. 
  This period may differ based on the trigger that initiated the
  readout}

\newduneabbrev{zs}{ZS}{zero-suppression}{Used to delete some portion of a
  data stream that does not significantly deviate from zero or
  intrinsic noise levels. 
  It may be applied at different granularity from per-channel to per
  \dword{detunit}}

\newduneabbrev{rc}{RC}{run control}{The system for configuring,
  starting and terminating the DAQ}

\newduneabbrev{snb}{SNB}{supernova neutrino burst}{A prompt 
  increase in the flux of low-energy neutrinos emitted in the first few seconds of a core-collapse supernova.  It can also refer to a trigger command type that may be due to an SNB,
  or detector conditions that mimic its interaction signature}

\newduneabbrev{snble}{SNB/LE}{supernova neutrino burst and low
  energy}{Supernova neutrino burst and low-energy physics program}

\newduneabbrev{snews}{SNEWS}{SuperNova Early Warning System}{A global
  supernova neutrino burst trigger formed by a coincidence of SNB
  triggers collected from participating experiments}

\newduneabbrev{pps}{1PPS signal}{one-pulse-per-second signal}{An
  electrical signal with a fast rise time and that arrives in real
  time with a precise period of one second}

\newduneabbrev{sls}{SLS}{spill location system}{A system residing at
  the DUNE far detector site that provides information, possibly
  predictive, indicating periods of time when neutrinos are being
  produced by the \fnal Main Injector beam spills}

\newduneabbrev{wib}{WIB}{warm interface board}{Digital electronics
  situated just outside the \single cryostat that receives digital data
  from the FEMBs over cold copper connections and sends it to the RCE
  FE readout hardware}

\newduneabbrev{nic}{NIC}{network interface controller}{Hardware for controlling the interface to a communication network.  Typically, one that obeys the Ethernet protocol}

\newduneabbrev{wiec}{WIEC}{warm interface electronics crate}{Crates mounted on the signal flanges that contain the warm interface boards}

\newduneabbrev{ptc}{PTC}{power and timing cards}{Cards that provide further processing and distribution of the signals entering and exiting the \single cryostat}

\newduneabbrev{sipm}{SiPM}{silicon photomultiplier}{A solid-state
  avalanche photodiode sensitive to single \phel signals}

\newduneabbrev{cisc}{CISC}{cryogenic instrumentation and slow controls}{A DUNE
  consortium responsible for the cryogenic instrumentation and slow controls components}

\newduneword{consortium}{consortium}{A unit of organization in the
  DUNE project focused on one major component of the far detector}

\newduneword{art}{\textit{art}}{A software framework implementing an
  event-based execution paradigm} 

\newduneabbrev{sam}{SAM}{sequential
  access via metadata}{A data-handling system to store and retrieve
  files and associated metadata, including a complete record of the
  processing that has used the files}

\newduneword{artdaq}{\textit{art}DAQ}{A general-purpose \fnal data acquisition framework for distributed data combination and processing. 
It is designed to exploit the parallelism that is possible with
modern multi-core and networked computers}

\newduneword{order}{$\mathcal{O}(n)$}{of order $n$}

\newduneword{beamline}{beamline}{A sequence of control and monitoring devices used for the formation of a directed collection of particles}
\newduneabbrev{cdr}{CDR}{conceptual design report}{A formal project
  document 
   that describes the experiment
  at a conceptual level}

\newduneabbrev{cf}{CF}{conventional facilities}{Pertaining to
  construction and operation of buildings or caverns and conventional infrastructure}

\newduneabbrev{cp}{CP}{charge parity}{Product of charge and parity
  transformations}

\newduneword{cpt}{CPT}{product of charge, parity
  and time-reversal transformations}

\newduneabbrev{cpv}{CPV}{charge-parity symmetry violation}{Lack of
  symmetry in a system before and after charge and parity
  transformations are applied}

\newduneword{doe}{DOE}{U.S. Department of Energy}

\newduneword{dune}{DUNE}{Deep Underground Neutrino Experiment}

\newduneword{esh}{ES\&H}{Environment, Safety and Health}

\newduneabbrev{fgt}{FGT}{fine-grained tracker}{A near detector module ??}

\newduneabbrev{fscf}{FSCF}{far site conventional facilities}{The
  \gslpl{cf} at the DUNE far detector site, \surf}
  
\newduneabbrev{nscf}{NSCF}{near site conventional facilities}{The
  \gslpl{cf} at the DUNE near detector site, \fnal}

\newduneabbrev{gut}{GUT}{grand unified theory}{A class of theories that
  unifies the electro-weak and strong forces}

\newduneabbrev{lar}{LAr}{liquid argon}{The liquid phase of argon}

\newduneabbrev{lartpc}{LArTPC}{liquid argon time-projection chamber}{A
  class of detector technology that forms the basis for the DUNE far
  detector modules. 
  It typically entails observation of ionization activity by
  electrical signals and of scintillation by optical signals}

\newduneabbrev{lbl}{LBL}{long-baseline}{Refers to the distance between the 
  neutrino source  and the far detector.  It can also refer to the distance between the near and far detectors. 
  The ``long'' designation is an approximate and relative distinction. For DUNE, this distance  (between \fnal and \surf) is approximately \SI{1300}{km}}

\newduneabbrev{lbnf}{LBNF}{Long-Baseline Neutrino Facility}{The
  organizational entity responsible for developing the neutrino beam, the cryostats
  and cryogenics systems, and the conventional facilities for DUNE}

\newduneabbrev{mh}{MH}{mass hierarchy}{Describes the separation
  between the mass squared differences related to the solar and
  atmospheric neutrino problems}

\newduneabbrev{mi}{MI}{\fnal Main Injector}{An accelerator at
  \fnal that provides a beam of high-energy protons that upon
  striking a target produce secondaries that decay to provide the
  neutrinos directed toward the DUNE far detector}

\newduneabbrev{pot}{POT}{protons on target}{Typically used as a unit
  of normalization for the number of protons striking the neutrino
  production target}

\newduneabbrev{qa}{QA}{quality assurance}{The process by which quality is maintained so as to preserve
high availability and precise function}

\newduneabbrev{qc}{QC}{quality control}{A system of maintaining
  quality through testing products against a specification}

\newduneabbrev{sm}{SM}{Standard Model}{Refers to a theory describing
  the interaction of elementary particles}

\newduneabbrev{tdr}{TDR}{technical design report}{A formal project
  document 
  that describes the experiment at a technical level}

\newduneabbrev{tp}{IDR}{interim design report}{An intermediate
milestone on the path to a full \gls{tdr}} 

\newduneabbrev{ckm}{CKM matrix}{Cabibbo-Kobayashi-Maskawa
  matrix}{Refers to the matrix describing the mixing between mass and
  weak eigenstates of quarks}

\newduneabbrev{cl}{CL}{confidence level}{Refers to a probability
  used to determine the value of a random variable given its
  distribution}

\newduneabbrev{pmns}{PMNS matrix}{Pontecorvo-Maki-Nakagawa-Sakata
  matrix}{Describes the mixing between mass and weak eigenstates of
  the neutrino}

\newduneabbrevs{cpa}{CPA}{cathode plane assembly}{cathode plane assemblies}{The component of the \single detector module that provides the drift HV cathode}

\newduneabbrev{fc}{FC}{field cage}{The component of a LArTPC that contains and shapes the applied \efield}

\newduneabbrev{topfc}{top FC}{top field cage}{The horizontal portions of the \single FC on the top}

\newduneabbrev{botfc}{bottom FC}{bottom field cage}{The horizontal portions of the \single FC on the bottom}

\newduneabbrev{ewfc}{endwall FC}{endwall field cage}{The vertical portions of the \single FC near the wall}

\newduneabbrev{gp}{GP}{ground plane}{An electrode that is held to be
  electrically neutral relative to Earth ground voltage}

  \newduneword{gg}{ground grid}{An electrode that is held to be
  electrically neutral relative to Earth ground voltage  ??}

\newduneabbrev{alara}{ALARA}{as low as reasonably
  achievable}{Typically used with regard management of radiation
  exposure but may be used more generally. It means making every
  reasonable effort to maintain e.g., exposures, to as far below the
  limits as practical, consistent with the purpose for that the
  activity is undertaken}

\newduneabbrev{ecal}{ECAL}{electromagnetic calorimeter}{A detector
  component that measures energy deposition of traversing particles}

\newduneabbrev{hv}{HV}{high voltage}{Generally describes a voltage
  applied to drive the motion of free electrons through some media}

\newduneword{spmod}{SP module}{single-phase detector module}
\newduneword{dpmod}{DP module}{dual-phase detector module}

\newduneabbrev{tc}{TC}{technical coordination}{A dedicated DUNE
  project effort responsible for assuring proper interfaces between
  the collaboration consortia}

\newduneabbrev{cc}{CC}{charged current}{Refers to an interaction
  between elementary particles where a charged weak force carrier
  ($W^+$ or $W^-$) is exchanged}

\newduneabbrev{dis}{DIS}{deep inelastic scattering}{Refers to
  interaction of an elementary charged particle with a nucleus in an
  energy range where the interaction can be modeled as being with
  individual nucleons}

\newduneabbrev{fsi}{FSI}{final-state interactions}{Refers to
  interactions between elementary or composite particles subsequent to
  the initial, fundamental particle interaction, such as may occur as
  the products exit a nucleus}

\newduneabbrev{geant4}{Geant4}{GEometry ANd Tracking, version 4}{A
  software toolkit for the simulation of the passage of particles
  through matter using Monte Carlo methods}

\newduneabbrev{genie}{GENIE}{Generates Events for Neutrino Interaction
  Experiments}{Software providing an object-oriented neutrino
  interaction simulation resulting in kinematics of the products of
  the interaction}

\newduneabbrev{mc}{MC}{Monte Carlo}{Refers to a method of numerical
  integration that entails the statistical sampling of the integrand
  function. 
  Forms the basis for some types of detector and physics simulations}

\newduneabbrev{qe}{QE}{quasi-elastic}{Refers to interaction between
  elementary particles and a nucleus in an energy range where the
  interaction can be modeled as occurring between constituent quarks
  of one nucleon and resulting in no bulk recoil of the resulting
  nucleus}

\newduneabbrev{mou}{MOU}{memorandum of understanding}{A document
  summarizing an agreement between two or more parties}

\newduneabbrev{pip2}{PIP-II}{Proton Improvement Plan II}{A plan for
  improving the protons on target delivered for the DUNE neutrino
  production beam. 
  This is revision two of this plan and is planned to be followed by a PIP-III}

\newduneabbrev{sdsta}{SDSTA}{South Dakota Science and Technology
  Authority}{The legal entity that manages the Sanford Underground
  Research Facility, \surf, in Lead, S.D}

\newduneabbrev{wbs}{WBS}{work breakdown structure}{An organizational
  project management tool by which the tasks to be performed are
  partitioned in a hierarchical manner}

\newduneabbrev{br}{BR}{branching ratio}{A fractional probability for a
  decay of a composite particle to occur into some specified set or
  sets of products}

\newduneabbrev{dm}{DM}{dark matter}{The term given to the unknown
  matter or force that explains measurements of motion of galaxies
  that are otherwise inconsistent with the amount of mass associated
  with observed amount of photon production}

\newduneabbrev{dsnb}{DSNB}{Diffuse Supernova Neutrino Background}{The
  term describing the pervasive, constant flux of neutrinos due to all
  past supernova neutrino bursts}

\newduneabbrev{globes}{GLoBES}{General Long-Baseline Experiment
  Simulator}{A software package for simulating energy spectra of
  neutrino flux, interaction and measured (after application of some
  model of a detector response) energy spectra}

\newduneabbrev{nc}{NC}{neutral current}{Refers to an interaction
  between elementary particles where a neutrally charged weak force carrier
  ($Z^0$) is exchanged}

\newduneabbrev{nh}{NH}{normal hierarchy}{Refers to the neutrino mass
  eigenstate ordering whereby the sign of the mass squared difference
  associated with the atmospheric neutrino problem is positive}

\newduneabbrev{ih}{IH}{inverted hierarchy}{Refers to the neutrino mass
  eigenstate ordering whereby the sign of the mass squared difference
  associated with the atmospheric neutrino problem is negative}

\newduneabbrev{nsi}{NSI}{nonstandard interactions}{A general class of
  theory of elementary particles other than the Standard Model}

\newduneabbrev{msw}{MSW}{Mikheyev-Smirnov-Wolfenstein effect}{Explains
  the oscillatory behavior of neutrinos produced inside the sun as
  they traverse the solar matter}

\newduneabbrev{susy}{SUSY}{supersymmetry}{Theoretical symmetry between a fermion and a boson}

\newduneabbrev{wimp}{WIMP}{weakly-interacting massive particle}{A
  hypothesized particle that may be a component of dark matter}

\newduneabbrev{ce}{CE}{cold electronics}{Refers to readout electronics that operate at cryogenic temperatures}

\newduneabbrev{crp}{CRP}{charge-readout plane}{In the \dual technology, a  collection of
  electrodes in a planar arrangement placed at a particular voltage
  relative to some applied \efield such that drifting electrons
  may be collected and their number and time may be measured}

\newduneabbrev{dram}{DRAM}{dynamic random access memory}{A computer memory technology}

\newduneword{fermilab}{\fnal}{Fermi National Accelerator Laboratory (in Batavia, IL, the Near Site)}
\newduneword{fnal}{FNAL}{see \fnal}

\newduneabbrev{fs}{FS}{full stream}{Relates to a data stream that has not undergone selection, compression or other form of reduction}

\newduneabbrev{lem}{LEM}{large electron multiplier}{A micro-pattern detector suitable for use in ultra-pure argon vapor; LEMs consist of copper-clad PCB boards with sub-millimeter-size holes through which electrons undergo amplification}
\newduneabbrev{lng}{LNG}{liquefied natural gas}{Pertaining to natural gas in its liquid phase}

\newduneabbrev{mip}{MIP}{minimum ionizing particle}{Refers to a
  momentum traversing some medium such that the particle is losing
  near the minimum amount of energy per distance traversed} 

\newduneabbrev{pd}{PD}{photon detector}{Refers to the detector
  elements involved in measurement of number and arrival times of
  optical photons produced in a detector module} 

\newduneabbrev{pmt}{PMT}{photomultiplier tube}{A device that makes use
  of the photoelectric effect to produce an electrical signal from the
  arrival of optical photons}

\newduneabbrev{ppm}{ppm}{parts per million}{A number equal to $10^{-6}$}
\newduneabbrev{ppb}{ppb}{parts per billion}{A number equal to $10^{-9}$}
\newduneabbrev{ppt}{ppt}{parts per trillion}{A number equal to $10^{-12}$}

\newduneword{rio}{RIO}{reconfigurable input output}

\newduneword{rpc}{RPC}{resistive plate chamber}
\newduneword{s/n}{S/N}{signal-to-noise (ratio)}
\newduneword{ssp}{SSP}{SiPM signal processor}
\newduneword{sbn}{SBN}{Short-Baseline Neutrino program (at \fnal)}
\newduneword{stt}{STT}{straw tube tracker}
\newduneword{tr}{TR}{transition radiation}
\newduneword{wire board}{wire board}{At the head end of the APA in the \single TPC, stacks of electronics boards referred to as ``wire boards'' are arrayed to anchor the wires.  They also provide the connection between the wires and the cold electronics} 

\newduneabbrev{wls}{WLS}{wavelength shifting}{A material or process by
  which incident photons are absorbed by a material and photons are
  emitted at a different, typically longer, wavelength}

\newduneabbrev{tpb}{TPB}{tetra-phenyl butadiene}{A type of wavelength shifting material}

\newduneabbrev{sft}{SFT}{signal feedthrough}{A cryostat penetration allowing for the passage of cables or other extended parts}
\newduneabbrev{sftchimney}{SFT chimney}{signal feedthrough chimney}{In the \dual technology, a volume above the cryostat penetration used for a signal feedthrough}

\newduneword{catiroc}{CATIROC}{charge and time integrated readout chip}

\newduneabbrev{wr}{WR}{White Rabbit}{A component of the timing system that forwards clock signal and time-of-day reference data to the master timing unit}

\newduneabbrev{mch}{MCH}{MicroTCA Carrier Hub}{An network switching device}

\newduneabbrev{wrmch}{WR-MCH}{White Rabbit \gls{utca} Carrier Hub}{add def} 

\newduneword{cmp}{CMP}{configuration management plan}

\newduneword{qap}{QAP}{quality assurance plan} 

\newduneword{ieshp}{IESHP}{integrated environmental, safety and health plan}

\newduneword{dmp}{DMP}{data management plan} 

\newduneword{qam}{QAM}{quality assurance manager} 

\newduneabbrev{dss}{DSS}{detector support system}{The system used to support the \single detector within the cryostat}

\newduneabbrev{itf}{ITF}{integration and test facility}{A facility where various detector components will be tested prior to installation}

\newduneabbrev{tco}{TCO}{temporary construction opening}{An opening in the side of a cryostat through which detector elements are brought into the cryostat; utilized during construction and installation}

\newduneabbrev{uit}{UIT}{underground installation team}{An organizational unit responsible for installation in the underground area at the \surf site}

\newduneabbrev{pdr}{PDR}{preliminary design review}{A project management device by which an early design is reviewed}
\newduneabbrev{fdr}{FDR}{final design review}{A project management device by which a final design is reviewed}
\newduneabbrev{prr}{PRR}{production readiness review}{A project management device by which the production readiness is reviewed}
\newduneabbrev{orr}{ORR}{operational readiness review}{A project management device by which the operational readiness is reviewed}
\newduneabbrev{ppr}{PPR}{production progress review}{A project management device by which the progress of production is reviewed}
\newduneabbrev{edms}{EDMS}{electronic document management system}{A computerized system used at CERN by which documents are managed}

\newduneword{wrgm}{WR grandmaster}{White Rabbit grandmaster}

\newduneword{larsoft}{\larsoft}{Liquid Argon Software (\larsoft),  a shared base of physics software across \lartpc experiments}
\newduneword{nova}{\nova}{The \nova off-axis neutrino oscillation experiment at \fnal }
\newduneword{minerva}{\minerva}{The \minerva neutrino cross sections experiment at \fnal }
\newduneword{microboone}{\microboone}{The \lartpc-based \microboone neutrino oscillation experiment at \fnal }
\newduneword{wirecell}{wire-cell}{A tomographic automated \threed neutrino event reconstruction method for \lartpc{}s}
\newduneword{ftslite}{F-FTS-lite}{Light-weight version of the \fnal File Transfer system used for rapid data transfers out of the online systems}
\newduneword{fts}{FTS}{File Transfer System developed at \fnal to catalog and move data to permanent storage}

\newduneword{35t}{35 ton prototype}{The 35 ton prototype cryostat and \gls{sp} detector built at \fnal before the \gls{protodune} detectors}

\newduneabbrev{cuc}{CUC}{central utility cavern}{The central underground cavern  containing utilities such as central cryogenics and other systems, and the underground data center and control room}
\newduneabbrev{cfd}{CFD}{computational fluid dynamics}{High performance computer-assisted modeling of fluid dynamical systems}
\newduneword{vuv}{VUV}{vacuum ultra-violet}
\newduneword{tallbo}{TallBo}{A cylindrical cryostat at \fnal primarily used for developing scintillation light collection technologies for \lartpc detectors}

\newduneabbrev{eos}{EOS}{EOS}{The XRootD based distributed file system developed by CERN}
\newduneabbrev{ehn1}{EHN1}{Experiment Hall North One}{Location at CERN of the ProtoDUNE experiments}
\newduneword{led}{LED}{Light-emitting diode}
\newduneword{rtd}{RTD}{Cryostat level and temperature monitoring devices}
\newduneword{swc}{SWC}{Software \& Computing}
\newduneabbrev{las}{LAS}{LEM-anode Sandwich}{In the \dual technology, a \gls{lem} and its corresponding anode are mounted together in a module called a LEM-anode sandwich}

\newduneword{roi}{ROI}{region of interest}
\newduneword{hpc}{HPC}{high-performance computing facilities; generally computing facilities emphasizing parallel computing with aggregate power of more than a teraflop}

\newduneword{comfund}{common fund}{The shared resources of the collaboration}
\newduneabbrev{ims}{IMS}{integrated master schedule}{A project management device consisting of linked tasks and milestones}

\newduneword{hvdb}{HVDB}{dual-phase HV divider board}
\newduneword{sas}{SAS}{A pass-through chamber used to ensure safe transfer of materials, avoiding contamination in both directions}

\newduneword{fea}{FEA}{finite element analysis}

\newduneword{fss}{FSS}{field shaping strips}
\newduneword{lvds}{LVDS}{low-voltage differential signaling}

\hypersetup{
    pdftitle={\expshort TDR \thedocsubtitle},
    pdfauthor={\expshort Collaboration},
    final=true,
    colorlinks=false,
    linktocpage=true,
    linkbordercolor=blue,
    citebordercolor=green,
    urlbordercolor=magenta,
    filecolor=black,
    pdfpagemode=UseOutlines,
    pdfborderstyle={/S/U},  
}

\renewcommand\thedoctitle{\volexecsumm} 
\begin{document}

\pagestyle{titlepage}
\includepdf[pages={-}]{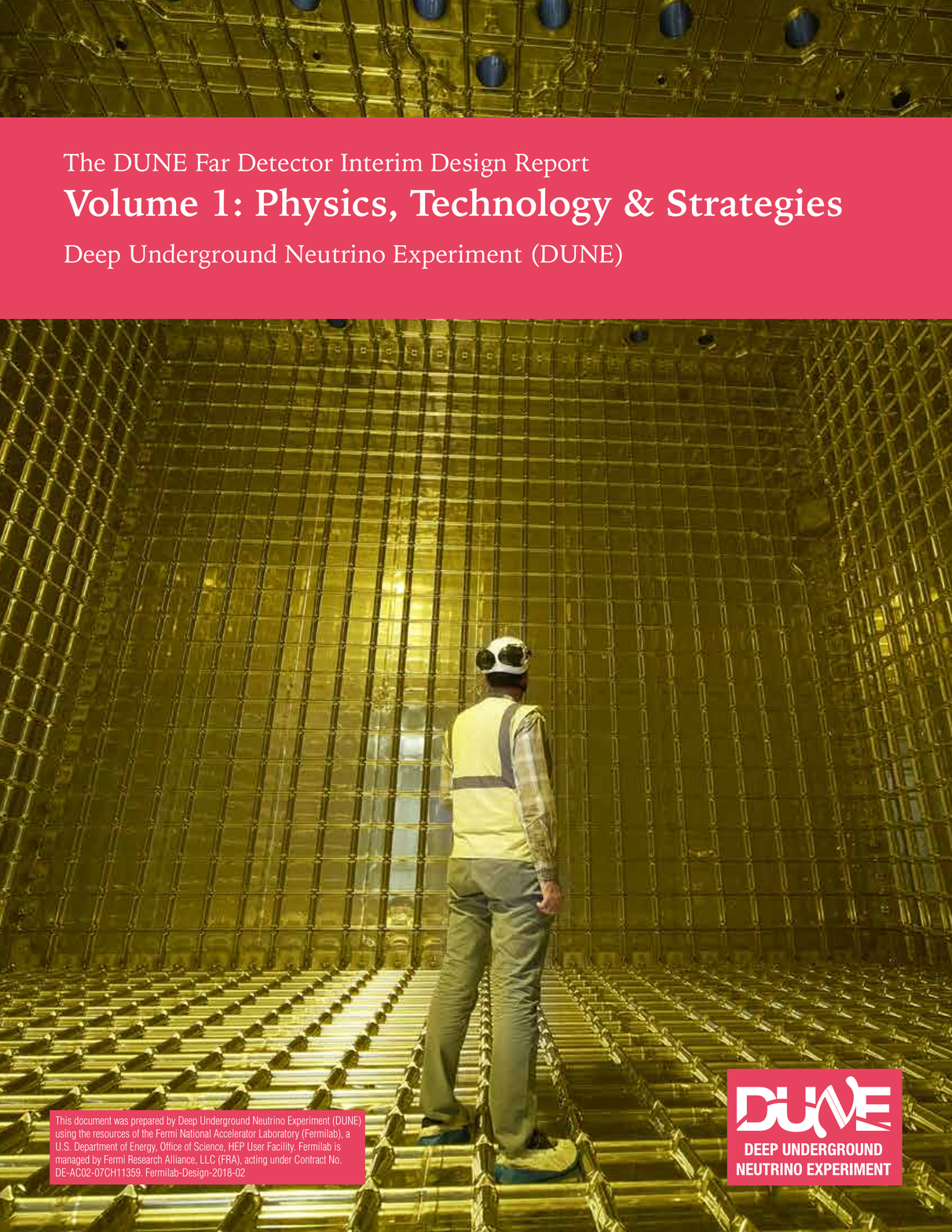}
\cleardoublepage



\cleardoublepage

\includepdf[pages={-}]{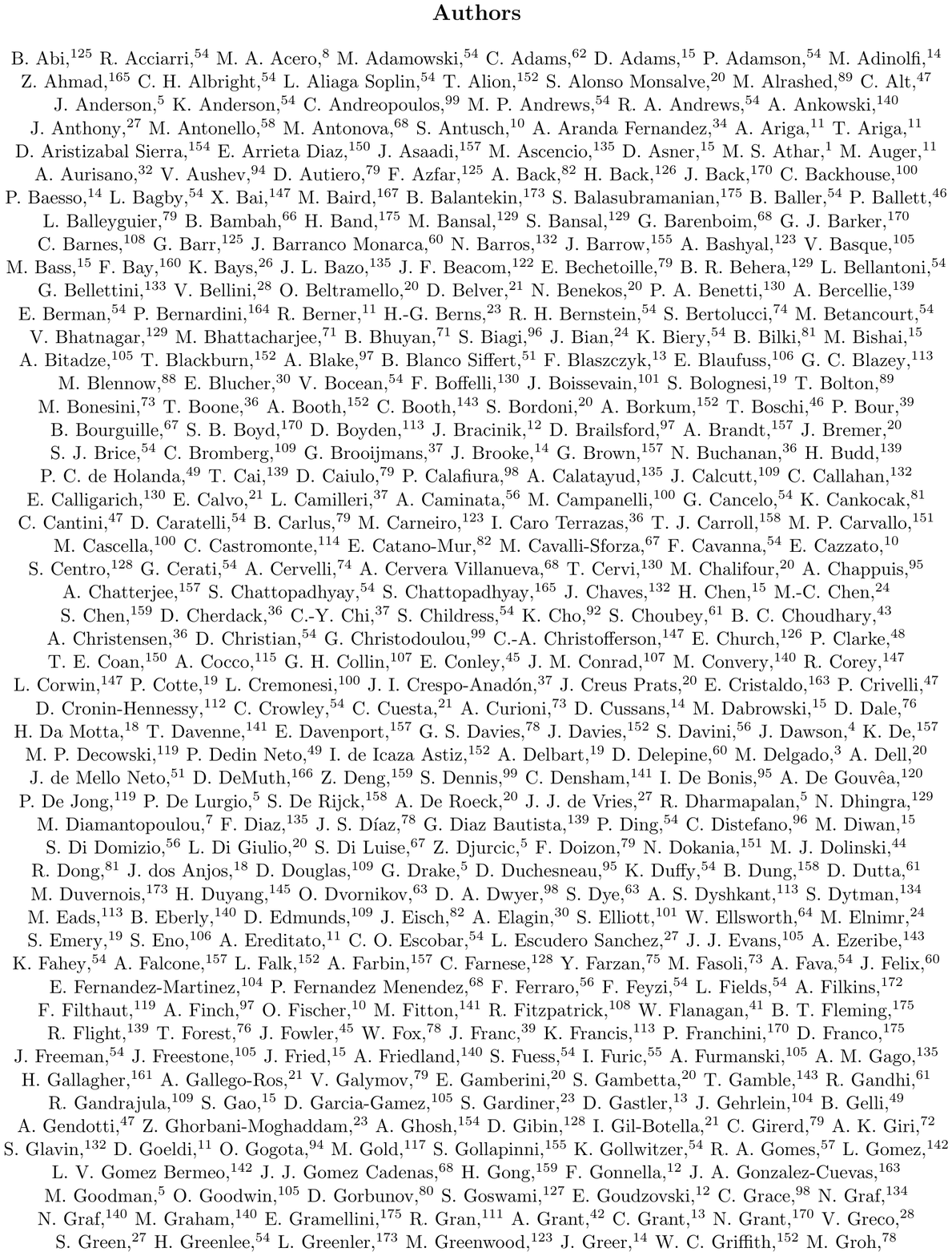}

\renewcommand{\familydefault}{\sfdefault}
\renewcommand{\thepage}{\roman{page}}
\setcounter{page}{0}

\pagestyle{plain} 


\textsf{\tableofcontents}

\textsf{\listoffigures}

\textsf{\listoftables}



\renewcommand{\thepage}{\arabic{page}}
\setcounter{page}{1}

\pagestyle{fancy}

\renewcommand{\chaptermark}[1]{%
\markboth{Chapter \thechapter:\ #1}{}}
\fancyhead{}
\fancyhead[RO,LE]{\textsf{\footnotesize \thechapter--\thepage}}
\fancyhead[LO,RE]{\textsf{\footnotesize \leftmark}}

\fancyfoot{}
\fancyfoot[RO]{\textsf{\footnotesize The DUNE Far Detector Interim Design Report}}
\fancyfoot[LO]{\textsf{\footnotesize \thedoctitle}}
\fancypagestyle{plain}{}

\renewcommand{\headrule}{\vspace{-4mm}\color[gray]{0.5}{\rule{\headwidth}{0.5pt}}}

\nocite{CD0}


\chapter{Executive Summary }
\label{ch:project-overview}
\section{Overview}

The Deep Underground Neutrino Experiment (DUNE) will be a world-class neutrino observatory and nucleon decay detector designed to answer fundamental questions about the nature of elementary particles and their role in the universe. The international DUNE experiment, hosted by the U.S. Department of Energy's \fnal{}, will consist of a far detector to be located about \SI{1.5}{km} underground at the Sanford Underground Research Facility (\surf) in South Dakota, USA, at a distance of  \SI{1300}{\km} from \fnal{}, and a near detector to be located at \fnal in Illinois. The far detector will be a very large, modular liquid argon time-projection chamber (\lartpc) with a \fdfiducialmass (\SI{40}{\giga\gram}) fiducial mass. This \lar technology will make it possible to reconstruct neutrino interactions with image-like precision and unprecedented resolution. 

The far detector will be exposed to the world's most intense neutrino beam originating at \fnal{}. A high-precision near detector, located \SI{575}{m} from the neutrino source on the \fnal site, will be used to characterize the intensity and energy spectrum of this wide-band beam. The Long-Baseline Neutrino Facility (LBNF), also hosted by \fnal, provides the infrastructure for this complex system of detectors at the Illinois and South Dakota sites. LBNF assumes the responsibility for the neutrino beam, the deep-underground site, and the infrastructure for the DUNE detectors.

\begin{dunefigure}[DUNE collaboration global map]{fig:mhexec}{The international DUNE
collaboration. Countries with DUNE membership are shown in orange.}
\includegraphics[width=0.9\textwidth]{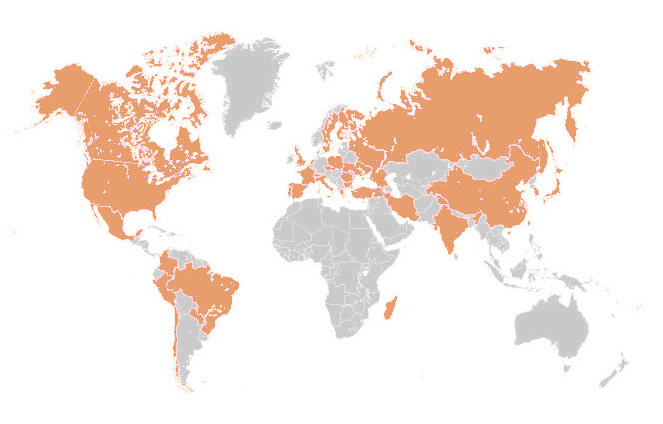}  
\label{fig:map}
\end{dunefigure}

The DUNE collaboration is a truly global organization including more than \num{1000} scientists and engineers from \num{32} countries (Figure~\ref{fig:map}). It represents the culmination of several worldwide efforts that developed independent paths toward a next-generation long-baseline (LBL) neutrino experiment over the last decade. It was formed in April 2015, combining the strengths of the LBNE project in the USA and the LBNO project in Europe, adding many new international partners in the process. DUNE thus represents the convergence of a substantial fraction of the worldwide neutrino-physics community around the opportunity provided by the large investment planned by the U.S. Department of Energy (DOE) and \fnal to support a significant expansion of the underground infrastructure at \surf in South Dakota, and to create a megawatt neutrino-beam facility at \fnal by 2026. The Proton Improvement Plan-II (PIP-II) upgrade at \fnal~\cite{pip2-2013} will enable the accelerator to drive the new neutrino beamline with a \SI{80}{\GeV} primary proton beam at a beam power 
up to \pipiibeampower{}. A further planned upgrade 
of the accelerator complex will enable it to provide up to \SI{2.4}{\MW} of beam power by 2030. 

The LBNF/DUNE project (the \textit{project}) strategy presented in this \dword{tp} has been developed to meet the requirements set out in the report of the Particle Physics Project Prioritization Panel (P5 in 2014). It also takes into account the recommendations of the European Strategy for Particle  Physics (ESPP) adopted by the CERN Council in 2013, which classified the \dword{lbl} neutrino program as one of the four scientific objectives that require international infrastructure.

The P5 report~\cite{p5report} set the goal of reaching a sensitivity to \dword{cpv} of better than three standard deviations (\num{3}$\sigma$) over more than $75\%$ 
of the range of possible values of the unknown \dshort{cp}-violating phase \deltacp.
Based partly on this goal, they stated that ``the 
minimum requirements to proceed are the identified capability to reach an exposure 
of \num{120}~\ktMWyr{} by the 2035 time frame, the far detector situated underground 
with cavern space for expansion to at least \fdfiducialmass \lar fiducial volume, and \SI{1.2}{MW} 
beam power upgradeable to multi-megawatt power.
The experiment should have the demonstrated 
capability to search for \dwords{snb} and for proton decay, providing a significant 
improvement in discovery sensitivity over current searches for the proton lifetime.'' The strategy and design presented in this \dword{tp} meet these requirements.

This document serves as the \dword{tp} for the DUNE \dword{fd}.  The \dword{tp}  is intended to provide a clear statement of the physics goals and methods of the DUNE experiment, and to describe the detector technologies that have been designed to achieve these goals. Introductions to each chapter are intended to be useful and informative to technical specialists who serve in national science agencies. The body of the document is intended to be useful and informative to members of the international high energy physics community.  The \dword{tp} deliberately emphasizes the connections between the physics and technologies of the DUNE \dword{fd} modules. Very important project related tasks are presented in summary form. No information about cost is included, and schedule information appears only in the form of high-level milestones. The \dword{tp} forms the nucleus of the \dword{tdr}, which will be presented to international science  agencies and the high energy physics (HEP) community in 2019.


%

\section{Primary Science Goals} 

The DUNE experiment will combine the world's most intense neutrino beam, a deep underground site, and massive LAr detectors to enable a broad science program addressing some of the most fundamental questions in particle physics. 


The primary science goals of DUNE, described in detail in Chapter~\ref{ch:exec-summ-physics}, are to: 
\begin{itemize}

\item Carry out a comprehensive program of neutrino oscillation measurements using \numu and \anumu beams from \fnal. This program includes measurements of the  \dword{cp} phase, determination of the neutrino mass ordering (the sign of \dm{31}$ \equiv m_3^2-m_1^2$), measurement of the mixing angle $\theta_{23}$ and the determination of the octant in which this angle lies,
and sensitive tests of the three-neutrino paradigm. Paramount among these is the search for \dword{cpv} in neutrino oscillations, which may give insight into the origin of the matter-antimatter asymmetry, one of the fundamental questions in particle physics and cosmology. 

\item Search for proton decay in several important decay modes. The observation of proton decay would represent a ground-breaking discovery in physics, providing a key requirement for grand unification of the forces. 

    \item Detect and measure the $\nu_\text{e}$ flux from a core-collapse supernova within our galaxy, should one occur during the lifetime of the DUNE experiment. Such a measurement would provide a wealth of unique information about the early stages of core-collapse, and could even signal the birth of a black hole.
    
\end{itemize}

The intense neutrino beam from LBNF, the massive DUNE \lartpc far detector, and the high-resolution
DUNE near detector will also provide a rich ancillary science program, beyond the primary goals of the experiment. The ancillary science program includes
\begin{itemize}
     \item other accelerator-based neutrino flavor transition measurements with sensitivity to beyond the standard model (BSM) physics, such as non-standard interactions (NSIs), Lorentz violation,  \dword{cpt} violation, sterile neutrinos, large extra dimensions, heavy neutral leptons;
 and measurements of tau neutrino appearance;
     \item measurements of neutrino oscillation phenomena using atmospheric neutrinos;
     \item a rich neutrino interaction physics program utilizing the DUNE near detector, including a wide-range of measurements of neutrino cross sections, studies of nuclear effects; 
     \item  searches for dark matter.
\end{itemize} 
Further advancements in the \lartpc 
technology during the course of the DUNE far detector construction may open up the opportunity
to observe very low-energy phenomena such as solar neutrinos or even the diffuse supernova neutrino flux.

\section{The LBNF Facility} 

The Long-Baseline Neutrino
Facility (LBNF), hosted by Fermilab, is separate from the DUNE collaboration and is intended to enable the construction and operation of the DUNE detectors in South Dakota and Illinois.
The DUNE collaboration will construct a deep-underground neutrino observatory in South Dakota based on four independent \nominalmodsize \lartpc{}s. 
LBNF will provide facilities in Illinois and South Dakota to enable the scientific program of DUNE.
These facilities are geographically separated into the near site facilities, those to be constructed
at \fnal, and the far site facilities, located at \surf. Figure~\ref{fig:lbnf} shows
a schematic of the facilities at the two sites, and Figure~\ref{fig:caverns} shows the cavern layout. 

\begin{dunefigure}[ 	
LBNF/DUNE project: beam from Illinois to South Dakota]{fig:lbnf}{ 	
LBNF/DUNE project: beam from Illinois to South Dakota.}
\includegraphics[width=0.9\textwidth]{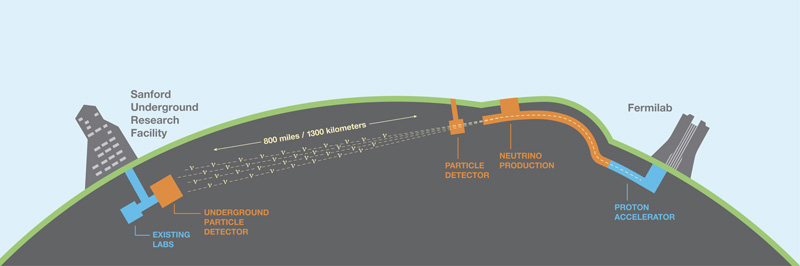}
\end{dunefigure}

Specifically, the Long-Baseline Neutrino Facility (LBNF) provides
\begin{itemize}

\item  the  technical and conventional facilities for a powerful \MWadj{1.2} neutrino beam utilizing the PIP-II upgrade of the \fnal accelerator 
complex, to become operational by \beamturnon  
at the latest, and to be upgradable to \SI{2.4}{\MW} with the proposed 
PIP-III upgrade;

\item  the civil construction, or \dword{cf}, for the near detector systems at \fnal; (see Figure~\ref{fig:beamline}); 

\item the excavation of four underground caverns at \surf, planned to be completed 
by 2021 
under a single contract, with each cavern to be capable of housing a cryostat with 
a minimum \nominalmodsize fiducial mass \lartpc; and

\item surface, shaft, and underground infrastructure to support 
the outfitting of the caverns with four free-standing, steel-supported cryostats 
and the required cryogenics systems. The first cryostat will be available for filling, after installation of the detector components, by
2023, enabling a rapid deployment of the first two \nominalmodsize far detector modules. 
The intention is to install the third and fourth cryostats as rapidly as funding will 
allow.

\end{itemize}

\begin{dunefigure}[ 	
Underground caverns for DUNE in South Dakota]{fig:caverns}{Underground caverns for DUNE far detectors and cryogenic systems at \surf{}, in South Dakota.}
\includegraphics[width=0.95\textwidth]{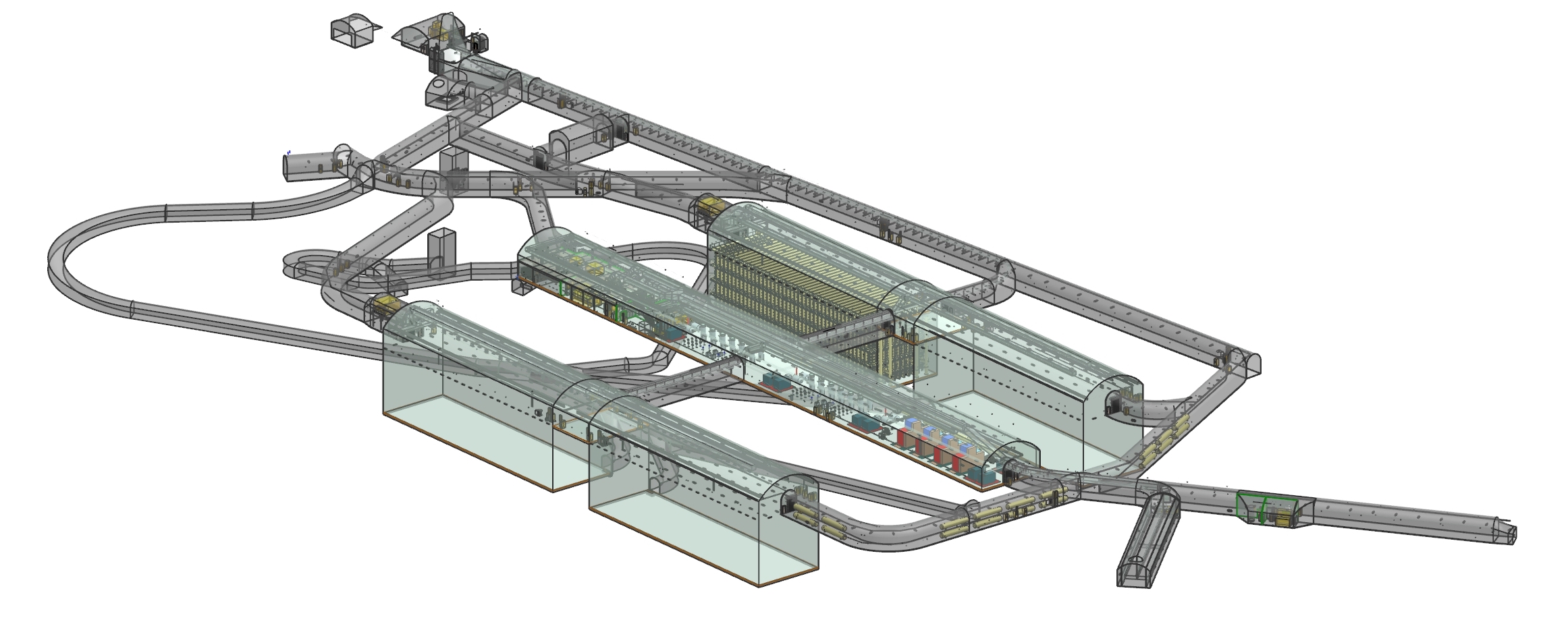}
\end{dunefigure}

\begin{dunefigure}[Neutrino beamline and DUNE near detector hall in Illinois
]{fig:beamline}{Neutrino beamline and DUNE near detector hall at Fermilab, in Illinois}
\includegraphics[width=0.95\textwidth]{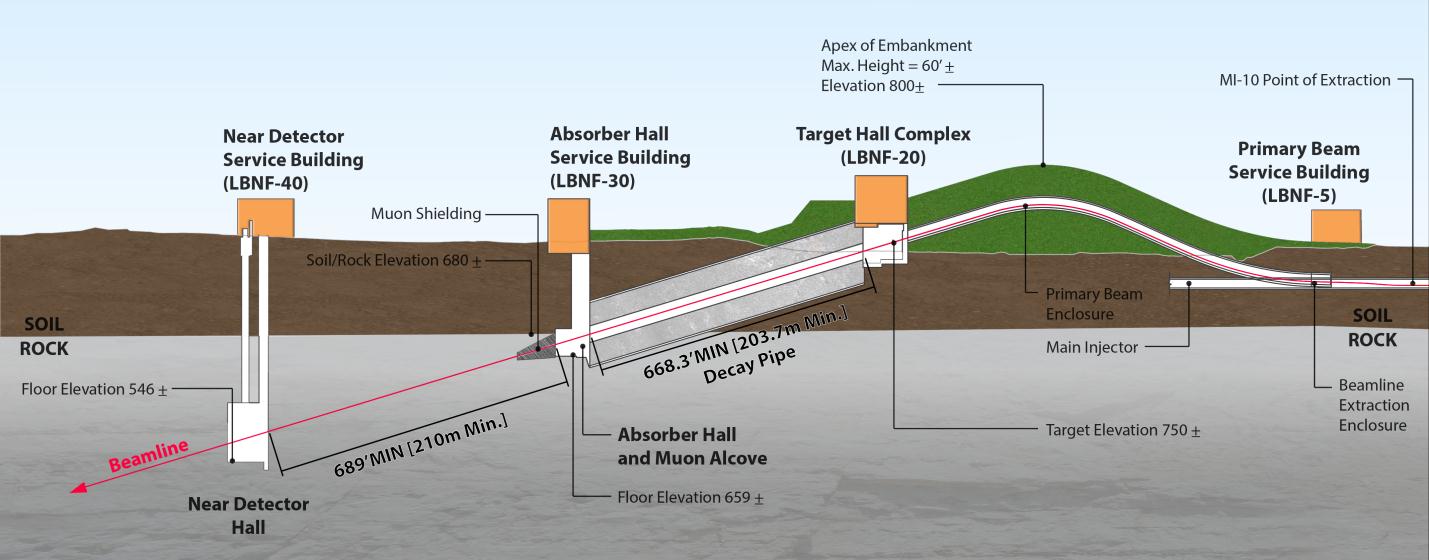}
\end{dunefigure}

The success of the DUNE project depends on the successful realization of the LBNF facilities.
This \dword{tp} focuses on the DUNE physics program that is enabled by
the first three \dword{fd} modules, which are expected to be based on the \dword{sp} and \dword{dp} \lar technologies. 

\section{The DUNE Experiment} 


The DUNE experiment includes a precision near detector at the edge of the \fnal site, in Batavia, Illinois, and a very large, modular far detector about \SI{1.5}{km} underground at \surf in Lead, South Dakota, \SI{1300}{km} (\SI{800}{miles}) from \fnal. The DUNE far detector is the focus of this \dword{tp}. 

The near detector will be located \SI{575}{m} from the target. It will consist of a \lartpc followed by a fine-grained magnetic spectrometer. The \lartpc will use pixel readout to deal with the high occupancy from neutrino events in the intense LBNF beam. The details of the magnetic spectrometer will be resolved in the near future. The goal for the near detector group is to produce a \dword{cdr} at the time of the \dword{fd} \dword{tdr}. The near detector \dword{cdr} will provide information critical to establishing the physics reach for the primary neutrino oscillation program of DUNE. The near detector \dword{tdr} will follow the \dword{fd} \dword{tdr} by approximately one year, consistent with the near detector construction schedule.

The DUNE \dword{fd} will consist of four similar \lartpc{}s, each with fiducial mass of at least \nominalmodsize, installed about \SI{1.5}{km} underground. Each detector will be installed in a cryostat with internal dimensions
\cryostatwdth (W) $\times$ \cryostatht (H) $\times$ \cryostatlen (L), and will contain a total \lar{} mass of about \larmass{}.
The \lartpc technology provides
excellent tracking and calorimetry performance, making it an ideal
choice for the DUNE far detectors. The four identically sized cryostats give flexibility for staging and evolution of the \lartpc technology.

DUNE is planning for and prototyping two \lartpc technologies:
\begin{itemize}
\item Single-phase (SP): This technology was pioneered by the ICARUS project, and after several decades of worldwide R\&D, is now a mature technology. It is the technology used for \fnal{}'s currently operating \microboone, and the planned SBND. In the \single technology, ionization charges are drifted horizontally in \lar and read out on wires in the liquid. The maximum drift length in the DUNE \dword{spmod} is \spmaxdrift and the nominal drift field is \spmaxfield, corresponding to a cathode high voltage of \sptargetdriftvoltpos. There is no signal amplification in the liquid, so readout with good signal-to-noise requires very low-noise electronics.

\item Dual-phase (DP): This technology was pioneered at large scale by the \dword{wa105} collaboration. It is less established than the \single technology but offers a number of potential advantages and challenges. Here, ionization charges are drifted vertically in \lar and transferred into the gas above the liquid. The signal charges are then amplified in the gas phase using large electron multipliers (LEMs). This gain reduces the requirements on the electronics, and makes it possible for the \dword{dpmod} to have a longer drift, which requires a correspondingly higher voltage.
The maximum drift length in the \dword{dpmod} is \dpmaxdrift and the nominal drift field is \dpnominaldriftfield, corresponding to a cathode high voltage of \dptargetdriftvoltpos. 

\end{itemize}
The plans for the single and dual-phase TPCs are described in detail in Volumes 2 and 3 of this \dword{tp}.

The DUNE collaboration is committed to deploying both technologies. 
For planning purposes, DUNE assumes the first \dword{detmodule} to be
\single and the second to be \dual.
The actual sequence of \dword{detmodule} installation will depend on results from the prototype detectors, described below, and on available resources.

The collaboration is in now in the final stages of constructing two large prototype detectors (called \dwords{protodune}), one employing \single readout (\dword{pdsp}) and the other employing \dual readout (\dword{pddp}). Each is approximately one-twentieth of a DUNE \dword{detmodule}, but uses components identical in size to those of the full-scale module. \dword{pdsp} has the same \spmaxdrift maximum drift length as the full \dword{spmod}. \dword{pddp} has a \SI{6}{m} maximum drift length, half of that planned for the \dword{dpmod}. 

These large-scale prototypes will allow us to validate key aspects of the TPC designs, test engineering procedures, and collect valuable calibration data using a hadron test beam. The following list includes the key goals of the \dword{protodune} program:
\begin{enumerate}
\item Test production of components:
\begin{itemize}
\item stress testing of the production and quality
assurance processes of detector components,
\item mitigation of the associated risks for the far detector.
\end{itemize}
\item Validate installation procedures:
\begin{itemize}
\item test of the interfaces between the detector elements,
\item mitigation of the associated risks for the far detector.
\end{itemize}
\item Detector operation with cosmic rays:
\begin{itemize}
\item validation of the detector designs and
performance.
\end{itemize}
\item Collection of test beam data:
\begin{itemize}
\item measurements of essential physics response of the detector.
\end{itemize}
\end{enumerate}

Items \numrange{1}{3} are required as input to the \dword{tdr}. Item 4, collection and the corresponding analysis of test beam data, will be vital to DUNE's physics program, but is not required for the TDR.

The full DUNE far detector requires four modules. For the \dword{tdr}, we will describe plans for at least the first two of these modules. Based on our current expectations, we hope to present a plan for two \dwords{spmod}, one of which will be the first module installed, and one \dword{dpmod}. At the time of the \dword{tdr}, it is likely that resources for the fourth \dword{detmodule} will remain to be identified.

\section{International Organization and Responsibilities}

DUNE is the first science project of this scale in the USA that will be built with large
international participation and as an international collaboration. This requires a new organizational and governance model that takes into account the international nature of the project.
The
model used by CERN for managing the construction and exploitation of the Large Hadron Collider (LHC) and its experiments served as a starting point for the joint management of LBNF and the DUNE experimental program. 
LBNF, which is responsible for the facilities, comprising the neutrino beam, the near site at \fnal and the far site at \surf, is organized as a
DOE-\fnal project incorporating international partners. 
DUNE is a fully international project
organized by the DUNE collaboration with appropriate oversight from all international stakeholders.
The DUNE collaboration is responsible for
\begin{itemize}
\item the definition of the scientific goals and corresponding scientific and technical requirements on the detector systems and neutrino beamline;
\item the design, construction, commissioning, and operation of the detectors; and
\item the scientific research program conducted with the DUNE detectors. 
\end{itemize}

A set  of  organizational structures  has been established  to  provide
coordination  among  the  participating  funding agencies;
oversight of the LBNF and DUNE projects;
and coordination and communication between the 
two. These structures and the relationships among them are shown 
in Figure~\ref{fig:org}. They comprise the following committees:
\begin{itemize}
\item International Neutrino Council (INC)

The INC is composed of regional representatives, such as CERN, and representatives of funding agencies making major contributions to LBNF infrastructure and to DUNE. The INC acts as the highest-level international advisory body to the U.S. Department of Energy (DOE) and the \fnal directorate. The INC facilitates high-level global coordination across the entire enterprise (LBNF and DUNE). The INC is chaired by the DOE Office of Science associate director for high energy physics and includes the \fnal director in its membership. The council meets and provides pertinent advice to the LBNF and DUNE projects through the \fnal director as needed.
\item Resources Review Board (RRB)

The RRB is composed of representatives of all funding agencies that sponsor LBNF, DUNE, and PIP-II, and the \fnal management. The RRB provides focused monitoring and detailed oversight of the DUNE collaboration, and also monitors the progress of LBNF and PIP-II. The \fnal director,  in consultation with the international funding partners for the projects, defines the membership of the RRB. A representative from the \fnal directorate chairs the RRB and organizes regular meetings to facilitate coordination and to monitor the progress of the projects. The management teams from the DUNE collaboration and the LBNF project participate in the RRB meetings and make regular reports to the RRB on technical, managerial, financial and administrative matters, as well as reporting on the status and progress of the DUNE collaboration.

\item Long-Baseline Neutrino Committee (LBNC)

The LBNC is composed of internationally prominent scientists with relevant expertise. It provides regular external scientific peer review of DUNE, and provides regular reports to the \fnal  directorate and the RRB. The LBNC reviews the scientific, technical and managerial decisions of the DUNE experiment. The LBNC will review the \dword{tdr} for DUNE and will provide a recommendation to the \fnal directorate and the RRB on whether to endorse the \dword{tdr}.


Upon request from the Fermilab director, the LBNC may employ additional DUNE and LBNF scrutiny groups for more detailed reports and evaluations. 

\item Neutrino Cost Group (NCG)

Like the LBNC, the NCG is composed of internationally prominent scientists with relevant experience.  The NCG reviews the cost, schedule, and associated risks for the DUNE experiment, and provides regular reports to the \fnal directorate and the RRB.  The NCG will review the \dword{tdr} for DUNE and will provide a recommendation to the \fnal directorate and the RRB on whether to endorse the \dword{tdr}.

\item Experiment-Facility Interface Group (EFIG)

Close and continuous coordination between DUNE and LBNF is required to ensure the success of the combined enterprise. The EFIG  oversees the interfaces between the two projects and ensures the required coordination during the design and construction phases and the operational phase of the program. This group covers areas including interfaces between the near and far detectors and the corresponding conventional facilities; interfaces between the detector systems provided by DUNE and the technical infrastructure provided by LBNF; design of the LBNF neutrino beamline and neutrino beamline operational issues that impact both LBNF and DUNE.  

\end{itemize}

\begin{dunefigure}[Structure for oversight of the DUNE and LBNF projects.]	
{fig:org}{Top-level organization structure for oversight of the DUNE and LBNF projects.}
\includegraphics[width=0.68\textwidth]{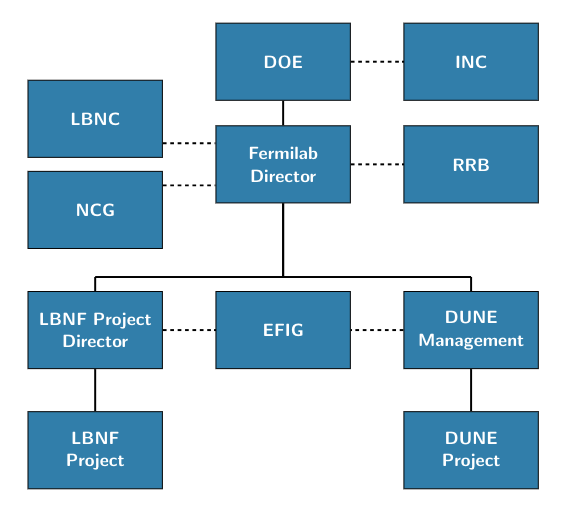}  
\end{dunefigure}

\section{DUNE Organization and Management}

All aspects of DUNE are organized and managed by the DUNE collaboration.  Stakeholders include the collaborating institutions, the funding agencies participating in DUNE, and \fnal as the host laboratory.  All collaborating institutions have a representative on the DUNE Institutional Board (IB). The collaboration is responsible for the design, construction, installation, commissioning, and operation of the detectors and prototypes used to pursue the scientific program. The DUNE Executive Board (EB), described below, is the main management body of the collaboration and approves all significant strategic and technical decisions.

The top-level DUNE management team consists of two elected co-spokespersons, the technical coordinator (TC), and the resource coordinator (RC). The TC and RC are selected jointly by the co-spokespersons and the \fnal director. The management team is responsible for the day-to-day management of the collaboration, and for developing the overall collaboration strategy, which is presented for approval to the executive board. The executive board consists of the leaders of the main collaboration activities. The composition of the EB, currently including the DUNE management team, IB chair, physics coordinator, beam interface coordinator, computing coordinator, near detector coordinator, and leaders of the \dword{fd} consortia, described below, is intended to ensure
that all stakeholders in the collaboration have a voice in the decision-making process. 
In the post-TDR phase of DUNE, the intention is that the consortium leaders and the coordinators of the other major collaboration activities will become elected positions.

To carry out design and construction work for the DUNE far detector, DUNE has  formed consortia of institutions that have taken responsibility for different detector subsystems. A similar structure will be formed for the DUNE near detector once the detector concept is selected. For the DUNE \dword{fd}, there are currently nine consortia, including three specific to \single, three specific to \dual, and three common to both technologies:
\begin{itemize}
\item (\single) \dwords{apa}, 
\item (\single) TPC \dword{ce}, 
\item (\single) \dword{pds}, 
\item (\dual) \dwords{crp}, 
\item (\dual) TPC electronics, 
\item (\dual) \dlong{pds} (PDS), 
\item \dword{hv} system, 
\item \dword{daq} system, 
\item \dword{cisc} system. 
\end{itemize}
It is possible that additional consortia may be added depending on the organization of computing and calibration systems. Each consortium has an overall leader, a technical lead, and a consortium board with representatives from each consortium institution. The consortia have full responsibility for their subsystems, including developing a full \dword{wbs}, understanding and documenting all interfaces with other systems, preparing final technical designs, and writing their respective sections of the \dword{tp} and the \dword{tdr}. Following approval of the  \dword{tdr}, they will be responsible for construction of their detector systems. 

\section{Schedule and Milestones} 

LBNF and DUNE are working toward three international project milestones:
\begin{itemize}
\item \maincavernstartexc{}: Start main cavern excavation in South Dakota; 
\item \firstfdmodstartinstall{}: Start installation of first \dword{fd} module; 
\item \beamturnon{}: Beam operation with two \dwords{detmodule}.
\end{itemize}
It is expected that these dates will be adjusted when the project baseline is defined. The key milestones to reach baseline status are:
\begin{itemize}
\item 2018 - Collect data with both \dword{protodune} detectors.
\item April 2019 - Submit \dword{tdr} for far \dwords{detmodule}.
\item July 2019 - Complete LBNC and NCG review of \dword{tdr}.
\item September 2019 - Present \dword{tdr} to RRB.
\item October 2019 - Conduct conceptual design (DOE CD-2/3b) review of LBNF and the USA scope of DUNE.
\end{itemize}
The \dword{tdr} for the near detector is expected to follow the \dword{fd} \dword{tdr} by approximately one year.

The schedule for the design and construction work for LBNF and DUNE has two critical parallel paths: one for the 
far site (South Dakota) 
another for the 
near site (Illinois). 
The schedule for the initial work is driven by the \dword{cf} design and construction at each site.

During the initial phase of the project, the far site \dword{cf} is advanced first. The Ross Shaft rehabilitation
work at \surf was recently halted at the 4850 level due to safety concerns, which have led to delays of two to four months. Early site preparation is timed to be completed 
in time to start excavation when the Ross Shaft rehabilitation work finishes. As each detector 
 cavern is excavated and sufficient utilities are installed, the cryostat and cryogenics system work proceeds, followed by detector installation, filling and commissioning. 
The first \dword{detmodule} is to be operational by 2024, with the second and third modules completed one and two years later, respectively.

The DOE project management process requires approvals at critical decision (CD) milestones that allow the LBNF/DUNE project to move to the next step. In spring 2018 LBNF near site \dword{cf} will seek CD-3b construction approval for Advanced Site Preparation to build the embankment. In 2020 LBNF and DUNE will seek to baseline the LBNF/DUNE scope of work, cost and schedule, as well as construction approval for the balance of the project scope of work. 

The project concludes with CD-4 approval to start operations.

\section{The DUNE Interim Design Report Volumes}


The DUNE \dword{tp} describes the proposed physics program and 
technical designs of the far detector in preparation for the full \dword{tdr} 
to be published in 2019.  
It is intended as an intermediate
milestone on the path to a full \dword{tdr}, justifying the technical choices that flow down from the high-level physics goals through requirements at all levels of the Project. These design choices will enable the DUNE experiment to make the ground-breaking discoveries that will help to  answer 
fundamental physics questions.

The  \dword{tp} is composed of three volumes. Volume 1 contains this executive summary, which describes 
the general aims of this document. The remainder of this first volume provides a more detailed description of the DUNE physics program that drives the choice of detector technologies. It also includes concise outlines of two overarching systems that have not yet evolved to consortium structures:  computing and calibration. 
Volumes 2  and 3 describe, for the \single and \dual, 
respectively, each module's subsystems, the technical coordination required for its design, construction, installation, and integration, and its organizational structure.

This \dword{tp}  represents the state of the design for the first three DUNE far detector modules at this moment
in time. The fourth module could employ a different \lartpc technology, taking into account  potential advances in technology to further optimize the sensitivity for physics discoveries. It is beyond the scope of this proposal. Also not covered in this proposal is the DUNE Near Detector, which constitutes an integral but less well developed portion of the experiment's  physics program. Separate \dwords{tdr}  for these detector systems will be
provided in the future.

\cleardoublepage

\chapter{DUNE Physics}
\label{ch:exec-summ-physics}

DUNE will address fundamental questions key to our understanding of the universe. These include:
\begin{itemize}
   \item {\bf What is the origin of the matter-antimatter asymmetry in the universe?}
      Immediately after the Big Bang, matter and antimatter were created equally, but 
      now matter dominates.  By studying the properties of neutrino and antineutrino oscillations, 
      LBNF/DUNE will pursue the current most promising avenue for understanding this asymmetry.
   \item {\bf What are the fundamental underlying symmetries of the universe?} 
      The patterns of mixings and masses between the particles of the standard model 
      is not understood. By making precise measurements of the mixing between the neutrinos 
      and the ordering of neutrino masses and comparing these with the quark sector, 
      LNBF/DUNE could reveal new underlying symmetries of the universe.
  \item{\bf  Is there a Grand Unified Theory of the Universe?} 
      Results from a range of experiments suggest that the physical forces observed today 
      were unified into one force at the birth of the universe.  Grand Unified Theories (GUTs), 
      which attempt to describe the unification of forces, predict that protons should decay, 
      a process that has never been observed. DUNE will search for proton decay in the range of 
      proton lifetimes predicted by a wide range of GUT models.
   \item{\bf How do supernovae explode and what new physics will we learn from a neutrino burst?}
      Many of the heavy elements that are the key components of life were created in the 
      super-hot cores of collapsing stars.  DUNE would be able to detect the neutrino bursts 
      from core-collapse supernovae within our galaxy (should any occur).  Measurements of the 
      time, flavor and energy structure of the neutrino burst will be critical for understanding 
      the dynamics of this important astrophysical phenomenon, as well as bringing information 
      on neutrino properties and other particle physics.
\end{itemize}

\section{Introduction: Scientific Goals}
\label{sec:exec-summ-physics-goals}

The DUNE scientific objectives are categorized into: the \textit{primary science program}, addressing the key science questions 
highlighted by the particle physics project prioritization panel (P5); 
a high-priority \textit{ancillary science program} that is 
enabled by the construction of LBNF and DUNE; and \textit{additional scientific objectives}, that may require further developments 
of the LArTPC technology. A detailed description of the physics objectives of DUNE is provided in Volume 2 of the DUNE \dword{cdr}\footnote{ \url{http://arxiv.org/abs/1512.06148}.}.

\subsection{The Primary Science Program}

The primary science program of DUNE  focuses on fundamental open issues in neutrino and astroparticle physics: 
\begin{itemize}
  \item Precision measurements of the parameters that govern $\nu_{\mu} \rightarrow \nu_\text{e}$ and
           $\overline{\nu}_{\mu} \rightarrow \overline{\nu}_\text{e}$ oscillations with the goal of
  \subitem -- measuring the charge-parity (CP) violating phase $\delta_\text{CP}$, where a value differing from zero or $\pi$ would represent the discovery of CP violation in the leptonic sector, providing a possible explanation for the matter-antimatter asymmetry in the universe;
  \subitem -- determining the neutrino mass ordering (the sign of $\Delta m^2_{31} \equiv m_3^2-m_1^2$), often referred to as the neutrino \textit{mass hierarchy}; and
  \subitem -- precision tests of the three-flavor neutrino oscillation paradigm through studies of muon neutrino disappearance 
    and electron neutrino appearance in both $\nu_\mu$ and $\overline{\nu}_{\mu}$ beams, including the 
    measurement of the mixing angle $\theta_{23}$ and the determination of the octant in which this angle lies.
    \item Search for proton decay in several important decay modes. The observation of proton decay would represent a ground-breaking discovery in physics, providing a portal to Grand Unification of the forces; and
    \item Detection and measurement of the $\nu_\text{e}$ flux from a core-collapse supernova within our galaxy, should one occur during the lifetime of the DUNE experiment.
\end{itemize}

\subsection{The Ancillary Science Program}

The intense neutrino beam from LBNF, the massive DUNE \lartpc far detector and the high-resolution
  DUNE near detector provide a rich ancillary science program, beyond the primary mission of the experiment. The ancillary science program includes
\begin{itemize}
     \item other accelerator-based neutrino flavor transition measurements with sensitivity to the beyond the standard model (BSM) physics, such as: non-standard interactions (NSIs); Lorentz violation,  \dword{cpt} violation, the search for sterile neutrinos at both the near and far sites, large extra dimensions, heavy neutral leptons;
 and measurements of tau neutrino appearance;
     \item measurements of neutrino oscillation phenomena using atmospheric neutrinos;
     \item a rich neutrino interaction physics program utilizing the DUNE near detector, including: a wide range of measurements of neutrino cross sections and studies of nuclear effects; and
     \item  the search for signatures of dark matter.
\end{itemize} 

Furthermore, a number of previous breakthroughs in particle physics have been serendipitous, in the sense that they were beyond the
original scientific objectives of an experiment. The intense LBNF neutrino beam and novel capabilities for both 
the DUNE near and far detectors will probe new regions of parameter space for both the accelerator-based and astrophysical frontiers, 
providing the opportunity for discoveries that are not currently anticipated.

\subsection{Context for Discussion of Science Capabilities in this Document}

The sections that follow highlight the projected capabilities of DUNE to realize the science program 
summarized above. These are documented in detail in \href{http://arxiv.org/abs/1512.06148}{Volume 2 
of the DUNE Conceptual Design Report} and in the following section.  Since publication of the CDR in late 2015, the DUNE science collaboration has undertaken a campaign to develop data analysis tools and strategies to aid 
in detector design optimization as well as to obtain a more rigorous understanding of experimental 
sensitivity.  This campaign is in progress as of this writing, and the outcomes will be reported 
as a component of the DUNE \dword{tdr} now in development.  Additionally, with 
currently-operating experiments beginning to reach peak fractional rates of integrated exposure, 
the rapid evolution of the world-wide experimental landscape in neutrino physics is particularly acute 
at present.  Thus, for the purposes of the present report, the discussion of capabilities here 
will reflect what is documented within the \dword{cdr} unless otherwise noted. In addition, the following section describes the status of the simulation and reconstruction strategies used to assess the physics requirements for DUNE.

\subsection{Strategies}
\label{sec:exec-summ-physics-ryan}
\subsubsection{Simulation and Reconstruction Strategies}
\label{sec:exec-summ-strat-simreco}

Liquid argon time projection chambers (\lartpc{}s) provide a robust and elegant method for measuring the properties of neutrino interactions above a few tens of MeV by providing three-dimensional (\threed) event imaging with excellent spatial and energy resolution.  The state of the art in \lartpc event reconstruction and particle identification is evolving rapidly and will continue to do so for many years.  The adoption of the common framework \larsoft{}\footnote{\larsoft, \url{http://inspirehep.net/record/1598096/export/hx}.} by several \lartpc experiments facilitates the exchange of tools and ideas.

The DUNE experimental design and physics program to be presented in the \dword{tdr} will be, in the main, based on a realistic end-to-end simulation and reconstruction chain.  This is in contrast to the highly parametrized methods used in the \dword{cdr}.  Note that the science case summarized in Chapter~\ref{ch:exec-summ-physics} of this \dword{tp} is still based on \dword{cdr}-era studies, as we intend to carry out the full refresh of the DUNE science case using our modern tools on the \dword{tdr} timeline (2019).  The primary exception to this strategy are sensitivity studies for BSM physics, which will largely continue to use parametrized analyses with updated assumptions to reflect our latest understanding.  A full description of the DUNE simulation and reconstruction tools will be included in the \dword{tdr}.  In this section, we give a brief summary of the techniques now in use.

\subsubsubsection{Simulation Chain}
Simulated events are created in four stages: event generation, \dword{geant4} tracking, TPC and \dword{pds} signal simulation, and digitization.  The first step is unique to each sample type while the remaining steps are common for all samples. Beam neutrino, atmospheric neutrino, and nucleon decay events are generated using \dword{genie} appropriately configured for each.  Supernova events are generated using the new low-energy, argon-specific MARLEY generator~\cite{marley}.  Cosmogenic events at depth are generated using MUSIC (Muon Simulation Code) \cite{MUSICPaper} and MUSUN (Muon Simulations Underground) \cite{Kudryavtsev:musun}.

Particle 4-vectors generated in the event generator step are passed to a {\sc GEANT4}-based detector simulation.  Energy depositions are converted to ionization electrons and scintillation photons, with recombination, electron attenuation, and diffusion effects included.  The response of the photon detectors is simulated using a ``photon library'' that has precalculated the likelihood for the propagation of photons from any point in the detector to any \dword{pds} element.  The response of the TPC induction and collection wires is based on a detailed GARFIELD~\cite{garfield} simulation.  Throughout, measurements from test stands or from operating \lartpc experiments such as ICARUS, LArIAT, and MicroBooNE are used to establish simulation parameters, where possible.

The raw signals on each wire are converted into \dword{adc} versus time traces by convolution with the field response and electronics response.  \dword{asic} electronics response is simulated with the BNL SPICE~\cite{spice} simulation.  The photon detector electronics simulation separately generates waveforms for each channel of a photon detector that has been hit by photons, with dark noise and line noise added.  The raw data are passed through hit finding algorithms that handle deconvolution and disambiguation to produce the basic data used by the downstream event reconstruction. \dword{pds} signals are reconstructed by searching for peaks on individual channels and then forming coincidences across channels. Techniques for matching the correct \dword{pds} signal to TPC signals to reconstruction $t_0$ are being developed, and early results from these tools can be seen in \voltitlespfd Chapter 5. 

\subsubsubsection{Reconstruction and Event Identification}
Several approaches to \lartpc reconstruction are under active development in DUNE and in the community at large.  En route to the \dword{tdr}, efforts on all fronts have been supported.  One  reconstruction path (``TrajCluster'' + ``Projection Matching'') forms two-dimensional (\twod) trajectory clusters in each detector view and then stitches these together into \threed objects.  Resulting objects are further characterized by, for instance, extracting $dE/dx$ information or comparing to electromagnetic shower profiles.  An alternative approach is provided by the Pandora reconstruction package~\cite{Marshall:2015rfa}, in which the reconstruction and pattern recognition task is broken down into a large number of decoupled algorithms, where each algorithm addresses a specific task or targets a particular topology.  Two additional algorithms (``WireCell'' and ``SpacePointSolver'') take a different approach and create \threed maps of energy depositions directly by solving a constrained system of equations governed by the geometry of the TPC wires.  Finally, several analyses are using deep learning and convolutional neural networks with promising early success, as these techniques are well suited to the type of data produced by \lartpc{}s.

Energy reconstruction is based on electron-lifetime-corrected calorimetry except in the case of muons where energy is determined from track range or (for uncontained muons) multiple Coulomb scattering.  Moving forward, more particle-specific energy estimators will be developed.

The output from all reconstruction algorithms is processed into standard ``ntuple files'' for use by analysis developers.  In the special case of long-baseline oscillation measurements, the CAFAna fitting toolkit developed originally for \nova is used to combine far detector and near detector information, to assess the impact of systematic uncertainties, and to ultimately produce neutrino oscillation sensitivities, discussed next.

\section{Long-Baseline Neutrino Oscillation physics program}
\label{sec:exec-summ-physics-osc}

Precision neutrino oscillation measurements lie at the heart of the DUNE scientific program.  The strengths of DUNE are (1) its discovery potential for \dword{cpv} in the neutrino sector, (2) its ability to resolve the neutrino mass ordering unambiguously, regardless of values of all other parameters governing neutrino oscillations, and (3) its unique ability to make high precision measurements of neutrino oscillations all within a single experiment.

\subsection{Experimental Context: Baseline, Configuration and Staging Scenario}

The \SIadj{1300}{\km} baseline, coupled with the wide-band
high-intensity neutrino beam from LBNF, establishes one of DUNE's key
strengths, namely sensitivity to the matter effect. This effect leads to a
discrete asymmetry in the \numu $\to$ \nue versus \anumu $\to$ \anue
oscillation probabilities, the sign of which depends on the presently
unknown mass hierarchy (\dword{mh}).  At \SI{1300}{\km}, the asymmetry,
\begin{equation}
\mathcal{A} = \frac{ P(\nu_\mu \rightarrow \nu_e)-P(\bar{\nu}_\mu \rightarrow \bar{\nu}_e)}{P(\nu_\mu \rightarrow \nu_e)+P(\bar{\nu}_\mu \rightarrow \bar{\nu}_e)}
\end{equation}
is approximately $\pm 40\%$ in the region of the peak flux in the
absence of CP-violating effects. This is larger than the maximal
possible CP-violating asymmetry associated with the CP-violating
phase, \deltacp, of the three-flavor PMNS mixing matrix in the region of
the peak flux. The CP asymmetry is larger in the energy regions below the peak
flux while the matter asymmetry is smaller. As a result, the LBNF
wide-band beam will allow unambiguous determination of both the \dword{mh} and
\deltacp with high confidence \textit{within the same experiment}, i.e., DUNE.   

The DUNE far detector will be built as four \ktadj{10} modules, which will
come online sequentially over the course of several years. 
This staged program enables an early scientific output from DUNE, 
initially focused on the observation of natural
sources of neutrinos, searches for nucleon decay and 
measurements of backgrounds. 
Two years after commissioning the first two detector modules, 
the LBNF neutrino
beam at Fermilab will  
begin sending neutrinos over the \kmadj{1300}
baseline, commencing the LBL oscillation physics program with a beam power of up to \SI{1.2}\MW{}. Upgrades to increase the beam power to \SI{2.4}\MW{} are planned to be in place six years later.
The early physics program
will be statistically limited and constraints from comparison of the $\nu_\mu$
disappearance spectrum with that from $\nu_e$ appearance will partially mitigate systematic uncertainties. The near detector is expected to come online in a timescale similar to that of the initial beam and will provide powerful constraints on the beam flux and neutrino interaction model, providing the
necessary control of systematic uncertainties for the full exploitation of LBNF/DUNE. 


The evolution of the projected DUNE sensitivities as a function of real time is estimated based on an assumed deployment plan
with the following assumptions:
\begin{itemize}
\item Year 1: \SI{20}\kt{} far detector fiducial mass, \MWadj{1.07} \GeVadj{80}
  proton beam with $1.47 \times 10^{21}$ protons-on-target per year 
  and initial near detector constraints;
\item Year 2: Addition of the third \ktadj{10} far detector module, for a total far detector fiducial mass of
  \SI{30}\kt;
\item Year 4: Addition of the fourth \ktadj{10} far detector module, for a total far detector fiducial mass of
  \SI{40}\kt, and improved systematic constraints from near detector analysis;
 \item Year 7: Upgrade of beam power to \SI{2.14}{\MW} for a \SIadj{80}{\GeV}
  proton beam.
\end{itemize}
With regard to the sensitivities reported here, it is assumed that the knowledge from the near detector can be
retroactively applied to previous data sets, such that each
successive improvement in the knowledge of systematic uncertainties
is applied to the full exposure
up to that point.

\subsection{Mass Hierarchy}
\label{sec:exec-summ-physics-mh-cpv}

As indicated above, unraveling the complex interplay of the matter effect with the degeneracies presented by multiple neutrino-mixing parameters with poorly known values will be a critical contribution from DUNE. Addressing the mass hierarchy question can be thought of as a first step toward this, one with intrinsic interest and import of its own.  While significant progress on this is expected from currently running experiments, DUNE's ability to resolve the \dword{mh} for all allowed values of mixing parameters is a key strength.

The discriminating power between the two \dword{mh} hypotheses is quantified
by the difference, denoted $\Delta \chi^2$, between the
$-2\log{\cal L}$ values calculated for the normal and inverted
hierarchies. As the sensitivity depends on the true value of the unknown
CP-violating phase, \deltacp, all possible values of \deltacp are
considered.  In terms of this test statistic\footnote{For the case of the \dword{mh} determination, the usual
  association of this test statistic with a $\chi^2$ distribution for
  one degree of freedom is not strictly correct; additionally the assumption of a
  Gaussian probability density 
  implicit in this notation is not exact.}, the \dword{mh}
sensitivity of DUNE for exposures of seven and ten years is
illustrated in Figure~\ref{fig:mhexec} for the case of normal
hierarchy and the NuFit 2016~\cite{Esteban:2016qun} best-fit value of \sinst{23} = 0.44. 
For this exposure, the DUNE determination of the \dword{mh} will be definitive for
the overwhelming majority of the  \deltacp and \sinst{23} parameter space.
Even for unfavorable combinations of the parameters, a statistically
ambiguous outcome is highly unlikely.  
\begin{dunefigure}[Summary of mass hierarchy sensitivities]{fig:mhexec}{The
    square root of the mass hierarchy discrimination metric $\Delta
    \chi^2$ is plotted as a function of the unknown value of \deltacp
    for exposures of seven and ten years  
    (left).  The minimum significance
    --- the lowest point on the curve on the left --- with which the mass
    hierarchy can be determined for all values of \deltacp and the significance for a true value of \deltacp=-$\pi$/2 as a
    function of years of running under the staging plan described in the text (right).
    The shaded regions represent the range in sensitivity corresponding to
    different true values of $\theta_{23}$.}
\includegraphics[width=0.49\textwidth]{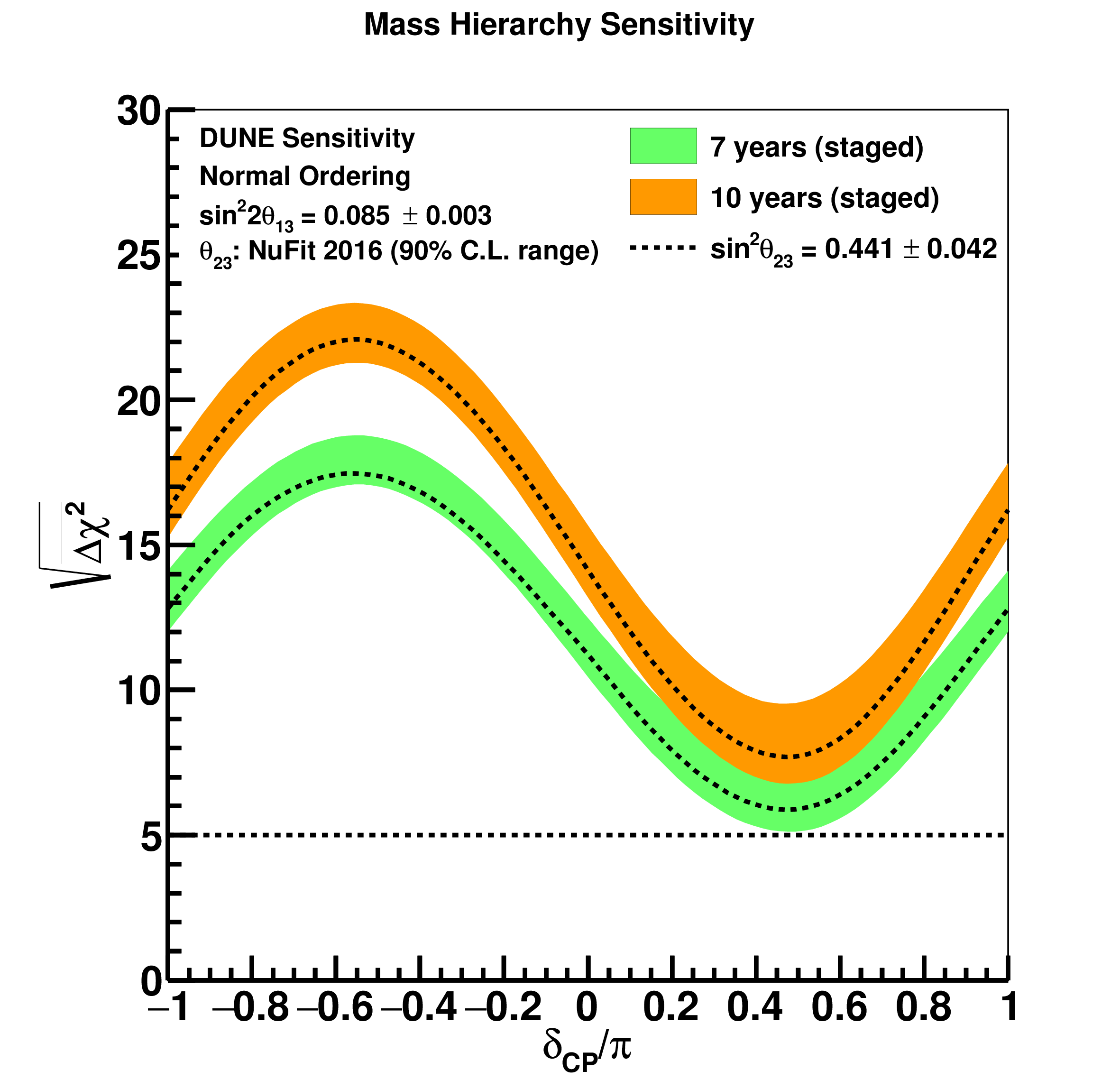}
\includegraphics[width=0.49\textwidth]{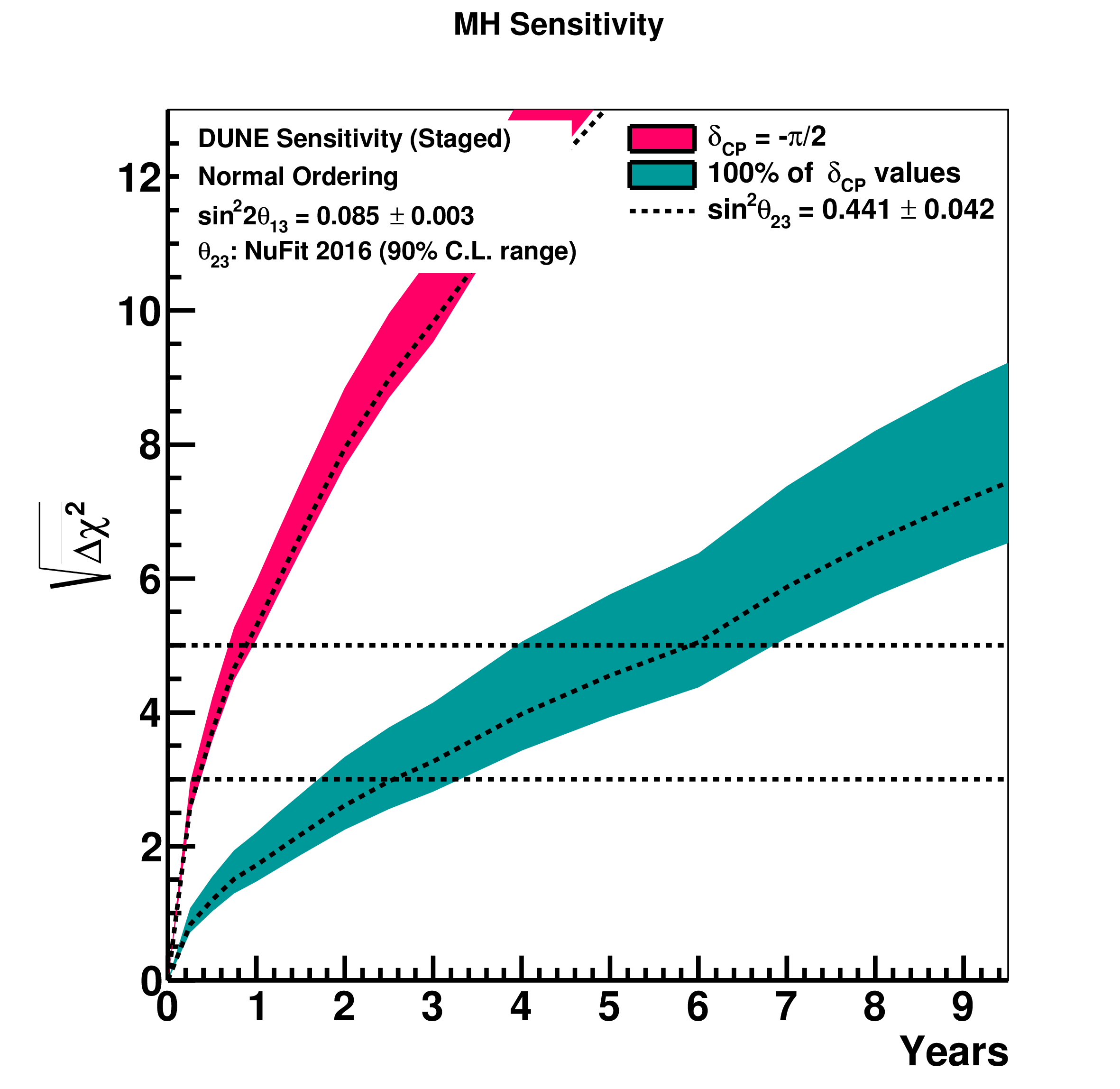}
\end{dunefigure}

Figure~\ref{fig:mhexec} shows the evolution of the sensitivity to the \dword{mh} determination as a function
of years of operation, for the least favorable scenario (blue band), corresponding to the case in which the \dword{mh} asymmetry is
maximally offset by the leptonic CP asymmetry. An exposure of \SI{209}~\ktMWyr{}  
(which corresponds to approximately five years of operation) is required to distinguish
between normal and inverted hierarchy with $|\Delta \chi^2| =
\overline{|\Delta \chi^2|} = 25$.  This corresponds to a $\geq
99.9996\%$ probability of determining the correct hierarchy. 
The dependence of the mass
hierarchy sensitivity on systematics is still under evaluation, but
current studies indicate only a weak dependence on the assumptions for 
the achievable systematic uncertainties. This indicates that a measurement of the unknown
neutrino mass hierarchy with very high precision can be carried out
during the first few years of operation.
Concurrent analysis of the corresponding atmospheric-neutrino
samples in an underground detector may improve the precision and
speed with which the \dword{mh} is determined.

\subsection{CP Violation}

DUNE will search for CP violation using the \numu to \nue and \anumu
to \anue oscillation channels, with two objectives.  First, DUNE aims
to observe a signal for leptonic CP violation independent of the
underlying nature of neutrino oscillation phenomenology. Such a signal
will be observable in comparisons of $\nu_\mu \rightarrow \nu_e$ and
$\bar{\nu}_{\mu} \rightarrow \bar{\nu}_e$ oscillations of the LBNF
beam neutrinos in a wide range of neutrino energies over the
\SIadj{1300}{\km} baseline.
Second,
DUNE aims to make a precise determination of the value of \deltacp
within the context of the standard three-flavor mixing scenario
described by the PMNS neutrino mixing matrix. Together, the pursuit of
these two goals provides a thorough and unprecedented test of the standard three-flavor
scenario.

\begin{dunefigure}[CP-violation sensitivity and $\delta_{\rm CP}$
  resolution as a function of exposure]{fig:execsummaryCP}{The
    significance with which CP violation can be determined for 75\% and 50\% of
    \deltacp values and for \deltacp=-$\pi$/2 (left) and the expected 1$\sigma$ resolution
    (right) as a function of exposure in years using the proposed
    staging plan outlined in this chapter. The shaded regions
    represent the range in sensitivity corresponding to
    different true values of $\theta_{23}$. The plots assume normal mass hierarchy.}
\includegraphics[width=0.49\textwidth]{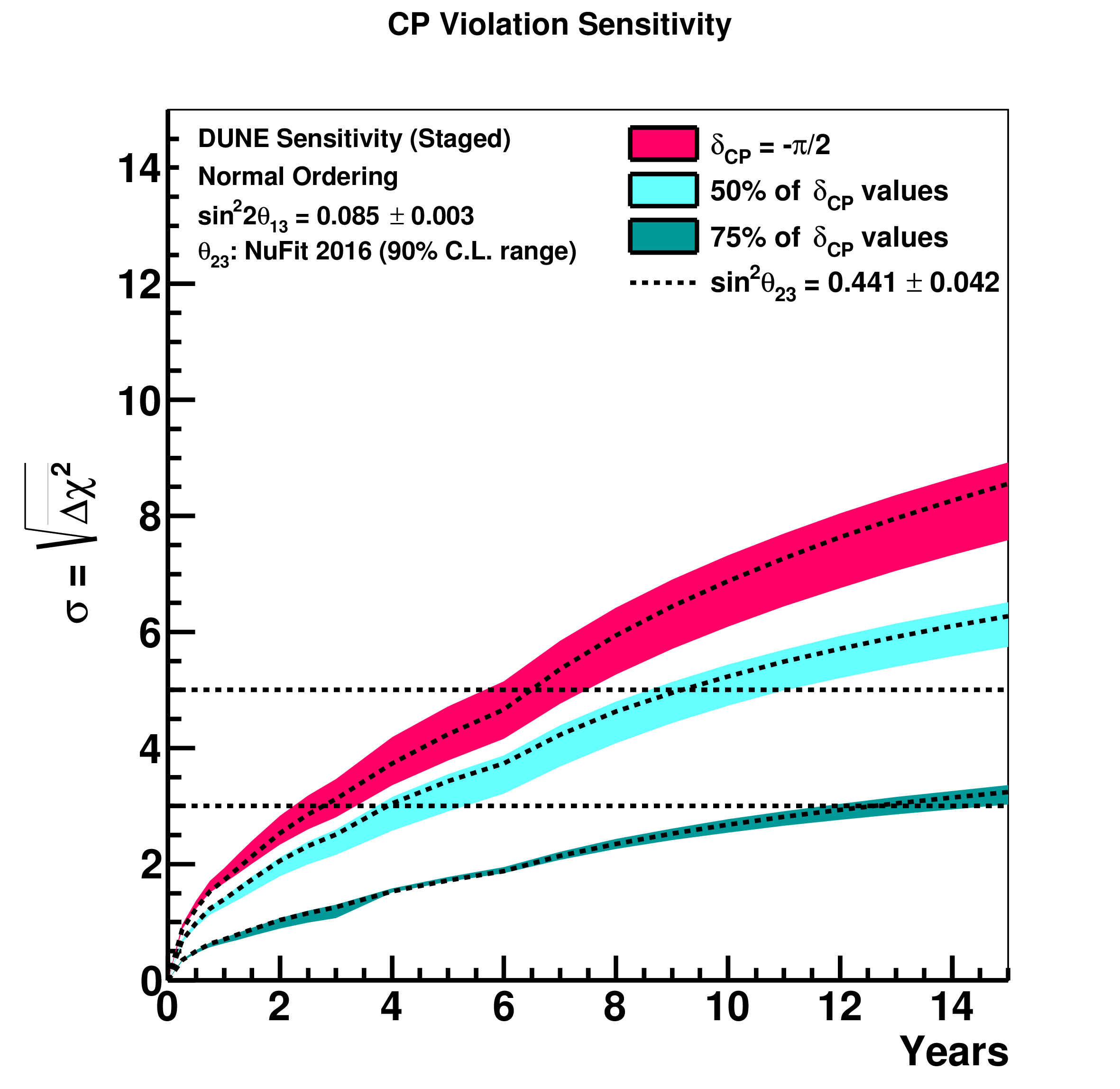}
 \includegraphics[width=0.49\textwidth]{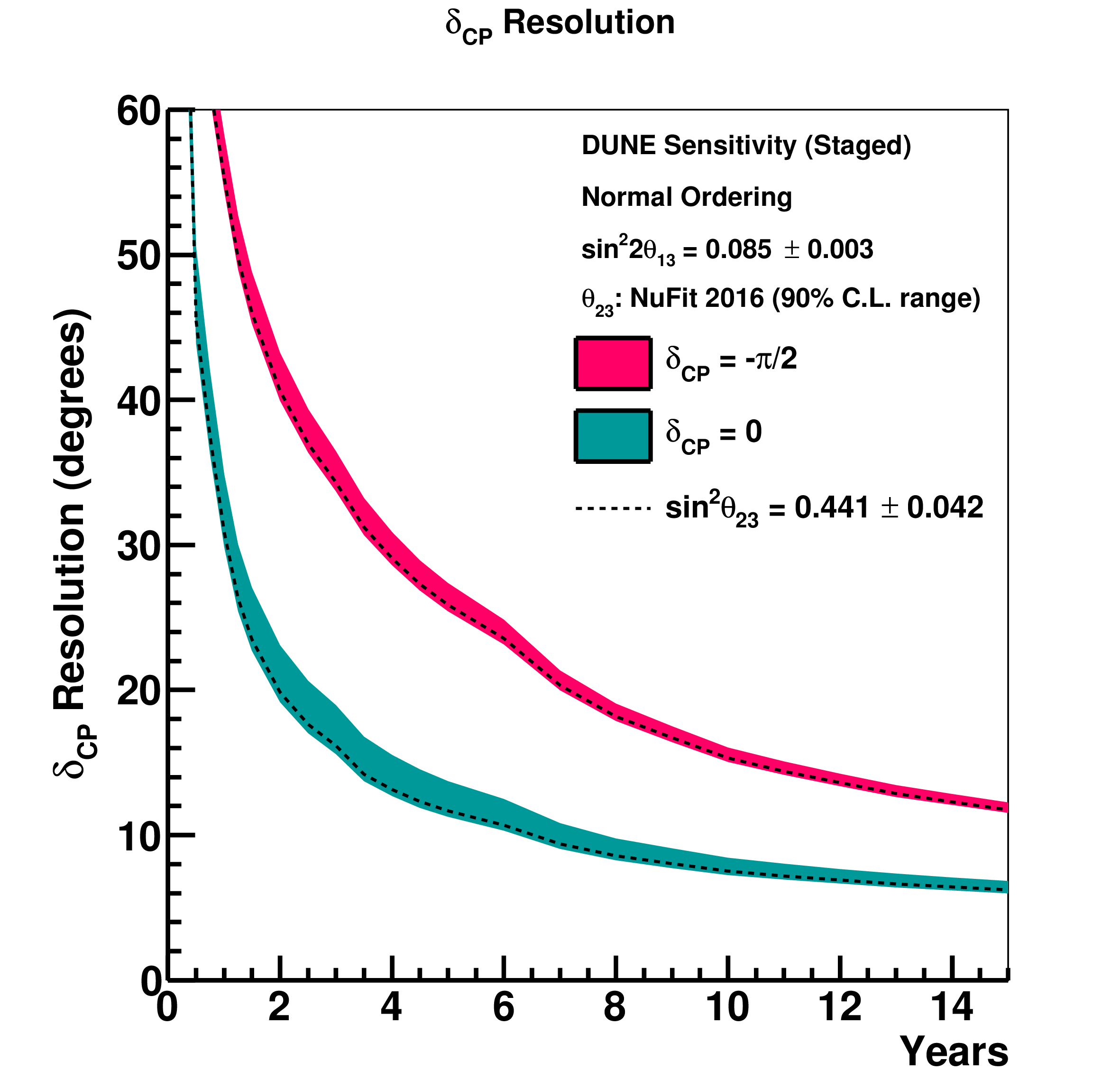}
\end{dunefigure}
Figure~\ref{fig:execsummaryCP} shows, as a function of time, the
expected sensitivity to CP violation expressed as the minimum significance
with which CP violation can be determined for 75\% and 50\% of
\deltacp values as well as the sensitivity when the true value of \deltacp=-$\pi$/2.
Also shown is the 1$\sigma$ resolution for \deltacp as a
function of time for $\delta_{\rm CP}=0$ (no CP violation) and
$\delta_{\rm CP}=-90^\circ$ (maximal CP violation). In both figures the staging scenario
described previously is assumed.  The exposure required to measure
$\delta_{\rm CP} = 0 $ with a precision better than $10^\circ$ is \SI{250}~\ktMWyr{} or about  
six and a half years of operation.  
A full-scope LBNF/DUNE operating with 
multi-megawatt 
beam power can in time achieve a precision 
comparable to the current precision on the CP phase in the
CKM matrix in the quark sector (5\%).

Table~\ref{tab:execosctable} summarizes the exposures needed to
achieve specific oscillation physics milestones, calculated 
for the current best-fit values of the known neutrino mixing parameters. 
For example, to reach $3\sigma$ sensitivity 
for 75\% of the range of \deltacp, a
DUNE exposure of \SI{775}~\ktMWyr{} or 12 years is needed. 
Changes in the assumed true value of
$\theta_{23}$ impact CP-violation and \dword{mh} sensitivities and can either reduce or increase the 
discovery potential for CP violation, as seen in Figure~\ref{fig:execsummaryCP}. To reach this level of sensitivity 
a highly capable near neutrino detector is required to control systematic uncertainties at a level lower than
the statistical uncertainties in the far detector. No experiment can provide coverage at 100\% of all 
\deltacp values, since CP-violating effects vanish as \mdeltacp\ $\to$ 0
or $\pi$.
\begin{dunetable}[Required exposures to reach oscillation physics
  milestones]{lcc}{tab:execosctable}{The exposure in mass (kt) $\times$ proton beam power
    (MW) $\times$ time (years) and calendar years assuming the staging plan described in this chapter needed to reach certain oscillation physics
    milestones. The numbers are for normal hierarchy using the NuFit 2016 best fit values of the known oscillation parameters.  }
Physics milestone & Exposure  & Exposure \\ \rowtitlestyle
  & (\ktMWyr{}) & (years)  \\ \toprowrule 
  $1^\circ$ $\theta_{23}$ resolution ($\theta_{23} = 42^\circ$) & 29  &  1\\ \colhline
  CPV at $3\sigma$ ($\delta_{\rm CP} = -\pi/2$)  & 77 &  3\\ \colhline
  \dword{mh} at  $5\sigma$ (worst point) & 209 & 6 \\ \colhline
  $10^\circ$ $\delta_{\rm CP}$ resolution ($\delta_{\rm CP} = 0$) & 252 & 
  6.5 \\ \colhline
  CPV at $5\sigma$ ($\delta_{\rm CP} = -\pi/2$)  & 253 & 
  6.5 \\ \colhline
  CPV at $5\sigma$ 50\% of \deltacp & 483 & 
  9 \\ \colhline
  CPV at $3\sigma$ 75\% of \deltacp & 775 & 12.5\\  \colhline
  Reactor $\theta_{13}$ resolution & 857 & 13.5 \\   
 ($\sin^2 2 \theta_{13} = 0.084 \pm 0.003$) &  &  \\  
\end{dunetable}

\subsection{Precision Measurement of Mixing Parameters}

In long-baseline experiments with \numu beams, the
magnitude of \numu disappearance and \nue appearance signals is
proportional to \sinstt{23} and \sinst{13},
respectively, in the standard three-flavor mixing scenario.  Current
\numu disappearance data are consistent with close to maximal
mixing, $\theta_{23} = 45^\circ$.  To obtain the best sensitivity to
both the magnitude of its deviation from $45^\circ$ as well the 
$\theta_{23}$ octant, a combined analysis of the two channels
is needed~\cite{Huber:2010dx}.  A DUNE detector with sufficient exposure will be able to
resolve the $\theta_{23}$ octant at the $3\sigma$ level or better for
$\theta_{23}$ values less than $43^\circ$ or greater than $48^\circ$.
The full LBNF/DUNE scope will allow $\theta_{23}$ to be measured with a precision of
$1^\circ$ or less, even for values within a few degrees of
$45^\circ$. 

To summarize, DUNE has great prospects to discover CP violation or, in the absence of the
effect, set stringent limits on the allowed values of \deltacp. 
DUNE will also determine the neutrino mass hierarchy with better
than a $5\sigma$~C.L. and provide precision measurements
of the mixing angles $\theta_{23}$ and $\theta_{13}$.

\section{Nucleon Decay and the GeV Scale Non-Accelerator Physics Program}

\subsection{Nucleon Decay}

Unification of three of the fundamental forces in the universe, the strong, 
electromagnetic and weak interactions, is a central paradigm for the current 
world-wide program in particle physics. Grand Unified Theories (GUTs), aiming 
at extending the standard model of particle physics to include a unified force 
at very high energies  (above $10^{15}$ GeV), predict a number of observable 
effects at low energies, such as nucleon 
decay \cite{Pati:1973rp,Georgi:1974sy,Dimopoulos:1981dw,Langacker:1980js,deBoer:1994dg,Nath:2006ut}. 
Several experiments have been searching for signatures of nucleon decay, with the best limits 
for most decay modes set by the Super-Kamiokande experiment \cite{Nishino:2012bnw}, 
which features the largest sensitive mass to date. 

The DUNE far detector, as the largest active volume of argon, 
will be highly sensitive to a number of possible nucleon decay modes, 
in many cases complementing the capabilities of large water detectors.  
In particular, the \lartpc technology is expected to be well-suited for observing 
nucleon decays into charged kaons, which can be identified with redundancy 
from their distinctive $dE/dx$ signature as well as by their decays.
A particularly interesting mode for the proton decay search with DUNE is 
$p\to K^+ \bar{\nu}$, which is expected to have a lifetime of the order of 
$>10^{33}$ years in SUSY models. This decay can be tagged in a \lartpc if a 
single kaon within a proper energy/momentum range can be reconstructed with 
its point of origin lying within the fiducial volume. 
Background events initiated by cosmic-ray muons can be controlled  by requiring 
no activity close to the edges of the TPCs and by stringent single kaon identification 
within the energy range of interest. 
Atmospheric neutrino-induced background with real kaon production will either have an 
associated strange baryon (for reactions with $\Delta S = 0$) whose decay can be 
reconstructed, or an identifiable charged lepton (for reactions with $\Delta S = 1$). 
Atmospheric neutrino-induced background may also arise from misidentification of protons 
from abundant quasielastic interactions.  Work is ongoing (see below) to understand 
in detail how to fully exploit the capabilities of the DUNE \lartpc FD modules to suppress the above 
backgrounds while maintaining the high acceptance necessary for discovery-level sensitivity.
Similarly, the DUNE FD is expected to have good sensitivity to other compelling 
nucleon decay modes, 
such as $n\to K^+ e$, $p\to l^+ K^0$, and $p\to \pi^0 e^+$, which are also under study. 

As is the case for the entire non-accelerator based physics program of DUNE, nucleon decay 
searches require efficient triggering and event localization (within the far detector) 
capabilities.  Given the 1-GeV energy release, the requirements on tracking and calorimetry 
capabilities are similar to those for the beam-based neutrino oscillation program described 
in the previous section.  
Experimental challenges such as particle identification to separate protons from kaons, 
the impact of final state interactions (FSI) on proton decay kinematics, and full control 
of the potential background processes, are presently under study with realistic detector simulations.
This includes opportunities for enhanced background rejection 
by using convolutional neural networks, as well as efforts to understand the 
uncertainty associated with the intra-nuclear cascade model used to simulate FSI. 
We expect that \dword{protodune} data taken with charged particle beams at CERN will 
provide important sample of events to train and improve on reconstruction algorithms 
and the resulting $dE/dx$ resolution.

Baryon number non-conservation can also be manifested by neutron-antineutron oscillations leading to subsequent antineutron annihilation with a neutron or a proton. This annihilation event will have a distinct signature of a vertex with several emitted light hadrons, with total energy of twice the nucleon mass and net momentum zero. The ability to re-construct these hadrons correctly and measure their energies is key to the identification of the signal event. The main background for these $n\bar n$ annihilation events is caused by atmospheric neutrinos. Most commonly mis-classified events are neutral current deep inelastic scattering events without a lepton in the final state. As above, nuclear effects and final state interactions make the picture more complicated and are probably the major component of the systematic uncertainty for the sensitivity studies carried out thus far.  Initial signal vs background discrimination studies have been performed using convolutional neural networks resulting in an equivalent sensitivity for the $n\rightarrow \bar n$ oscillation lifetime of $1.6 \times 10^9$~s at 90\% confidence level, a factor of 5 improvement on the current limit from Super-Kamiokande.
More information about particle identification and energy measurements will be provided by the \dword{protodune} experiment with charged particle beams. 


\subsection{Atmospheric neutrinos}

Atmospheric neutrinos are a unique tool to study neutrino oscillations: the oscillated flux contains all flavors of neutrinos and antineutrinos, is very sensitive to matter effects and to both $\Delta m^2$ parameters, and covers a wide range of $L/E$. In principle, all oscillation parameters could be measured, with high
complementarity to measurements performed with a neutrino beam. In addition, atmospheric neutrinos are available all the time, in particular before the beam becomes operational. The DUNE far detector, with its large mass and the overburden to protect it from atmospheric muon background, is an ideal tool for these studies.  Given the strong overlap in event topology and energy scale with beam neutrino interactions, most requirements will necessarily be met by the far detector design. Additional requirements include the need to self-trigger since atmospheric neutrino events are asynchronous with respect to accelerator timing, and a more stringent demand on neutrino direction reconstruction.

The sensitivity to neutrino oscillation parameters has been evaluated with a dedicated, but simplified, simulation, reconstruction and analysis chain. The fluxes of each neutrino species at the far detector location were computed taking into account oscillations. Interactions in the LAr medium were simulated with the GENIE
event generator. Detection thresholds and energy resolutions based on full simulations were applied to the outgoing particles, to take into account detector effects. Events were classified as fully contained (FC) or partially contained (PC) by placing the vertex at a random position inside the detector and tracking the lepton until it reached the detector edges. The number of events expected for each flavor and category is summarized in Table \ref{tab:atmnu-rates}.


\begin{dunetable}[Atmospheric neutrino rates]{cc}{tab:atmnu-rates}{Atmospheric neutrino event rates per year in \fdfiducialmass fiducial mass of the DUNE FD.}
Sample & Yearly Event Rate \\ \toprowrule
Fully contained atmospheric $e$-like & $1.6\times10^{3}$ \\ \colhline
Fully contained atmospheric $\mu$-like & $2.4\times10^{3}$ \\ \colhline
Partly contained atmospheric $\mu$-like & $7.9\times10^{2}$ \\
\end{dunetable}%

When neutrinos travel through the Earth, the MSW resonance influences electron neutrinos in the few-GeV energy range. More precisely, the resonance occurs for $\nu_e$ in the case of normal mass hierarchy (NH, $\Delta m^2 > 0$), and for $\bar \nu_e$ in the case of inverted mass hierarchy (IH, $\Delta m^2 < 0$). 
The mass hierarchy (\dword{mh}) sensitivity can thus be greatly enhanced if neutrino and antineutrino events can be separated. The DUNE detector will not be magnetized; however, its high-resolution imaging offers possibilities for tagging features of events that provide statistical discrimination between neutrinos
and antineutrinos. Two tags can be used to discriminate $\bar \nu$  and $\nu$ events: a proton tag (a signature of a likely neutrino interaction) and a positive muon decay tag (a signature of an antineutrino interaction since only 25\% of negative muons will decay).

Unlike for beam measurements, the sensitivity to \dword{mh} with atmospheric neutrinos is nearly independent of the CP-violating phase. The sensitivity comes
from both electron neutrino appearance as well as muon neutrino disappearance, and is strongly dependent on the true value of $\theta_{23}$. Despite the much smaller mass, DUNE would have comparable sensitivity to the \hyperk atmospheric neutrino analyses due to better event reconstruction.

These analyses will provide a complementary approach to beam neutrinos. Atmospheric neutrinos can help to lift degeneracies that can be present in beam analyses, for instance, through the fact that the \dword{mh} sensitivity is essentially independent of $\delta_{CP}$. Atmospheric neutrino data will be acquired even in the absence of the beam, and will provide a useful sample for the development of reconstruction software and analysis methodologies. Atmospheric neutrinos provide a window into a range of new physics scenarios, and can place limits on CPT violation \cite{Kostelecky:2003cr}, non-standard interactions, mass-varying neutrinos \cite{Abe:2008zza}, sterile neutrinos \cite{Abe:2014gda}, and Lorentz invariance violation \cite{Kostelecky:2011gq}.


\section{Supernova-Neutrino Physics and Astrophysics}

The neutrinos from a core-collapse supernova are emitted in a burst of
a few tens of seconds duration, with about half the signal emitted in the first
second. The neutrino energies are mostly in the range \numrange{5}{50}{MeV}, and the 
flux is divided roughly equally between the three known neutrino
flavors.  Current water and scintillator detectors are sensitive primarily to
electron antineutrinos ($\bar{\nu}_e$), with detection through the inverse-beta decay
process on free protons, 
 which dominates the interaction rate in these detectors.  Liquid argon has a unique sensitivity to
the electron-neutrino ($\nu_e$) component of the flux, via the absorption
interaction on $^{40}$Ar,
\begin{eqnarray*}
\nu_e +{}^{40}{\rm Ar} & \rightarrow & e^-+{}^{40}{\rm K^*}.
\end{eqnarray*} 
This interaction can in principle be tagged via the coincidence of the emitted
electron and the accompanying photon cascade from the $^{40}{\rm K^*}$
de-excitation.  About \num{3000} events would be expected in a \ktadj{40}
fiducial mass liquid argon detector for a supernova at a distance of
\SI{10}{\kilo\parsec}.  In the neutrino channel, the oscillation
features are in general more pronounced, since the $\nu_e$ spectrum is almost
always significantly different from the $\nu_\mu$ ($\nu_\tau$) spectrum 
in the initial core-collapse stages, to a larger degree than is the
case for the corresponding $\bar{\nu}_e$ spectrum.  
While $\nu_e$ absorption should represent $\sim$90\% of the signal, there are in addition other channels of interest, including $\bar{\nu}_e$ charged current, elastic scattering on electrons (which provides pointing information) and neutral-current
interactions which result in final-state deexcitation $\gamma$'s. 
Each channel has a distinctive signature in the detector, but in all cases, events appear as small (tens of cm scale) tracks and blips.   Figure~\ref{fig:snb} shows an example of a simulated event.  Section~\ref{sec:exec-summ-strat-simreco} describes reconstruction and calibration challenges for detecting these events.

Observation of the core-collapse neutrino burst in DUNE
will provide critical information on key
astrophysical phenomena~\cite{Mirizzi:2015eza}.  These include the neutronization burst, for which the initial sharp, bright flash
of $\nu_e$ from  $p+e^- \rightarrow n + \nu_e$
heralds the formation of a compact neutron star remnant.
The collapse of the proto-neutron star into a black hole would be signaled by a sharp cutoff in neutrino flux.
Shock wave effects, shock instability oscillations, turbulence effects, and transitions to quark stars could all produce observable features in the energy, flavor and time structure of the neutrino burst.
Furthermore, detection of the supernova burst 
neutrino signal in DUNE will provide information on neutrino properties: see reference~\cite{Mirizzi:2015eza}.  Most notably, several features offer multiple signatures of mass ordering~\cite{Scholberg:2017czd}, likely the most robust being the level of suppression of the  neutronization burst: see Figure~\ref{fig:snb}.  Because the neutronization burst is $\nu_e$-rich, this mass ordering signature is especially clean in DUNE.
``Collective effects'', due to self-induced transitions driven by \textit{neutrino-neutrino interactions} in the dense matter of the supernova, result in a rich phenomenology with multiple observables primarily at later times.

\begin{dunefigure}[Characteristics of the neutrino signal from core-collapse supernovae]{fig:snb}{Top: Event display of a \SI{30}{MeV} neutrino event simulated using MARLEY. Bottom: Expected event rates as a function of time for the electron-capture \dword{snb} model in~\cite{Huedepohl:2009wh} for \SI{40}{kt} of argon during early stages of the event -- the neutronization burst and early accretion phases, for which self-induced effects are unlikely to be important.  Shown is the event rate for the unrealistic case of no flavor transitions (blue), the event rate including the effect of matter transitions for the normal (red)  and inverted (green) hierarchies.  Error bars are statistical, in unequal time bins.}
\includegraphics[width=0.6\textwidth]{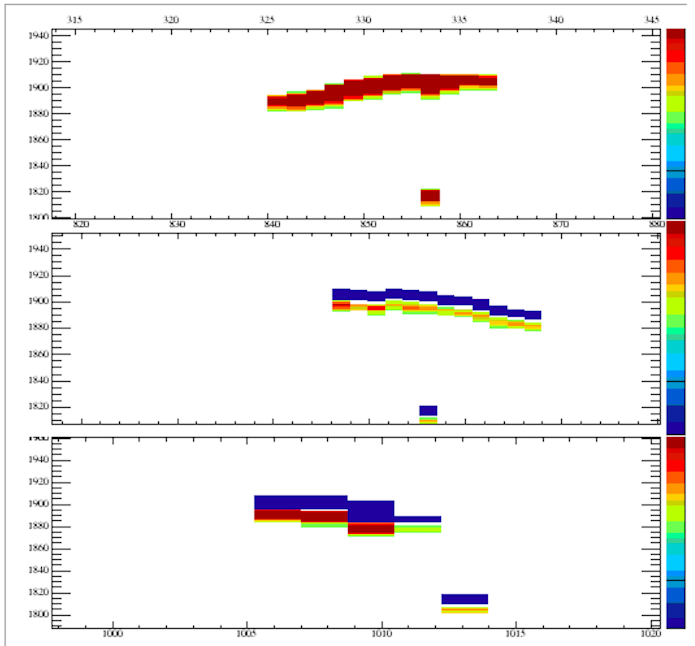}
\includegraphics[width=0.8\textwidth]{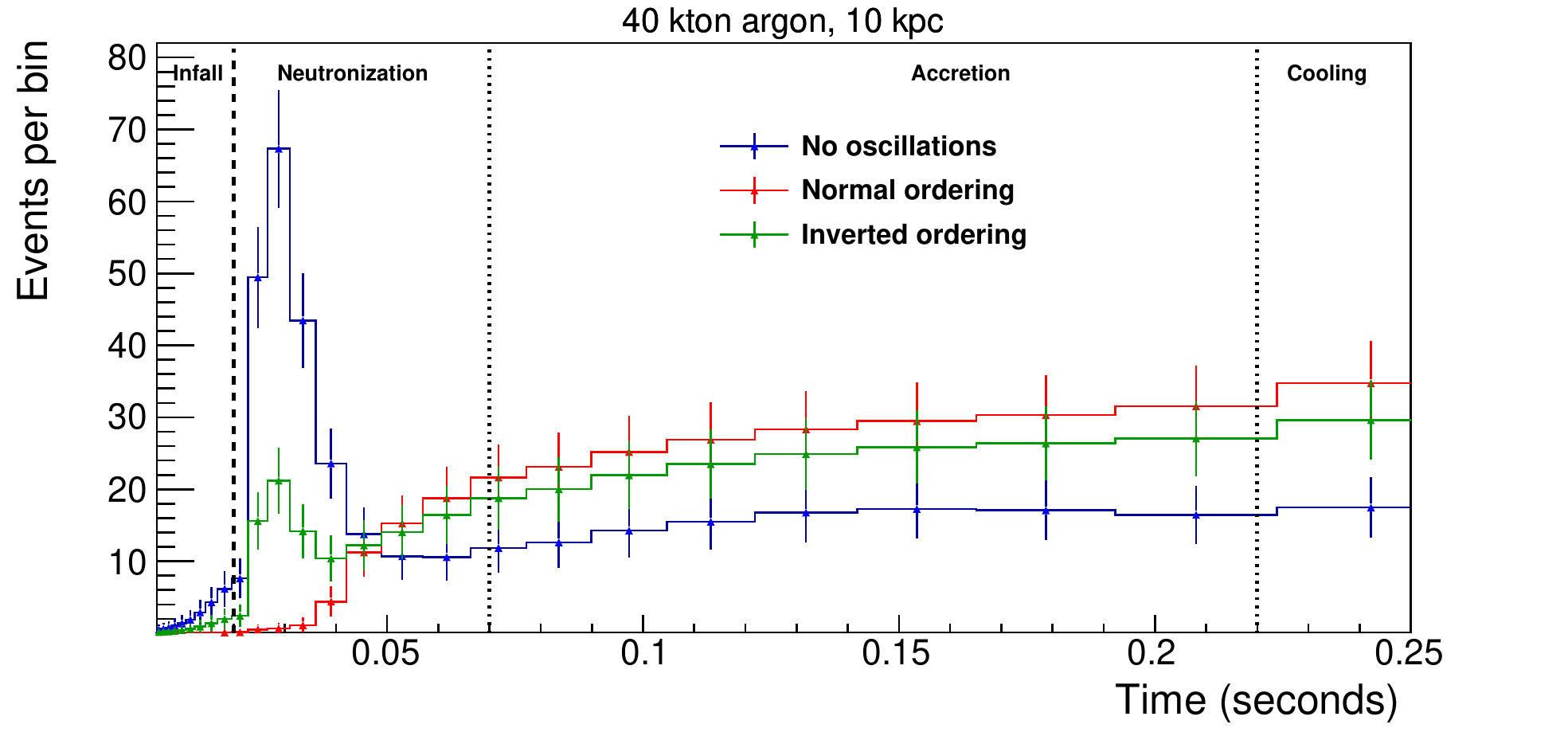}
\end{dunefigure}

Because no beam trigger is available for a supernova, efficient triggering and continuous data collection is critical for supernova neutrino burst physics. To fully capitalize on the physics opportunities, the DUNE far detector must provide event timing capability at the sub-millisecond level, must have spatial readout granularity sufficient to track electrons down to 5 MeV with good energy resolution, and must operate at noise levels and thresholds that allow detection and energy measurement for deexcitation gammas and nucleons at the MeV level.  The \lartpc technologies underlying the DUNE far detector conceptual designs is capable of meeting these requirements.

We note that information from DUNE will be highly complementary with neutrino burst information from other detectors, and furthermore multi-messenger astronomy information (from gravitational waves and a broad range of electromagnetic wavelengths) will combine to provide a full picture of a core-collapse event.

\section{Precision Measurements with the DUNE Near Detector Complex}
\label{sec:exec-summ-nd-precision-physics}

The DUNE near detector
will provide precision measurements of
neutrino interactions that are essential
for controlling the systematic uncertainties in the long-baseline neutrino 
oscillation physics program.  The near detector 
will include argon targets and will measure the absolute flux and energy-dependent
shape of all four neutrino species, \numu, \anumu, \nue and \anue,
to accurately predict for each species the
far/near flux ratio as a function of energy.  It will also measure the
four-momenta of secondary hadrons, such as charged and neutral mesons,
produced in the neutral- and charged-current interactions that
constitute the dominant backgrounds to the oscillation signals.

The near detector will also be the source of data for a rich program
of neutrino-interaction physics in its own right. For an integrated
beam intensity of \num{1e20} 
protons-on-target at \SI{120}{GeV}, the expected number of events per
ton is \num{170000} (\num{59000}) 
\numu (\anumu) charged-current and \num{60000} (\num{25000}) neutral-current interactions in the $\nu$ ($\overline\nu$) beam. With PIP-II, the integrated protons-on-target per year is
  expected to be around $1.1\times 10^{21}$ at \SI{120}\GeV. The mass
  of the argon target in the low-mass tracker option for the DUNE near detector is expected to be approximately
  80 tons. 
  These numbers correspond to \num{e5} neutrino interactions
on argon per year for the range of beam configurations and near detector
designs under consideration.  Measurement of fluxes, cross sections
and particle production over a large energy range of
\SIrange{0.5}{50}{\GeV} are the key elements of this program.  These
data will also help constrain backgrounds to proton-decay signals
from atmospheric neutrinos.  Furthermore, very large samples of events
will be amenable to precision reconstruction and analysis, and will be
exploited for sensitive studies of electroweak physics and nucleon
structure, as well as for searches for new physics in unexplored
regions, such as heavy sterile neutrinos, high-$\Delta m^2$
oscillations, and light Dark Matter particles. 

\section{Opportunities in Beyond the Standard Model Physics}
\label{sec:exec-summ-physics-bsm}
The unique combination of the high-intensity LBNF neutrino beam with DUNE's near detector and massive \lartpc far detector modules at a 1300 km baseline enables a variety of probes of BSM physics, either novel or with unprecedented sensitivity. This section describes a selection of such topics, and briefly summarizes how DUNE can make leading contributions in this arena.

\subsection{New Particle Searches}
{\bf Search for Low-mass Dark Matter.}
Various cosmological and astrophysical observations strongly support the existence of dark matter (DM) representing $\approx$27$\%$ of the mass-energy of the universe, but its nature and potential non-gravitational interactions with regular matter remain undetermined. 
The lack of evidence for weakly interacting massive particles (WIMP) at direct detection and the LHC experiments has resulted in a reconsideration of the WIMP paradigm. For instance, if  dark matter has a mass which is much lighter than the electroweak scale (e.g., below the GeV level), it motivates theories for  dark matter candidates that interact with ordinary matter through a new ``vector portal'' mediator.
High flux neutrino beam experiments, such as DUNE, have been shown to provide coverage of DM+$mediator$ parameter space which cannot be covered by either direct detection or collider experiments. 
Dark
matter particles can be detected in the near detector through neutral-current-like interactions 
either with electrons or nucleons in the detector material.
The neutrino-induced backgrounds can be suppressed using timing and the kinematics of the scattered electron.
These enable DUNE's search for light dark matter be competitive and complementary to other experiments.\\

{\bf Search for Boosted Dark Matter}
Using its large far detector, DUNE will be able to search for boosted dark matter.
A representative model is composed of heavy and light  dark matter components and the lighter one can be produced from the annihilation of the heavier one in e.g., the nearby sun or galactic centers.
Due to the large mass difference between the two  dark matter components, the lighter one is produced relativistically.
The incoming energy of the lighter  dark matter component can be high enough above the expected energy thresholds of DUNE in a wide range of parameter space. 
A first attempt at observing the inelastic boosted dark matter signal with \dword{protodune} prior to running DUNE is proposed in Ref.~\cite{Chatterjee:2018mej} and the same analysis strategy can be used in DUNE.

{\bf Heavy Neutral Leptons.}
The high intensity of the LBNF neutrino beam and the production of charm and bottom mesons in the beam enables DUNE to search for a wide variety of lightweight long-lived, exotic particles, by looking for topologies of rare event interactions and decays in the fiducial volume of the DUNE near detector. These particles include weakly-interacting heavy neutral leptons -- right-handed partners of the active neutrinos, vector, scalar, or axion portals to the Hidden Sector, and light super-symmetric particles.  
Assuming these heavy neutral leptons are the lighter particles of their hidden sector, they will only decay into standard model particles.
The parameter space explored by the DUNE near detector extends into the cosmologically relevant region complementary to the LHC heavy-mass dark-matter searches through missing energy and mono-jets. 


\subsection{Searches for Deviations from the PMNS Neutrino Mixing Paradigm}
{\bf Non-Standard Neutrino Interactions.}
Non-standard neutrino interactions, affecting neutrino propagation through the Earth, can significantly modify the data to be collected by DUNE as long as the new physics parameters are large enough~\cite{Masud:2015xva}. Leveraging its very long baseline and wide-band beam, DUNE is uniquely sensitive to  these probes. If the DUNE data are consistent with standard oscillations for three massive neutrinos, interaction effects of order 0.1 $G_F$ can be ruled out at DUNE~\cite{deGouvea:2015ndi,Coloma:2015kiu}. We note that DUNE might improve current constraints on $\epsilon_{\tau e}$ and $\epsilon_{\mu e}$ by a factor 2-5~\cite{Farzan:2017xzy}.

{\bf Non-Unitarity.} 
A generic characteristic of most models explaining the neutrino mass
pattern is the presence of heavy neutrino states, additional to the
three light states of the standard model of particle
physics~\cite{Minkowski:1977sc,Mohapatra:1979ia,Yanagida:1979as,GellMann:1980vs}. This implies a deviation from unitary of the $3\times3$ PMNS matrix, which can be particularly sizable the lower the mass of the extra states are~\cite{Mohapatra:1986bd,Akhmedov:1995vm,Akhmedov:1995ip,Malinsky:2005bi}.
For values of the unitarity deviations of order $10^{-2}$, this would decrease the expected reach of DUNE to the standard parameters, although stronger bounds existing from charged leptons would be able to restore its expected performance~\cite{Blennow:2016jkn,Escrihuela:2016ube}.

{\bf Violations of Lorentz or CPT Symmetry.}
CPT symmetry, the combination of charge conjugation, parity and time reversal, is a cornerstone of our model building strategy and therefore the repercussions of its potential violation will severely threaten the standard model of particle physics. DUNE can improve the present limits on Lorentz and CPT violation by several orders of magnitude~\cite{Kostelecky:2003cr,Kostelecky:2011gq,Streater:1989vi,Barenboim:2002tz,Barenboim:2017ewj}, contributing
as a very important experiment to test one of the deepest results of quantum field theory.

{\bf Active-Sterile Neutrino Mixing.}
Experimental results in tension with the three-neutrino-flavor paradigm~\cite{Aguilar:2001ty,Aguilar-Arevalo:2013pmq,Acero:2007su,Mention:2011rk}, which may be interpreted as mixing between the known active neutrinos and one or more {\it sterile} states, have led to a rich and diverse program of searches for oscillations into sterile neutrinos.
DUNE is sensitive over a broad range of potential sterile neutrino mass splittings by looking for disappearance of CC and NC interactions over the long distance separating the near and far detectors, as well as over the short baseline of the near detector. 
With a longer baseline, a more intense beam, and a high-resolution large-mass far detector, compared to previous experiments, DUNE provides a unique opportunity to improve significantly on the sensitivities of the existing probes, and greatly enhance the ability to map the extended parameter space if a sterile neutrino is discovered.

{\bf Large Extra Dimensions.}
DUNE can search for or constrain the size of large extra-dimensions (LED) by looking for distortions of the oscillation pattern predicted by the three-flavor paradigm. These distortions arise through mixing between the right-handed neutrino Kaluza-Klein modes, which propagate in the compactified extra dimensions, and the active neutrinos, which exist only in the four-dimensional brane~\cite{Dvali:1999cn}. Searching for these distortions in, for instance, the $\nu_\mu$~CC disappearance spectrum should provide significantly enhanced sensitivity over existing results from the MINOS/MINOS+ experiment~\cite{Adamson:2016yvy}.

{\bf Neutrino Trident Production.}
The intriguing possibility that neutrinos may be charged under new gauge symmetries beyond the standard model $SU(3)_c\times SU(2)_L\times U(1)_Y$, and interact with the corresponding new gauge bosons can be tested with unprecedented precision by DUNE through near detector measurements of neutrino-induced di-lepton production in the Coulomb field of a heavy nucleus, also known as  neutrino trident interactions~\cite{Altmannshofer:2014pba}. Although this process is extremely rare (SM rates are suppressed by a factor of $\sim 10^{-5}-10^{-7}$ with respect to CC interactions), the CHARM-II collaboration \cite{Geiregat:1990gz} and the CCFR collaboration \cite{Mishra:1991bv} both reported detection of several trident events ($\sim 40$ events at CCFR) and quoted cross-sections in good agreement with the SM predictions. With a predicted annual rate of over 100 di-muon neutrino trident interactions at the near detector, DUNE will be able to measure deviations from the SM rates and test for the presence of new gauge symmetries.


\section{Summary}

In summary, the primary science goals of DUNE are drivers for the
advancement of particle physics. DUNE's physics program brings together the 
international neutrino community as well as leading experts in nucleon decay
and particle astrophysics to explore key questions at the forefront of
particle physics and astrophysics.

The questions being addressed are of wide-ranging consequence: the origin of flavor and the generation
structure of the fermions, the physical mechanism that provides the CP
violation needed to generate the baryon asymmetry of the universe, 
and the high-energy physics that would lead to the instability
of matter.  Achieving these goals requires a dedicated, ambitious and
long-term program.  
The staged implementation of
the far detector as four 10~kt modules will enable
exciting physics in the intermediate term, including a definitive mass
hierarchy determination and possibly a measurement of the CP phase 
(assuming neutrinos are CP-violating),  
while providing the fastest route toward achieving the
full range of DUNE's science objectives.  


\cleardoublepage

%
\chapter{DUNE Software and Computing }

\section{Overview}

Offline computing for  \dword{dune} faces new and considerable challenges due to the large scale and diverse physics goals of the experiment.  In particular, the advent of \dwords{lartpc}  with exquisite resolution and sensitivity, combined with enormous physical volumes, creates challenges in acquiring and storing large data volumes and in analyzing and reducing them.  The computing landscape is changing rapidly, with the traditional HEP architecture of individual cores running Linux being superseded by multi-core machines and GPUs. At the same time, algorithms for \dword{lar} reconstruction are still in their infancy and developing rapidly.  As a result, we have reason to be optimistic about the future, but we are not able to predict it accurately.  The \dword{protodune} single and dual phase tests at CERN in the fall of 2018 will provide a wealth of data that will inform the future evolution of  the \dword{dune} computing models.

The  \dword{dune} offline computing challenges can be classified in several ways.  We will start with the different detector and physics configurations that drive the large scale data storage and reconstruction. 
This discussion leans heavily on the \dword{daq} design described in \voltitlespfd and \voltitledpfd  of the \dword{dune} \dword{tp}. 

\subsection{Detectors}


The \dword{dune} experiment will consist of four \larmass \dword{fd} modules located at  \surf, using either \dword{sp} or \dword{dp} \lartpc{}s, and a not fully specified near detector at \fnal. The proposed \dword{fd} \dword{spmod} has an active mass of \spactivelarmass and the \dword{dpmod} has an active mass of \dpactivelarmass{}.

\subsubsection{Single-phase estimates}
Each \single module will consist of three rows of anode planes with a cathode plane between  each anode plane pair. The planes are spaced \SI{3.5}{m} 
apart and operated at \SI{180}{kV} for a \SI{500}{V/cm} drift field. 
The anode planes are made up of \dwords{apa} which are 6.3~m tall by 2.3~m wide and have 2,560 readout channels each. Each channel is sampled with 12-bit precision every 500 nsec. 
For modules of this size, drift times in the liquid argon are of order 2.5~ms and raw data sizes before compression are of order 6~GB per module per 5.4~ms readout window.  With no triggering and no zero suppression or compression, the raw data volume for four modules would be of order 145~EB/year.

\subsubsection{Dual-phase technology}

For \dual, electrons may traverse the full height of the cryostat, emerge from the liquid and be collected,  after gas amplification, on a grid of instrumented strips at the top of the \dword{detmodule}.  The \dword{wa105} test of this technology ran successfully in the summer of 2017\cite{Murphy:20170516}. 
Each \dpactivelarmass module will have \dpnumcrpch channels. Drift time in the \lar is \SI{7.5}{ms}. Given \num{20000} samples in an \SI{8}{ms} readout, the uncompressed event size is \SI{4.2}{GB} (for one  drift window).  Due to gas amplification, the \dword{s/n} ratio is quite high, allowing lossless compression to be applied at the front-end  with a compression factor of ten, bringing the event size/module to  \SI{0.42}{GB}. Recording the entire module drift window can be considered a pessimistic figure, since events are normally contained in smaller detector regions. A \dword{fd} module can be treated as \num{20} smaller detectors (witha similar number  of readout channels to the prototype currently being constructed at CERN), running in parallel, each one defining a \dword{roi}. For beam or cosmic events it is possible to record only the interesting \dwords{roi} with the compressed size of a single \dword{roi} being \SI{22}{MB}.

\subsubsection{Beam coincident rates}

Requiring  coincidence with the \dword{lbnf} beam will reduce the effective live-time from the full 1.2-1.5 sec beam cycle to a 5.4~ms (8~ms for DP)  readout window coincident with the 10 microsecond beam spill, leading to an uncompressed data rate for beam-coincident events of around 20~GB/sec for four 17~kT single-phase detector modules ($\sim$ 16~GB/s for dual-phase), still too high to record permanently.
Only a few thousand true beam interactions in the far detectors are expected each year.  Compression and conservative triggering based on photon detectors and ionization should reduce the data rate from beam interactions by several orders of magnitude without sacrificing efficiency.

\subsubsection{ Near detector} The near detector configuration is not yet defined  but we do have substantial experience from T2K and  \dword{microboone} at lower energies, and  \dword{minerva} at the  \dword{dune} beam energies on cosmic and beam interactions under similar conditions.  We can expect that a near detector will experience \numrange{5}{10} beam interactions per beam pulse and non-negligible rates of cosmic rays, spread over an area of a few square meters. \dword{microboone} experience and \dword{protodune} simulations indicate compressed event sizes of 100-1000 MB, leading to yearly data volumes of 2-20 PB.  Storing and disentangling this information will be challenging but comparable to the \dword{protodune} data expected in 2018.



%


\subsection{Physics Challenges}

 \dword{dune} physics will consist of several different processes with very different rates and event sizes.

\subsubsection{Long-baseline neutrino oscillations} Neutrino oscillation measurements will require a near detector operating in a high rate environment and far detectors in which beam-coincident events are rare but in time with the beam spill and of sufficient energy to be readily recognizable.  Studies discussed in the \dword{daq} section of  \dword{tp} Volumes 2 and 3 indicate that high efficiencies are achievable at an energy threshold of \SI{10}{MeV}, leading to event rates for beam-initiated  interactions of $\sim$\,\num{6400}/year and an uncompressed data volume of around \SI{30}{TB/year} per \larmass \dword{spmod}.

Tables \ref{tab:daq-data-rates-sp} and  \ref{tab:daq-data-rates-dp} summarize the event and data rates after appropriate filtering from the  \dword{daq} section of Volumes 2 and 3 of the \dword{tp}.

\begin{dunetable} [Uncompressed data volumes/year for one \dword{sp} module.]
  {p{0.30\textwidth}p{0.13\textwidth}p{0.4\textwidth}}
  {tab:daq-data-rates-sp} {Anticipated annual, uncompressed data rates
    for a single \dword{sp} module (from the \dword{spmod} \dword{tp} volume). The rates for normal (non-\dword{snb} triggers)
    assume a readout window of \SI{5.4}{\ms}. 
    In reality, lossless compression will be applied which is expected
    to provide as much as a $4\times$ reduction in data volume for each \dword{sp} module.}
  Event Type  & Data Volume \si{\PB/year} & Assumptions \\ \toprowrule
  Beam interactions & 0.03 & 800 beam and 800 rock muons; \SI{10}{\MeV} threshold in coincidence with beam time; include cosmics\\ \colhline
  Supernova candidates & 0.5 & 30 seconds full readout, average once per month \\ \colhline
 Cosmics and atmospherics & 10 &  \SI{10}{\MeV} threshold\\ \colhline
  Radiologicals  ($^{39}$Ar and others.& $\le$1 & fake rate of $\le$100 per year\\ \colhline
 Front-end calibration & 0.2 & Four calibration runs per year, 100 measurements per point \\ \colhline
 Radioactive source calibration & 0.1 & source rate $\le$10~Hz; single fragment readout; lossless readout \\ \colhline
 Laser calibration & 0.2 & 1$\times$10$^6$ total laser pulses, lossy readout \\ \colhline
 Random triggers & 0.06 & 45 per day\\ \colhline
 Trigger primitives & $\le$6 &  all three wire planes; 12 bits per primitive word; 4 primitive quantities; $^{39}$Ar-dominated\\ \colhline
\end{dunetable}

\begin{dunetable} [Uncompressed data volumes/year for one \dword{dp} module.]
{p{0.30\textwidth}p{0.13\textwidth}p{0.4\textwidth}}
{tab:daq-data-rates-dp} {Anticipated annual, uncompressed data rates
for  one \dword{dp} module. The rates for normal (non-\dword{snb} triggers)
assume a readout window \SI{7.5}{\ms}.
These numbers do not include lossless compression which  is expected
to provide as much as a  $10\times$ reduction in data volume. }
Event Type & Data Volume \si{\PB/year} & Assumptions \\ \toprowrule
Beam interactions (DP) & 0.007 & 800 beam and 800 rock muons; this becomes 700 GB/year if just 2 ROIs/event are dumped on disk \\
\colhline
Supernova candidates (DP) & 0.06 & 10 seconds full readout, all ROIs are dumped on disk \\
\colhline
Cosmics/atmospherics (DP)& 2.33 & This becomes 230 TB/year if two ROIs/event are dumped on disk \\
\colhline
Radiologicals ($^{39}$Ar and other).& $\le$1 & fake rate of $\le$100 per year\\ \colhline
Miscellaneous calibrations& 0.5 & similar to SP\\ \colhline
Random triggers & 0.02 & 45 per day\\ \colhline
Trigger primitives & $\le$6 &  similar to SP \\ \colhline
\end{dunetable}

\subsubsection{Processes not in synchronization with the beam spill} These include supernova physics, atmospheric neutrinos, proton decay, neutron conversion and solar neutrinos.  These processes are generally at lower energy, making triggering more difficult, and asynchronous, thus requiring an internal or external trigger.  In particular, supernovae signals will consist of a large number of low-energy interactions spread throughout the far detector volume over a time period of 1-30 seconds. Buffering and storing 10 seconds of data would require around 2000 readout windows, or around 50~TB per supernova readout.  At a rate of one such event/month, this is 600~TB of uncompressed data per module/year.

\subsubsection{
Calibration}
In addition to physics channels, continuous calibration of the detectors will be necessary.  It is likely that, for the far detectors, calibration samples will  dominate the data volume. Cosmic-ray muons and atmospheric neutrino interactions will provide a substantial sample for energy and position calibration.  Dedicated runs with radioactive sources and laser calibration will also generate substantial and extremely valuable samples. Table \ref{tab:daq-data-rates-sp} includes estimates for the single-phase far detector.   Cosmic ray and atmospheric neutrino signals collected for calibration make up the bulk of the uncompressed \dword{sp} data volume at $\sim$10~PB/year per 17~kT module and will dominate the rates from the far detectors.  

\subsubsection{Zero suppression}

The data volumes discussed above are for un-zero-suppressed data.  Efficient zero suppression mechanisms can substantially reduce the final data volume but previous experience in HEP indicates that signal processing must be done carefully and often happens well into data-taking when the data are well understood.  Experience from  \dword{microboone} and the \dword{protodune} experiments will aid us in developing these algorithms, but it is likely that they will be applied later in the processing chain for single-phase.  No zero-suppression is planned for dual-phase.

The constrained environment at \surf motivates a model where any further data reduction via zero-suppression is done downstream, either on the surface or after delivery to computing facilities at FNAL or elsewhere. This could be analogous to the HLT's used by LHC experiments. The relative optimization of data movement and processing location is an important consideration for the design of both the \dword{daq} and offline computing.

\subsection{Summary}
In summary, uncompressed data volumes will be dominated by calibration for the far detector modules ($\sim$10~PB/year/module \dword{sp} or $\sim$3 PB/year/module \dword{dp}) and by beam and cosmic ray interactions in the near detectors (2-20 PB/year).   With four \dword{fd} modules, but a conservative factor of four for compression, a total compressed data volume of 12-30~PB per year is anticipated.

Data transfer rate from the far detector to \fnal  is limited to \surffnalbw, which is consistent with projected network bandwidths in the mid 2020s and a limit of 30~PB/year raw data stored to tape.  

\section{Building the Computing Model}\label{sw:bld-cmp-mdl}

The  \dword{dune} computing model is a work in progress.  Major advances will take place over the next year on several fronts, with data from \dword{protodune} and the full incorporation of lessons from  \dword{microboone} into \dword{larsoft} .

The overall model can be divided into several major parts:  infrastructure, algorithms and adaption for the future.  These are in different stages of planning and completion.  An overarching theme is evaluating and using community codes and resources wherever possible.

\subsection{Infrastructure}
This category includes the wealth of databases, catalogs, storage systems, compute farms, and the software that drives them.  HEP fortunately has already developed much of this technology and our plan is to adopt pre-existing systems wherever possible.  As  \dword{dune} is a fully global experiment, integrating the resources of multiple institutions is both an opportunity and a logistical challenge.

Current plans have the primary raw data repository at \fnal, with derived samples and processing distributed among collaborating data centers.  For \dword{protodune}, raw data will also be stored at CERN.  Data processing is being designed to run on HEP grid resources, with significant ongoing effort to containerize it so that DUNE can make use of heterogenous resources worldwide.

\subsubsection{Core HEP code infrastructure}
Shared HEP infrastructure will be used wherever possible, notably the ROOT\cite{root} and \dword{geant4}~\cite{geant4,Allison:2006ve} frameworks.   For event simulation, we plan to use and contribute to  the broad range of available generators (e.g., GENIE~\cite{Andreopoulos:2009rq}, NuWro~\cite{NuWro2012}) shared with the worldwide neutrino community.

In addition, we are using the infrastructure developed for the LHC and the Intensity Frontier experiments at Fermilab, notably grid infrastructure,  the \dword{art} framework and the \dword{sam} data catalog.  The \dword{nova} and \dword{microboone} experiments are already using these tools for distributed computing and the \dword{protodune} data challenges are integrating CERN and Fermilab storage and CPU resources.  We are now extending this integration to the  institutions within the collaboration who have access to substantial storage and CPU resources.

\subsection{Algorithms}
This category includes the simulations, signal processing and reconstruction algorithms needed to reconstruct and understand our data. Algorithms are currently under development but are  informed by existing general codes (for example GENIE and \dword{geant4}) and the experience of other liquid argon experiments as encoded in the shared \dword{larsoft}  project.  Simulations are quite advanced but full understanding of reconstruction algorithms will need real data from \dword{protodune}. 

\subsubsection{External products}
The image-like nature of TPC data allows us to make use of external machine-learning systems such as TensorFlow\cite{DBLP:journals/corr/AbadiABBCCCDDDG16}, 
Keras\cite{chollet2015keras} and Caffe\cite{Jia:2014:CCA:2647868.2654889}.  Many of these are being evaluated for pattern recognition. While they encapsulate a wealth of experience, they are also somewhat volatile as they are driven by needs of non-HEP users.  We must have access to and must preserve the underlying source codes in order to maintain reproducibility.

\subsection{Adaptability}
As the experiment will be expected to run at least two decades past the present we must be prepared for the inevitable and major shifts in the underlying technologies that will occur. The ability to keep operating over decades almost requires that we emphasize open source over proprietary technologies for most applications.  DUNE should also plan to be able to utilize and support a large range of compute architectures in order to fully utilize the resources available to the collaboration.

Table \ref{sw:computingTasks} summarizes the responsibilities of the software and computing group and reconstruction and algorithms groups for both \dword{dune} and \dword{protodune}.

\begin{dunetable}[Computing tasks]
{p{0.5\textwidth}p{0.3\textwidth}} 
{sw:computingTasks}
{Computing Tasks - see the \dword{protodune} section for details on current status.}
Task & Status \\
\toprowrule

Code management & in place  \\ \colhline
Documentation and logging of \dword{daq} and detector configurations & in design \\ \colhline
Data movement & design rates achieved for short periods \\ \colhline
Grid processing infrastructure & early version in use for data challenges \\ \colhline
Data catalog & sam, in place \\ \colhline
Beam instrumentation and databases & ifbeam, in test \\ \colhline
Calibration and Alignment processing & needs development \\ \colhline
Calibration and Alignment databases & needs development \\ \colhline
Noise reduction & tested in simulation \\ \colhline
Hit finding & tested in simulation \\ \colhline
Pattern recognition algorithms & tested in simulation \\ \colhline
Event simulation & use existing software \\ \colhline
Analysis formats & no common format \\ \colhline
Distribution of analysis samples to collaborators& needs development \\ \colhline
\end{dunetable}

\subsection{Downstream Activities}

The previous sections have concentrated on movement and recording of raw data, as that is most time-critical and drives the primary data storage requirements. Basic simulation and reconstruction algorithms are in place, but other components, in particular physics analysis models, are in a much earlier stage of development. 

\subsubsection{Simulation}  Our simulation efforts will build on the combined experience of multiple neutrino experiments and theory groups for input.  DUNE already has a solid foundation of event and detector simulation codes thanks to prior work by the \dword{larsoft} and event generator teams.  However,   even with good software in place, detector simulation in detectors of this high resolution is highly CPU and memory intensive and we are actively following projects intended to exploit \dword{hpc}s for more efficiency.  As simulation is much less I/O and database intensive than raw data reconstruction, (due in part to our ability to trigger efficiently on signal), we can anticipate resource contributions to this effort being distributed across the collaboration and grid resources worldwide. Simulation sample sizes orders of magnitude larger than the number of beam events  in the  far detector will be reasonably easy to achieve while near detector samples would need to be prohibitively large to equal the millions of events that will be collected every year. 

\subsubsection{Reconstruction} DUNE has working frameworks for large-scale reconstruction of simulated and real data in place thanks to the \dword{larsoft} effort.  These, and the simulations, have been exercised in large scale data challenges. Optimization of algorithms awaits data from \dword{protodune}. 
 
 \subsubsection{Data Analysis}
 The  data analysis framework has not been defined yet.  We are working to build a distributed model, where derived data samples are available locally and regionally, similar to the LHC experiments.   Provision of samples  of \dword{protodune} data and simulated samples for the \dword{tdr} will help define the analysis models that are most useful to the collaboration. However,  previous experience on the Tevatron experiments indicates that data analysis methods are often best designed by end-users rather than imposed by central software mandates.

\section{Planning Inputs}

\subsection{Running Experiments}\label{sw:IF-input}

The Fermilab intensity frontier program experiments (MINOS\cite{minosNIM},  \dword{minerva}\cite{minerva}, \dword{microboone}\cite{microboone} and  \dword{nova}\cite{Adamson:2016xxw}) have developed substantial computing infrastructure for the storage, reconstruction and analysis of data on size scales of order 5\% that of full  \dword{dune} and comparable to the \dword{protodune} experiments. While the \lartpc technology requires unique algorithms, the underlying compute systems, frameworks and database structures already exist and are being adapted for use on both \dword{protodune} and  \dword{dune}.

For algorithms, the  \dword{microboone}\cite{Acciarri:2016smi} experiment has been running since 2015 with a \lartpc that shares many characteristics with the  \dword{dune} APAs.    \dword{microboone} has, over the past year, published studies of noise sources and signal processing \cite{Acciarri:2017sde,Adams:2018dra}, novel pattern recognition strategies \cite{Acciarri:2016ryt,Acciarri:2017hat} and calibration signatures such as Michel electrons and cosmic rays \cite{Acciarri:2017sjy,Acciarri:2017sde}.   \dword{dune} shares both the \larsoft software framework and many expert collaborators with  \dword{microboone} and is taking direct advantage of their experience in developing simulations and reconstruction algorithms.

\subsection{ProtoDUNE}\label{sw:PD-planning}

The \dword{protodune} single and dual-phase experiments will run in the Fall of 2018.  While the detectors themselves have only 4-5\% of the channel count  of the final far detectors, the higher beam rates (up to 100~Hz) and the presence of cosmic rays make the expected instantaneous data rates of 2.5~GB/sec from these detectors comparable to those from the full far detectors and similar to those expected for a near detector. 

In addition, the entire suite of issues in transferring, cataloging, calibrating, reconstructing and analyzing these data are the same as for the full detectors and are driving the design and development of a substantial array of computing services necessary for both \dword{protodune} and  \dword{dune}.

Substantial progress is already being made on the infrastructure for computing, through a series of data challenges in late 2017 and early 2018. Development of reconstruction algorithms is currently restricted to simulation but is already informed by the experience with  \dword{microboone} data.

In summary, most of the important systems are already in place or are in development for full \dword{protodune} data analysis and should carry over to the full DUNE.
We have indicated where infrastructure is in place in  table  \ref{sw:computingTasks}.

\subsubsection{Single-Phase ProtoDUNE}

\dword{pdsp} utilizes six prototype \dwords{apa} with the full drift length envisioned for the final far detector. In the \single detector module, the readout planes are immersed in the \lar  and no amplification occurs before the electronics.    \dword{pdsp} is being constructed in the NP04 test beamline at CERN and should run with tagged  beam for around six weeks in the fall of 2018.  In addition cosmic ray commissioning beforehand and cosmic running after the end of beam are anticipated.  Table \ref{sw:np04_data_rate}
shows the anticipated data rates and sizes.

\begin{dunetable}[Parameters for the \dword{pdsp} run at CERN]
{p{0.5\textwidth}p{0.3\textwidth}} 
{sw:np04_data_rate}
{Parameters for the \dword{pdsp} run at CERN}

Parameter & Value \\ \colhline
    Beam trigger rate & 25\,Hz \\ \colhline
    Cosmic trigger rate & 1\,Hz \\ \colhline
    Spill duration & 2$\times$4.5\,s\\ \colhline
    SPS cycle & 32\,s \\ \colhline
    Average trigger rate& 7.8\, Hz\\ \colhline
    Readout time window & 5.4\,ms \\ \colhline
    \# of APAs to be read out & 6 \\ \colhline
    Uncompressed single readout size (per trigger) & 276\,MB \\ \colhline
    Lossless compression factor & 4 \\ \colhline
    Instantaneous compressed data rate (in-spill) & 1728\,MB/s \\ \colhline
    Average compressed data rate & 536\,MB/s \\ \colhline
    Three-day buffer depth & 300\,TB \\ \colhline
    Planned total statistics of beam triggers in 42 beam days &18M\\ \colhline
    Planned overall storage size of beam events&   1.25 PB\\ \colhline
   Requested storage envelope for \dword{pdsp} &5~PB at \fnal, 1.5~PB at CERN \\ \end{dunetable}


\subsubsection{Dual-Phase ProtoDUNE}

The \dword{pddp} will either run in the NP02 beamline in late fall 2018, or run at high rate on cosmics soon thereafter.
Given the most recent construction schedule for \dword{pddp} is now likely that the collaboration will forgo beam data taking and focus on  detector performance  assessment with two \dwords{crp} read out and cosmics only. \dword{pddp} will then run with cosmics at a rate going from 20 to 100 Hz from late fall 2018 to at least April 2019. During six months of operation, with 50\% efficiency, \dword{pddp} is expected to collect about 300 million cosmic triggers at various rates, corresponing to a total data volume of 2.4 PB.

\begin{dunetable}[Parameters for a six month \dword{pddp} cosmic run at CERN]
{p{0.5\textwidth}p{0.3\textwidth}}
{sw:np02_data_rate}
{Parameters for a six month \dword{pddp} cosmic run at CERN}
Parameter & Value \\ \colhline
Trigger rate & 20-100\,Hz \\ \colhline
\dword{crp}s read out &2\\ \colhline
Uncompressed single readout size (per trigger) & 80\,MB \\ \colhline
Lossless compression factor & 10\\ \colhline
Maximum data rate & $\le$800\,MB/s \\ \colhline
Cosmic rays over a 6 month run& 300\,M\\ \colhline
Requested cosmic storage envelope for \dword{protodune}-DP&2.4\,PB \\ 
\end{dunetable}

\subsection{Data Challenges}

Computing and software is performing a series of data challenges to ensure that systems will be ready when the detectors become fully operational in the summer of 2018.  To date we have performed challenges using simulated single and double-phase data and real data from cold-box tests of single-phase electronics.   DUNE anticipates average rates of $\sim$ 600 MB/sec but have set our design criteria at 2.5~GB/sec for data movement from the experiments to CERN Tier-0 storage and from there to Fermilab. 

In data challenge 1.5 in mid-January 2018, dummy data based on non-zero-suppressed simulated events were produced at CERN's EHN1 and successfully transferred via 10-50 parallel transfers to the \dword{eos} disk systems in the CERN Tier-0 data center at a sustained rate of 2~GB/sec.    Transfers to dCache/Enstore at Fermilab achieved rates of 500~MB/sec.  

Data challenge 2.0 was performed in early April 2018 is still being analyzed but preliminary estimated  rates of 4~GB/sec from CERN's  \dword{ehn1} to the tier-0 were achieved over several days. Rates to Fermilab disk cache were 2~GB/sec.  Movement from FNAL disk cache to tape was substantially slower due to configuration for a lower number of drives than needed and contention for mounts with other running experiments.   Fermilab is in the process of upgrading their tape facilities but we may require additional offsite buffer space if data rates from the experiments exceed the $\sim$ 600~MB/sec expected. 

A subsample of the data was used for data quality monitoring at CERN and the full sample was reconstructed automatically on the grid using resources at multiple sites, including CERN. 
Our overall conclusion from this test is that most components for data movement and automated processing are in place.  Remaining issues are integration of beam information, detector configuration and calibrations into the main processing stream, and faster tape access. 

\subsection{Monte Carlo Challenges}
The collaboration has performed multiple Monte Carlo challenges to create samples for physics studies for the \dword{tp} and in preparation for the \dword{tdr} in early 2019.  In the last major challenge,  MCC10 in early 2018, 17M events, taking up 252~TB of space were generated and catalogued automatically using the central \dword{dune} production framework in response to requests by the Physics groups. 

\subsection{Reconstruction tests}
Reconstruction tests have been performed on simulated  single-phase \dword{protodune} test beam interactions with cosmic rays and an electronic noise simulation based on \dword{microboone} experience.  Hit finding and shaping is found to take around 2 minutes/event with a 2~GB memory footprint, leading to a reduction in data size of a factor of four.  Higher-level pattern recognition occupies 10-20 minutes/event with a 4-6~GB memory footprint.  For real data, calibration, electric field non-uniformities and other factors will likely raise the CPU needs per event. We will learn this when real data starts to arrive in late summer.

\section{Resource Planning and Prospects}

The  \dword{dune} computing effort  relies heavily on the human and hardware resources of  multiple organizations,  with the bulk of hardware resources at CERN, and national facilities worldwide.  The   \dword{dune} computing organization serves as an interface between the collaboration and those organizations.  Computational resources are currently being negotiated on a yearly basis, with additional resources available opportunistically. Human assistance is largely on a per-project  basis, with substantial support when needed but very few personnel as yet permanently assigned to the  \dword{dune} or \dword{protodune} efforts.  We are working with the laboratories and funding agencies to identify and solidify multi-year commitments of dedicated personnel and resources for \dword{protodune} and \dword{dune}, analogous to, but smaller than, those assigned to the LHC experiments.   In-kind contributions of computing resources and people can also  be an alternative way for institutions to make substantial contributions to \dword{dune}.

The \dword{protodune} efforts in 2018-2019 will exercise almost all computing aspects of DUNE, although at smaller scale.  Much of the infrastructure needed for full DUNE, in particular  databases, grid configurations and code management systems need to be fully operational  for \dword{protodune}.   We believe that the systems in place (and tested) will be adequate for that purpose.

However, \dword{protodune} represents only 4-5\% of the final volume of the far detectors and the near detector technology is, as yet, unknown.  At the same time, computing technology is evolving rapidly with increased need for flexibility and the ability to parallelize codes.  Liquid argon detectors, because of their reasonably simple geometry and image-like data, are already able to take advantage of parallelization and generic machine learning techniques.  We have good common infrastructure such as the \dword{larsoft} suite and \dword{geant4}, and will have an excellent testbed with the \dword{protodune} data,  but our techniques will need substantial adaption to scale to full \dword{dune} and to take full advantage of new architectures.  These scaling challenges are similar to those facing the LHC experiments as they move towards the High Luminosity LHC on a similar timescale. We look forward to working with them on shared solutions.  Achieving full scale will be one of our major challenges -- and  one of our prime opportunities for collaboration -- over the next five years.

\cleardoublepage

\chapter{DUNE Calibration Strategy}
\label{ch:exec-summ-calib}

The DUNE \dword{fd} presents a unique challenge for calibration in many ways. Not only because of its size---the largest \dword{lartpc} ever constructed -- but also because of its depth. It differs both from existing long-baseline neutrino detectors, and existing \dwords{lartpc} (e.g. the deep underground location). While DUNE is expected to have a \dword{lartpc} as ND, DUNE is unlike previous long baseline experiments (MINOS, \dword{nova}) in that the near detector will have significant differences (pile-up, readout) that may make extrapolation of detector characteristics non-trivial.  
Like any \dword{lartpc}, DUNE has the great advantage precision tracking and calorimetry, but,  fully exploiting this capability requires a detailed understanding of the detector response. This challenge is driven by the inherently highly convolved detector response model and strong correlations that exist between various calibration quantities. For example, the determination of energy associated to an event of interest will depend on the simulation model, associated calibration parameters, non-trivial correlations between the parameters and spatial and temporal dependence of those parameters. These variations in parameters occur since the detector is not static. Changes can be abrupt (e.g. noise, a broken resistor in the field cage), or ongoing (e.g. exchange of fluid through volume, ion accumulation). 

A convincing measurement of CP violation, or a resolution of the neutrino mass ordering, or  \dword{snb} detection will require a demonstration that the overall detector response is well understood. 
This chapter describes a strategy for detector calibration for both 
\dwords{spmod} and \dwords{dpmod} using dedicated \dword{fd} systems or existing calibration sources. A large portion of the calibration work reported here is done under the joint \dword{sp} and \dword{dp} calibration task force formed in August 2017.
Section~\ref{sec:calibstrat} summarizes a calibration strategy currently envisioned for DUNE. The systematic uncertainties for the long baseline and low energy (supernova) program will determine how precisely each calibration parameter needs to be measured. For example, how precisely will the drift velocity need to be measured to know fiducial volume better than \num{1}\%? In general, the calibration program must provide measurements at the few percent or better level stably across an enormous volume and a long period of time. The calibration strategy must also provide sufficient redundancy in the measurement program.

Existing calibration sources for DUNE include beam or atmospheric neutrino-induced samples, cosmic rays, argon isotopes, and instrumentation devices such as liquid argon purity and temperature monitors. It is important for calibration strategy to further separate these sources into those used to measure a response model parameter, and those used to test the response model. In addition to existing sources, external measurements prior to DUNE will validate techniques, tools and design of systems applicable to the DUNE calibration program; data from  \dword{protodune} and SBND are essential to the success of the overall calibration program. 
Section~\ref{sec:exis} describes calibration from existing source of particles and external measurements.

Section~\ref{sec:calibnew} describes dedicated external calibration systems currently under consideration for DUNE to perform calibrations that cannot be achieved fully from existing sources or external measurements. All the systems proposed are currently being actively discussed in the calibration task force and were agreed as important systems by the DUNE collaboration. Under current assumptions, the calibration strategy and proposed calibration systems described in this document are applicable to both \dwords{spmod} and \dwords{dpmod}. 
Finally, Section~\ref{sec:calibsum} provides a summary along with future steps for calibration and a path to the \dword{tdr}. 

\section{Calibration Strategy}
\label{sec:calibstrat} 
DUNE has a broad physics program that includes long baseline neutrino oscillation physics, supernova physics, nucleon decay, and other searches for new physics. The physics processes that lead to the formation of these signals and the detector effects that influence their propagation must be carefully understood in order to perform adequate calibrations, as they ultimately affect the detector's energy response. Several other categories of effects can impact measurements of physical quantities such as the neutrino interaction model or reconstruction pathologies. 
These other effects are beyond the scope of the \dword{fd} calibration effort and can only amplify the overall error budget.
In the reminder of this section, we briefly describe the physics-driven calibration requirements, including the calibration sources and systems required for the different stages of the experiment.

\subsection{Physics Driven Calibration Requirements}

\textbf{Long-baseline physics:} In the physics volume of the DUNE CDR~\cite{Acciarri:2015uup}, Figure~3.23 shows that increasing the uncertainties on the \nue event rate from \num{2}\% overall 
to \num{3}\% results in a \num{50}\% longer run period to achieve a 5$\sigma$ determination of CP violation for 50\% of possible values of \dword{cpv}. The CDR also assumes that the fiducial volume is understood at the 1\% level. Thus, calibration information needs to provide approximately 1-2\% understanding of normalization, energy, and position resolution within the detector. Later studies~\cite{ebias} expanded the simple treatment of energy  presented above. In particular, \num{1}\% bias on the lepton energy has a significant impact on the sensitivity to \dword{cpv}. 
A \num{3}\% bias in the hadronic state (excluding neutrons) is important, as the inelasticity  distribution for neutrinos and antineutrinos is quite different.  A different fraction of the antineutrino's energy will go into the hadronic state. Finally, while studies largely consider a single, absolute energy scale, relative spatial differences across the enormous DUNE \dword{fd} volume will need to be monitored and corrected; this is also true for changes that occur in time. A number of in-situ calibration sources will be required to address these broad range of requirements. Michel electrons, neutral pions and radioactive sources (both intrinsic and external) are needed for calibrating detector response to electromagnetic activity in the tens to hundreds of MeV energy range. Stopping protons and muons from cosmic rays or beam interactions form an important calibration source for calorimetric reconstruction and particle identification. \Dword{protodune}, as a dedicated test beam experiment, provides critical measurements to characterize and validate particle identification strategies in a 1~kt scale detector and will be an essential input to the overall program. Dedicated calibration systems, like lasers, will be useful to provide in-situ full volume measurements of electric field distortions. Measuring the strength and uniformity of the electric field is a key aspect of calibration, as  estimates of calorimetric response and particle identification depend on electric field through recombination. The stringent physics requirements on energy scale and fiducial volume also put similarly stringent requirements on detector physics quantities such as electric field, drift velocity, electron lifetime, and the time dependences of these quantities.

\textbf{Supernova burst and low-energy neutrino physics:} For this physics, the signal events present specific reconstruction and calibration challenges and observable energy is shared between different charge clusters and types of energy depositions. Some of the primary requirements here include calibration of absolute energy scale and 
understanding and improving the nominal 20\% energy resolution, important for resolving spectral features of \dword{snb} events,
calibration of time and and light response of optical photon detectors, absolute timing of events and understanding of detector response to radiological backgrounds. Potential calibration sources in this energy range include Michel electrons from muon decays (successfuly utilized by ICARUS and \dword{microboone}~\cite{Acciarri:2017sjy}), which have a well known spectrum up to $\sim$50~MeV. Radiological sources provide calibrations of photon, electron, and neutron response for energies below \SI{10}{\MeV}. It is more challenging to find ``standard candles'' between 50~MeV and \SI{100}{\MeV}, beyond cosmic-ray muon energy loss. \Dword{protodune} could potentially be a test bed for various calibration strategies. One can imagine also ancillary studies of detector response using detectors such as LArIAT~\cite{Cavanna:2014iqa}, \dword{microboone}~\cite{Acciarri:2016smi}, and SBND~\cite{Antonello:2015lea}. 

\textbf{Nucleon decay and other exotic physics:} The calibration needs for nucleon decay and other exotic physics are comparable to the \dword{lbl} program as listed before. Signal channels for light dark matter and sterile neutrino searches will be neutral current interactions that are background to the \dword{lbl} physics program. 
Based on the widths of $dE/dx$-based metrics of particle identification, qualitatively, we need to calibrate $dE/dx$ across all drift and track orientations at the few percent level or better, which is a similar target of interest as the \dword{lbl} effort.

\begin{dunefigure}[Categories of measurements provided by Calibration.]{fig:calibneeds}{Categories of measurements provided by Calibration.}
\includegraphics[width=.9\textwidth]{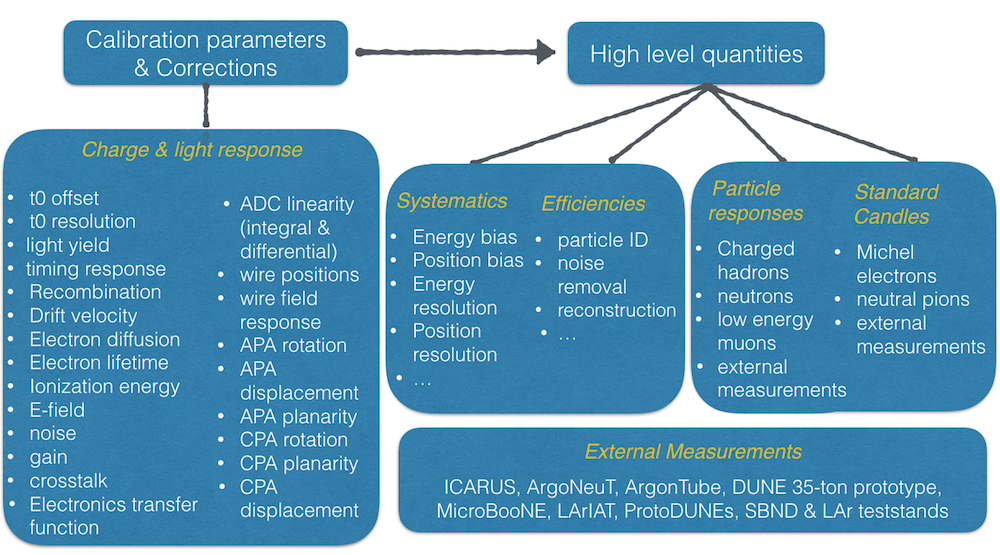}
\end{dunefigure}

\textbf{Calibration Sources and Systems:} Calibration sources and systems provide measurements of the detector response model parameters, or provide tests of the response model. 
Calibration measurements can also provide corrections applied to data, data-driven efficiencies, systematics and particle responses. Figure~\ref{fig:calibneeds} shows the broad range of categories of measurements calibrations can provide and lists the critical calibration parameters for DUNE's detector response model applicable to both \dword{sp} or \dword{dp}. Due to the significant interdependencies of many parameters (e.g. recombination, electric field, liquid argon purity), a calibration strategy will either need to iteratively measure parameters, or find sources that break these correlations. Table~\ref{tab:calibsystem} provides a list of various calibration sources and proposed calibration systems along with their primary usage which will comprise the currently envisioned nominal DUNE FD calibration design to adequately address the needs for physics. More details on each of the calibration sources and systems are provided in the next sections. \Dword{protodune} and previous measurements provide independent, tests of the response model, where the choice of parameterization and values correctly reproduces real detector data, however not all of the ex situ measurements will be directly extrapolatable to DUNE due to other detector effects and conditions. Only those that are believed to be universal (e.g., argon ionization energy) can be extrapolated. Also, while there are many existing calibration sources, each source comes with its own challenges. For example, electrons from muon decay (Michel electrons) are very useful to study the detector response to low-energy electrons (\SI{50}{\MeV}). However, low-energy electrons present reconstruction challenges due to the loss of charge from radiative photons as demonstrated in \dword{microboone}~\cite{Acciarri:2017sjy}. In terms of source category, Michel electrons are considered as an important, independent, and necessary test of the TPC energy response model, and not as a measurement of a particular response parameter.

\begin{dunetable}[Calibration systems and sources of the nominal DUNE FD calibration design]
{p{.4\textwidth}p{.55\textwidth}}{tab:calibsystem}{Primary calibration systems and sources that comprise the nominal DUNE FD calibration design along with their primary usage.} 
System & \textbf{Primary Usage}  \\ \toprowrule 
\textbf{Existing Sources} & \textbf{Broad range of measurements} \\ \toprowrule
$\mu$, predominantly from cosmic ray & Position (partial), angle (partial) velocity (timing),  electron lifetime, wire response\\ \colhline 
Decay electrons, $\pi^0$ from beam, cosmic, atm $\nu$ & Test of electromagnetic response model \\ \colhline
$^{39}$Ar &  electron lifetime (x,y,z,t), diffusion \\   \colhline 
\textbf{External Measurements} & \textbf{Tests of detector model, calibration techniques and systems} \\ \toprowrule
ArgoNeuT~\cite{Acciarri:2013met}, ICARUS~\cite{Amoruso:2004dy, Antonello:2014eha, Cennini:1994ha}, MicroBooNE & Model parameters (e.g. recombination, diffusion) \\ \colhline 
DUNE \dword{35t}~\cite{Warburton:2017ixr} & Alignment and \textit{t0} techniques\\ \colhline 
ArgonTUBE~\cite{Ereditato:2014tya}, MicroBooNE~\cite{Acciarri:2016smi}, SBND, ICARUS~\cite{Auger:2016tjc},  \Dword{protodune}~\cite{Abi:2017aow} & Test of systems (e.g. Laser, external muon taggers) \\ \colhline
ArgoNeuT~\cite{Acciarri:2015ncl}, MicroBooNE~\cite{bib:uBlifetime, MICROBOONE-NOTE-1018-PUB, MICROBOONE-NOTE-1028-PUB, Acciarri:2017sjy, Abratenko:2017nki, Acciarri:2013met}, ICARUS~\cite{Ankowski:2008aa,  Ankowski:2006ts,Antonello:2016niy},  \Dword{protodune} & Test of calibration techniques and detector model (e.g., electron lifetime, Michel electrons, $^{39}$Ar beta decays) \\ \colhline
\Dword{protodune}, LArIAT~\cite{Cavanna:2014iqa} & Test of particle response models and fluid flow models \\  \colhline
\dword{lartpc} test stands~\cite{Cancelo:2018dnf, Moss:2016yhb, Moss:2014ota, Li:2015rqa} & Light and LAr properties; signal processing techniques \\ \colhline 
\textbf{Monitoring Systems} & \textbf{Operation, Commissioning and Monitoring} \\ \toprowrule
Purity Monitors & Electron lifetime \\ \colhline
Photon Detection Monitoring System & \dword{pds} response \\ \colhline
Thermometers & Temperature, velocity; test of fluid flow model \\ \colhline
Charge injection & Electronics response \\ \colhline
\textbf{Proposed Systems} & \textbf{Targeted (near) independent, precision calibration}\\ \toprowrule
Direct ionization via laser & Position, angle, electric field (x,y,z,t) \\ \colhline
Photoelectric ejection via laser & Position, electric field (partial) \\ \colhline
Radioactive source deployment & Test of SN signal model \\ \colhline
Neutron injection & Test of SN signal, neutron capture model \\ \colhline
External Muon Tracker & Position, angle, muon reconstruction efficiency \\ \colhline
\end{dunetable}  


\subsection{A Staged Approach}
The calibration strategy for DUNE will need to address the evolving operational and physics needs at every stage of the experiment in a timely manner using the primary sources and systems listed in Table~\ref{tab:calibsystem}. At the TDR stage, a clear and complete calibration strategy with necessary studies will be provided to demonstrate 
how the existing and proposed systems meet the physics requirements.
 Post-TDR, once the calibration strategy is set, necessary designs for calibration hardware along with tools and methods to be used with various calibration sources will need to be developed. To allow for flexibility in this process, the physical interfaces for calibration such as flanges or ports on the cryostat should be designed for a wide range of possible uses to accommodate the calibration hardware. As described in section~\ref{sec:FTs}, the calibration TF has made necessary accommodations for calibration systems for the \dword{spmod} in terms of feedthrough penetration design and will soon start finalizing the design and accommodations for calibration penetrations for the \dword{dpmod}. 

As DUNE physics turns on at different rates and times, a calibration strategy at each stage for physics and data taking is described below. This strategy assumes that all systems are commissioned and deployed according to the nominal DUNE run plan. 

\textbf{Commissioning:} When the detector is filled, data from various instrumentation devices will be needed to validate the argon fluid flow model and purification system. When the detector is filled and at desired high voltage, the detector immediately becomes live for supernovae and proton decay signals (beam and atmospheric neutrino physics will require a few years of data accumulation) at which stage it is critical for early calibration to track the space-time dependence of the detector. Noise data (taken with wires off) and pulser data (taken with signal calibration pulses injected into electronics) will be needed to understand the detector electronics response. Essential systems at this stage include temperature monitors, purity monitors, HV monitors, robust front-end charge injection system for cold electronics, and a PDS monitoring system for light. Additionally, as the $^{39}$Ar data will be available immediately, readiness (in terms of reconstruction tools and methods) to utilize $^{39}$Ar decays will be needed, both for understanding low energy response and space-time uniformity. External calibration systems as listed in Table~\ref{tab:calibsystem} will be deployed and commissioned at this stage and commissioning data from these systems will be needed to verify expected configuration for each system and any possible adjustments needed to tune for data taking.

\textbf{Early data taking:} Since DUNE will not have all in-situ measurements of liquid argon properties at this stage, early calibration of the detector will utilize liquid argon physical properties from \Dword{protodune} or SBND, and \efield{}s from calculations tuned to measured HV. This early data will most likely need to be recalibrated at a later stage when other data sets become available. This is expected to improve from in-situ measurements as data taking progresses and with dedicated calibration runs. The detector response models in simulation will need to be tuned on \Dword{protodune} and\slash or SBND data during this early phase, and the mechanism for performing this tuning needs to be ready. This together with cosmic ray muon analysis will provide an approximate energy response model that can be used for early physics. Analysis of cosmic ray muon data to develop methods and tools for muon reconstruction from MeV to TeV and a well-validated cosmic ray event generator with data will be essential for early physics. Cosmic ray or beam induced muon tracks tagged with an external muon tracker system will be very useful in these early stages to independently measure and benchmark muon reconstruction performance and efficiencies in the FD. Dedicated early calibration runs from external calibration systems will be needed to develop and tune calibration tools to data taking and correct for any space-time irregularities observed in the TPC system for early physics. Given the expected low rate of cosmic ray events at the underground location (see Section~\ref{sec:exis}), calibration with cosmic rays are not possible over short time scales and will proceed from coarse-grained to fine-grained over the course of years as statistics is accumulated. The experiment will have to rely on external systems such as laser for calibrations that require an independent probe with reduced or removed interdependencies, fine-grained measurements (both in space and time) and detector stability monitoring in the time scales needed for physics. Additionally some measurements are not possible with cosmic rays (e.g. APA flatness or global alignment of all APAs). 

\textbf{Stable operations:} Once the detector conditions are stabilized and the experiment is under stable running, dedicated calibration runs, ideally before, during and after each run period, will be needed to ensure detector conditions have not significantly changed. As statistics are accumulated, standard candle data samples (e.g. Michel electrons and neutral pions) both from cosmic rays, beam induced and atmospheric neutrinos can be used to validate and improve the detector response models needed for precision physics.  As DUNE becomes systematics limited, dedicated calibration campaigns using the proposed external systems will become crucial for precision calibration to meet the stringent physics requirements both for energy scale reconstruction and detector resolution. For example, understanding electromagnetic (EM) response in the FD will require both cosmic rays and external systems. The very high energy muons from cosmic rays at that depth that initiate EM showers (which would be rare at \Dword{protodune} or SBND), will provide information to study EM response at high energies. External systems such as radioactive sources or neutron injection sources will provide low energy EM response at the precision required for low energy supernovae physics. Other calibration needs not addressed with existing sources and external systems, will be determined initially from the output of the \Dword{protodune} and\slash or SBND program, and later from the DUNE ND if the design choice will be a LArTPC. 






\section{Inherent Sources and External Measurements}
\label{sec:exis}

Existing sources of particles, external measurements and monitors are an essential part of the DUNE FD calibration program which we briefly summarize here. 

\textbf{Existing sources:} Cosmic rays and neutrino-induced interactions provide commonly used ``standard candles'' like electrons from muon and pion decays, and neutral pions, which have characteristic energy spectra. Cosmic ray muons are also used to determine detector element locations (alignment), timing offsets or drift velocity, electron lifetime, and channel-by-channel response. The rates for cosmic rays events are summarized in Table~\ref{tab:cosmic-ray-calib-rates}, and certain measurements (e.g. channel-to-channel gain uniformity and cathode panel alignment) are estimated to take several months of data. The rates for atmospheric $\nu$ interactions can be found in Table~\ref{tab:atmnu-rates} and are comparable to beam-induced events; both atmospheric and beam induced interactions do not have sufficient rates to provide meaningful spatial or temporal calibration and are expected to provide supplemental measurements only. Also, beam neutrinos may not contribute to the first module calibration during early data taking as the beam is expected to arrive later. The reconstructed energy spectrum of ${}^{39}$Ar beta decays can be used to make a spatially and temporally precise electron lifetime measurement. It can also provide other necessary calibrations, such as measurements of wire-to-wire response variations and diffusion measurements using the signal shapes associated with the beta decays.
The ${}^{39}$Ar beta decay rate in commercially provided argon is about \SI{1}{\becquerel\per\kilo\gram}, so $O(\mathrm{50k})$ ${}^{39}$Ar beta decays are expected in a single \SI{5}{\milli\s} event readout in an entire \SI{10}{\kt} \detmodule. The ${}^{39}$Ar beta decay cut-off energy is \SI{565}{\keV} which is close to the energy deposited on a single wire by a \dword{mip}. However, there are several factors that can impact the observed charge spectrum from ${}^{39}$Ar beta decays such as electronics noise, electron lifetime and recombination fluctuations.

\begin{dunetable}
[Annual rates for classes of cosmic-ray events useful for calibration]
{lrl}
{tab:cosmic-ray-calib-rates}
{Annual rates for classes of cosmic-ray events described in this section assuming 100\% reconstruction efficiency.  Energy, angle, and fiducial requirements
have been applied. Rates and geometrical features apply to the single-phase far detector design. }
Sample & Annual Rate & Detector Unit \\ \colhline
Inclusive & $1.3\times 10^6$ & Per 10 kt module \\ \colhline
Vertical-Gap crossing & 3300 & Per gap \\ \colhline
Horizontal-Gap crossing & 3600 & Per gap \\ \colhline
APA-piercing & 2200 & Per APA \\ \colhline
APA-CPA piercing & 1800 & Per active APA side \\ \colhline
APA-CPA piercing, CPA opposite to APA & 360 & Per active APA side \\ \colhline
Collection-plane wire hits & 3300 & Per wire \\ \colhline
Stopping Muons & 11000 & Per 10 kt module \\ \colhline
$\pi^0$ Production & 1300 & Per 10 kt module \\ \colhline
\end{dunetable}

\textbf{Monitors:} Several instrumentation and detector monitoring devices discussed in detail in Chapter 8 of \voltitlespfd{} and \voltitledpfd{} of the \dword{tp} will provide valuable information for early calibrations and to track the space-time dependence of the detector. The instrumentation devices include liquid argon temperature monitors, \lar purity monitors, gaseous argon analyzers, cryogenic (cold) and inspection (warm) cameras, and liquid level monitors. The computational fluid dynamics (CFD) simulations play a key role for calibrations initially in the design of the cryogenics recirculation system, and later for physics studies when the cryogenics instrumentation data is used to validate the simulations. Other instrumentation devices essential for calibration such as drift \dword{hv} current monitors and external charge injection systems are discussed in detail in Chapters 4 and 5, respectively, of \voltitlespfd{} and \voltitledpfd{} of the \dword{tp}, respectively.


\textbf{External measurements:} External measurements here include both past measurements (e.g., ArgoNeuT, DUNE \dword{35t}, \dword{microboone}, ICARUS, SBND, \lariat), anticipated measurements from ongoing and future experiments (e.g., \dword{microboone}, \dword{protodune}) as well as from small scale \dword{lartpc} test stands. External measurements provide a test bed for proposed calibration hardware systems and techniques which are applicable to the DUNE FD. In particular, \dword{protodune} will provide validation of the fluid flow model using instrumentation data. Early calibration for physics in DUNE will utilize liquid argon physical properties from \Dword{protodune} or SBN  for tuning detector response models in simulation. Table~\ref{tab:calibsystem} provides  references for specific external measurements. The usability of ${}^{39}$Ar has been demonstrated with \microboone data. 
Use of  ${}^{39}$Ar  and other radiological sources, including the DAQ readout challenges associated with their use, will be tested on the   \dword{protodune} detectors. Proposed systems for DUNE, including the laser system below, are part of the \microboone and SBND programs which will provide increased information of the use of the system and optimization of the design. Measurements from small-scale liquid argon test stands can also provide valuable information for DUNE. The liquid argon test stand planned at Brookhaven National Lab will provide important information for how field response is simulated and calibrated at DUNE.

\textbf{Remaining Studies:} In advance of the \dword{tdr}, studies will be done to clarify the physics use limitations of the various sources presented in this section. 
For example, quantification of what can be achieved for  electron lifetime measurements and the overall energy scale calibration from cosmic rays,  ${}^{39}$Ar beta decays, long baseline interactions and atmospheric neutrinos, in terms of spatial and temporal granularity using decay electrons, $\pi^0$ samples; determining the relative importance of electromagnetic shower photons below pair production threshold. It is expected that combinations of information from cosmic-ray events with proposed and existing systems (laser-based, neutrino-induced events, and dedicated muon systems) will reduce the total uncertainties on mis-alignment.  The impact of misalignments on the physics case needs to be studied, especially for alignment modes which are weakly constrained due to cosmic ray direction, as shown in Figure~\ref{fig:apacurtainalign}, including global shifts and rotations of all detector elements, and crumpling modes where the edges of the \dwords{apa} hold together but angles are slightly different from nominal. The impact of the fluid model on physics needs require quantification via CFD simulations (e.g., overall temperature variation in the cryostat and impact on drift velocity; overall impurity variation across the \detmodule{}, and impact on energy scale especially for \dword{dp} which has a \SI{12}{\m} long single drift path). The CFD studies will also be important in understanding how \lar flow can impact space charge from both ionization and non-ionization sources and ion accumulation (both positive and negative ions), separately for \dword{spmod} and \dword{dpmod} designs. 

\begin{dunefigure}[Sample distortion that may be difficult to detect with cosmic rays]{fig:apacurtainalign}{An example of a distortion that may be difficult to detect with cosmic rays.  The \dword{apa} frames are shown as
rotated rectangles, as viewed from the top.}
\includegraphics[width=0.8\textwidth]{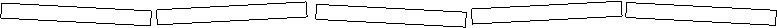}
\end{dunefigure}



\section{Proposed Systems}
\label{sec:calibnew} 

The nominal calibration design includes the existing sources, external measurements, and monitors listed in the previous section, and the following proposed systems: laser (Section~\ref{sec:laser}), radioactive source deployment (Section~\ref{sec:calibrs}), neutron injection (Section~\ref{sec:neutron}), and external muon tracker (Section~\ref{sec:calibemt}). While the systems described previously are necessary they are not sufficient for the entire DUNE calibration program. The proposed systems discussed here are motivated as they supply necessary information beyond the reach of the existing systems. 

\subsection{Laser Systems}\label{sec:laser} 

 None of the systems discussed in the previous section can  provide an independent, fine-grained estimate of the \efield in space or time, which is a critical parameter for physics signals as it ultimately impacts the spatial resolution and energy response of the detector. The primary purpose of a laser system is to provide such a measurement. There are multiple sources which may distort the electric field temporally  or spatially in the detector. Current simulation studies indicate that positive ion accumulation and drift (space charge) due to ionization sources such as cosmic rays or ${}^{39}$Ar is small in the DUNE \dword{fd};  however, not enough is known yet about the fluid flow pattern in the FD to exclude the possibility of stable eddies which may amplify the effect for both \single and \dual modules. This effect can get further amplified significantly in \dword{dpmod} due to ion accumulation at the liquid-gas interface. 
Additionally, other sources in the detector (especially detector imperfections) can cause \efield distortions. For example, field cage resistor failures, non-uniform resistivity in the voltage dividers, CPA misalignment, CPA structural deformations, and APA and CPA offsets and  deviations from flatness can create localized \efield distortions. Each individual \efield distortion may add in quadrature with other effects, and can reach 4\% under certain conditions. Understanding all these effects require in-situ measurement of \efield for proper calibration. 
Many useful secondary uses of laser include alignment (especially modes that are weakly constrained by cosmic rays; see Figure~\ref{fig:apacurtainalign}), stability monitoring, and diagnosing detector failures (e.g., \dword{hv}).  



Two laser-based systems have been considered to extract the electric field map. They fall into two categories: \phel from the \lartpc cathode and direct ionization of the \dword{lar}, both driven by a \SI{266}{\nano\m} laser system. The reference design uses direct ionization laser light  with multiple laser paths, as it can provide field map information in $(x, y, z, t)$; \phel only provides integral field across the drift. An ionization-based system has been used in the ARGONTUBE~\cite{Zeller:2013sva}, \dword{microboone}, CAPTAIN, and SBND experiments. Assuming multiple, steerable laser entry points as discussed in Section~\ref{sec:FTs}, the ionization-based system can characterize the electric field with fewer dependencies compared to other systems. 
Two ``laser tracks'' that cross in a detector volume element can be used to estimate the local \efield in that volume. If two tracks enter the same spatial voxel 
($10 \times 10 \times 10~\textrm{cm}^3$ volume) in the \dword{detmodule}, the relative position of the tracks provides an estimate of the local \threed \efield. A scan of the full detector using \SI{1}{L} volume elements would take a day, but it is expected that practically shorter runs could be done to investigate specific regions. 
The deviation from straighness of single ``laser tracks'' can also be used to constrain local \efield{}s. 
The direct ionizing laser system may also be used to create \phel{}s from the cathode, even under low power operation.

A \phel{}-based calibration system was used in the T2K gaseous (predominantly Ar), TPCs~\cite{Abgrall:2010hi}. 
Thin metal surfaces placed at surveyed positions on the cathode provided point-like and line sources of \phel{}s when illuminated by a laser. The T2K \phel system provided measurements of adjacent electronics modules' relative timing response, drift velocity with few \si{\nano\s} resolution of \SI{870}{\milli\m} drift distance, electronics gain, transverse diffusion, and an integrated measurement of the electric field along the drift direction. For DUNE, the system would be similarly used as on T2K to diagnose electronics or TPC response issues on demand, and provide an integral field measurement and relative distortions of $y$, $z$ positions with time, and of either $x$ or drift velocity. Ejection of \phel{}s from the direct ionization laser system has also been observed, so it is likely this is a reasonable addition to the nominal design.

The remaining studies for the laser systems to be done prior to the \dword{tdr} are: 
\begin{itemize}
\item Determine a nominal design for photoelectric thin metal surfaces on the cathode. A survey in cold conditions is not possible for the \single system, and the photoelectric system could provide both known positions in the detector and information complementary to a survey or cosmic data.
\item For the \dual system, quantify the additional benefit of a photoejection system since it will be possible to survey the \dwords{crp} externally under cold conditions.
\item  Determine whether the known classes of possible \efield distortions warrant a mechanical penetration of the \dword{fc} (versus reduced sampling from projecting laser light inward between \dword{fc} elements) and further understand sensitivity of the laser to realistic \efield distortions. 
\item Continue to study the range of possible \efield distortions in order to further refine the estimation of overall variation of the \efield  in the \dword{detmodule}. 
\end{itemize}

\subsection{Radioactive Source Deployment System}
\label{sec:calibrs}

Radioactive source deployment provides an in-situ source of the electrons and de-excitation gamma rays, which are directly relevant for physics signals from supernova or $^{8}B$ solar neutrinos. Secondary measurements from the source deployment include electromagnetic (EM) shower characterization for long-baseline $\nu_e$ CC events, electron lifetime as a function of \dword{detmodule} vertical position, and help determine radiative components of the electron energy spectrum from muon decays.




A composite source can be used that consists  of $^{252}$Cf, a strong neutron emitter, and $^{58}$Ni, which, via the $^{58}$Ni(n,$\gamma$)$^{59}$Ni process, converts one of the $^{252}$Cf decay neutrons, suitably moderated, to a monoergetic 9 MeV photon~\cite{Rogers:1996ks}. 
The source is envisaged to be inside a cylindrical teflon moderator with mass of about \SI{10}{kg} and a diameter of \SI{20}{\cm} such that it can be deployed via the multipurpose instrumentation ports discussed in Section~\ref{sec:FTs}. The activity of the radioactive source is chosen such that no more than one \SI{9}{\MeV} capture $\gamma$-event occurs during a single 
drift period. This allows one to use the arrival time of the measured light as a $t0$ and then measure the average drift time of the corresponding charge signal(s).
This restricts the maximally permissible rate of \SI{9}{\MeV} capture $\gamma$-events occurring inside the radioactive source to be less
than \SI{1}{\kilo\hertz}, given a spill-in efficiency into the active \dword{lar} of
less than \num{10}\%. 
The sources would be deployed outside the \dword{fc} within the cryostat to avoid regions with high electric field. Sources would be removed and stored outside the cryostat when not in use.




Assuming stable detector conditions, a radioactive source would be deployed every half year, before and after a given run period.
If stability fluctuates for any reason (e.g., electronic response changes over time) at a particular location, it is desirable to deploy the source at that location once a month, or more often, depending on how bad the stability is. It would take of order eight hours to deploy the system at one feedthrough location, and a full radioactive source calibration campaign might take a week.

For the \dword{tdr}, continued development of geometry and simulation tools of radioactive source system is necessary to demonstrate the usage of these sources, including studies of various radiological contaminants on detector response, and source event rate and methods to suppress them. In addition, a test will be performed at South Dakota School of Mines and Technology. A radioactive source deployment in a potential phase 2 of ProtoDUNE could be envisaged to demonstrate proof of principle of the radioactive source deployment. However, studies need to be performed to first understand how cosmic rays can be vetoed sufficiently well for a radioactive source measurement.

\subsection{Pulsed Neutron Source}  
\label{sec:neutron}

An external neutron generator system would provide a triggered, well defined neutron energy deposition that can be detected throughout the volume. Neutron capture is a critical component of signal processes for \dword{snb} and \dword{lbl} physics. 

A triggered pulse of neutrons can be generated outside the TPC, then injected via a dedicated opening in the insulation into the \dword{lar}, where it spreads through the entire volume to produce monoenergetic photons via the $^{40}$Ar(n,$\gamma$)$^{41}$Ar capture process. The uniform population of neutrons throughout the detector module volume exploits a remarkable property of argon -- the near transparency to neutrons with an energy near \SI{57}{\keV} due to a deep minimum in the cross section caused by the destructive interference between two high-level states of the \isotope{Ar}{40} nucleus. This cross section ``anti-resonance'' is about \SI{10}{\keV} wide, and 57 keV neutrons consequently have a scattering length of 859 m. For neutrons moderated to this energy the DUNE \dword{lartpc} is essentially transparent.
The 57 keV neutrons that do scatter quickly leave the anti-resonance and thermalize, at which time they capture. Each neutron capture releases exactly the binding energy difference between \isotope{Ar}{40} and \isotope{Ar}{41}, about \SI{6.1}{\MeV} in the form of gamma rays.  



The fixed, shielded deuterium-deuterium ($DD$) neutron generator would be located above a penetration in the hydrogenous insulation. Of order \SI{100}{$\mu$s} pulse width commercially available $DD$ generators exist that are about the size of a thermos bottle, and are cost competitive. Between the generator and the cryostat, layers of water or plastic and intermediate fillers will be included for sufficient degradation of the neutron energy. Initial simulations indicate that a single neutron injection point would illuminate the entire volume of one of the \dword{protodune} detectors and would be rapid (likely less than 30 min). 

The remaining studies for the \dword{tdr} for the external neutron source include an assessment of the full design: degrader materials, shielding, and the space and mounting (weight) considerations above the cryostat. Detailed simulation studies to understand the neutron transport process will be performed. In addition, the neutron capture gamma spectrum is also being characterized. In Nov 2017, the ACED\cite{aced-svoboda} Collaboration took several hundred thousand neutron capture events at the DANCE\cite{Reifarth:2013xny} facility at LANSCE which will be used to prepare a database of the neutron capture gamma cascade chain. 

\subsection{External Muon Tracker} 
\label{sec:calibemt}

A external muon tracker (EMT), a dedicated fast tracking system, would provide track position, direction, and time information independent of TPC and PDS systems.

Rock muons from beam interactions in the rock surrounding the cryostat have similar energy  and angular  spectrum as CC \numu 
events. A nominal design of the EMT would cover the front face of the detector (approximately $14\textrm{m} \times 12\textrm{m}$) to provide an estimate of the initial position, and the time for a subset of these events, independent of the TPC and PDS systems. A second, similarly sized panel, \SI{1}{\m} away from the cryostat would provide directional information. 
Additional measurements are possible elsewhere in the detector if the system is portable; it could be positioned on top of the cryostat to capture (nearly downward-going) cosmic rays during commissioning, or positioned along the side for rock muon-induced tracks along the drift direction. The EMT  pixelization will be small enough that rock-muon statistics will allow  determination of the center of each pixel to the same resolution as that expected for the detector (roughly \SI{1}{\cm}). So, for example, even with \SI{50} one-\si{\cm}-sized pixels,  with about \num{1000} rock muons per year passing through the EMT, the achievable precision  on average  for the incident position (before subsequent multiple scattering) is about \SI{5}{\milli\m}. 






The remaining studies for the EMT system prior to the \dword{tdr} include continued study of the precision with which the EMT (including panels on the sides and bottom) can determine biases or other problems with the detector model. Optimization of EMT size and pixelization and possible cost-saving options including re-use of existing scintillators (e.g. MINOS) or counter systems (e.g., \dword{protodune} or SBN) will be investigated. The available space for the EMT around the cryostat needs more investigation. A plan for surveying the EMT relative to the \dwords{apa} also needs to be developed, to be coordinated with the APA consortium; error in such a survey could be misinterpreted as a reconstruction bias. 



\subsection{Configuration of Proposed Systems} 
\label{sec:FTs}

The current cryostat design for the 
\spmod with multi-purpose cryostat penetrations for various sub-systems is shown in Figure~\ref{fig:ftmap}. The penetrations dedicated for calibrations are highlighted in black ovals. The placement of these penetrations is largely driven by the ionization track laser and radioactive source system requirements but would be usable for the neutron system as well.  The ports that are closer to the center of the cryostat are placed near the \dwords{apa} (similarly to what is planned for SBND) to minimize any risks due to the \dword{hv} discharge. For the far east and west ports, \dword{hv} is not an issue as they are located outside the \dword{fc} and the penetrations are located near mid-drift to meet radioactive source requirements. 


\begin{dunefigure}[Top view of the \dword{spmod} 
cryostat showing various penetrations]{fig:ftmap}{
Top view of the \dword{spmod} cryostat showing various penetrations. 
Highlighted in black ovals are multi-purpose calibration penetrations. 
The green ports are TPC signal cable penetrations. 
The orange ports are \dword{dss} penetrations. The blue ports are cryogenics penetrations.
The larger purple ports at the four corners of the cryostat are human access tunnels.}
\includegraphics[height=2.0in]{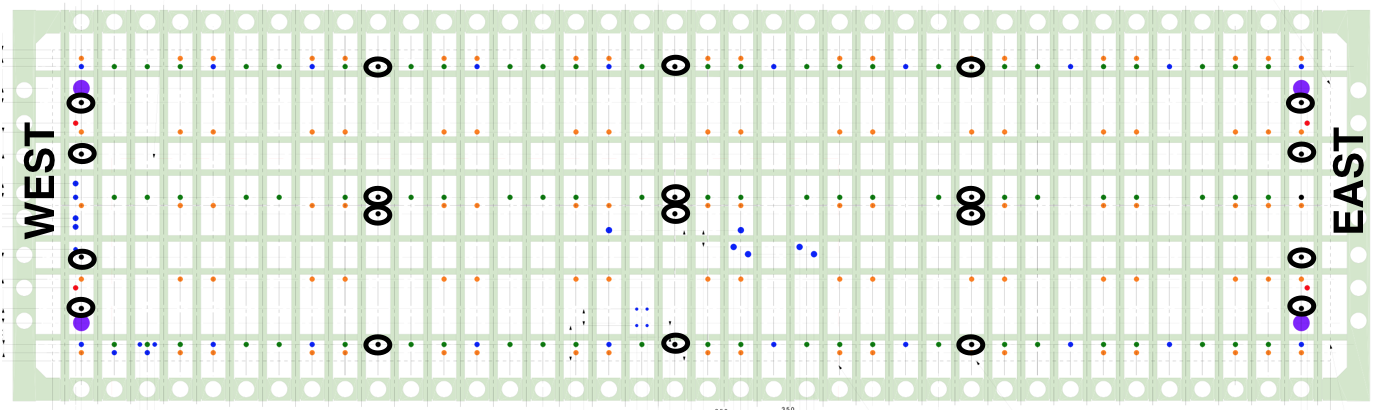}
\end{dunefigure}

Implementation of the ionization laser system, proposed in Section~\ref{sec:laser}, requires 20 feedthroughs to cover the four TPC drift volumes; this arrangement would provide (almost) full volume calibration of the electric field and associated diagnostics (e.g. HV). The crossing laser tracks are necessary to unambiguously construct the field map. A steerable plastic insulator laser head and fiber interface would be mounted on top of the cryostat in the feedthrough. Two options are under investigation: (1) the \dword{fc} (but not the ground plane or the active volume) is penetrated, and (2) the \dword{fc} is not penetrated. In the former case, the \dword{fc} penetration has been shown to create a small distortion to the \efield, for the benefit of full volume \efield mapping. When the \dword{fc} is not penetrated, the laser shines through the \dword{fc} tubes, producing some regions that are not mappable by the laser, and it will not be possible to map the position of the track start making the analysis more difficult. This is the case for laser system which use the far east and west ports. The necessity of penetrating the \dword{fc} has not been fully assessed yet. The \phel system would employ fixed fibers, and would not require a steering mechanism or mechnical FC penetrations.

The distance between any two consecutive feedthrough columns in Figure~\ref{fig:ftmap} is about \SI{15}{\m}, a plausible distance for the laser beam to travel. The maximum distance light would travel to the bottom corner of the detector, would be approximately \SI{20}{\m}.  Direct-ionization tracks have been demonstrated at a maximum possible distance in \microboone of \SI{10}{\m}. The Rayleigh scattering of the laser light is about \SI{40}{\m}, but additional optics effects, including self-focusing (Kerr) effects may limit the maximum practical range. Assuming these are not a limitation, this laser arrangement could illuminate the full volume with crossing track data. It is important to note that at this point in time, a maximum usable track length is unknown and it is possible that the full \SI{60}{\m} \detmodule length could be achieved by the laser system after optimization.

The calibration group focused on finalizing the cryostat penetrations for the \spmod driven by the cryostat design timeline. A similar exercise will be done to finalize \dpmod penetrations for calibrations in the near future.


\section{Summary} 
\label{sec:calibsum}

The physics requirements for the broad DUNE physics program places stringent requirements on calibration systems and sources. The aim of the upcoming TDR will be to demonstrate that the proposed calibration systems, in conjunction with existing sources, external measurements and monitors, will be sufficient for DUNE's calibration requirements. The proposed systems discussed here have been identified as important to DUNE's physics program. However, the multi-purpose ports also enable deployment of other possible calibrations systems in the future.

The calibration group has started establishing relevant connections to physics working groups and consortia as appropriate. The calibration group has started collaborating with the \lbl physics group to develop the necessary tools and techniques to propagate detector physics effects into oscillation analyses and similar effort will occur to connect with other physics groups. The calibration group has also started working closely with the consortia and identified liaisons for each 
to ensure that calibration needs are strongly considered as each consortium develops its designs. For example, a preliminary list of DAQ calibration requirements are already in process and  will be finalized in the coming months.

The calibration systems for DUNE, as presented in this document, will be further discussed and developed for the TDR within DUNE's management 
structure. Two options are currently favored for 
calibration, (1) formation of a new calibration consortium, or (2) inclusion of calibration in the \dword{cisc} consortium. This decision will depend on the scope of the proposed calibration systems presented in this document. The goal is to make and execute this decision in June 2018, shortly after the final \dword{tp} submission. 
The full design development for each calibration system, along with costing and risk mitigation, will follow. 

\begin{dunetable}[Key calibration milestones leading to first detector installation]{ll}{tab:TDRsteps}{Key calibration milestones leading to first detector installation.}
Date & \textbf{Milestone}\\ \toprowrule
May 2018 & \dword{tp} \\ \colhline
June 2018 & Finalize process of integrating calibration into consortium structure\\ \colhline
Jan. 2019 & Design validation of calibration systems using \dword{protodune}\slash SBN data  \\
&(where applicable) and incorporate lessons learned into designs \\ \colhline
Apr. 2019 & Technical design report \\ \colhline
Sep. 2022 & Finish construction of calibration systems for Cryostat \#1 \\ \colhline
May 2023 & Cryostat 1 ready for TPC installation \\ \colhline
Oct. 2023 & All calibration systems installed in Cryostat \#1 \\
\end{dunetable} 


\cleardoublepage







\cleardoublepage


\cleardoublepage
\printglossaries

\cleardoublepage
\cleardoublepage
\renewcommand{\bibname}{References}

\bibliographystyle{utphys} 
\bibliography{common/tdr-citedb}

\end{document}